\documentclass[reprint,amssymb, amsmath,aps,pre,cha,twocolumn]{revtex4-1}
\usepackage{CJK}
\usepackage{lineno,hyperref}

\usepackage{ulem}
\usepackage{color}
\usepackage{float}
\usepackage{graphicx}
\usepackage{subcaption}
\usepackage[autostyle]{csquotes}
\usepackage{epsfig}
\begin{document}
\begin{CJK*}{GB}{} 
\title{Observation of ion acceleration in nanosecond laser generated plasma on a nickel thin film under forward ablation geometry }%
\author{Jinto Thomas}%
\email{jinto@ipr.res.in}
\affiliation{Institute For Plasma Research,HBNI, Bhat, Gandhinagar,Gujarat, India, 382428}%
\author{Hem Chandra Joshi}
\affiliation{Institute For Plasma Research,HBNI, Bhat, Gandhinagar,Gujarat, India, 382428}%
\author{ Ajai Kumar}
\affiliation{Institute For Plasma Research,HBNI, Bhat, Gandhinagar,Gujarat, India, 382428}%
\author{Reji Philip}
\affiliation{Raman Research Institute, C.V. Raman Avenue, Sadashivanagar,
Bangalore 560080, India}%
\date{\today}

\begin{abstract}

 In this article we report acceleration observed for ions in a forward ablation geometry of 50 nm
  thick nickel film coated on a glass plate under nanosecond laser ablation. 
A detailed study with varying background pressure and laser energy of time of flight (TOF) spectra of ionic and other neutral transitions from plasma is undertaken. The results indicate that TOF spectra recorded for different neutral transitions exhibit different dynamical behavior (the slower peak of the neutral emission becomes faster with background pressure). Similar observation is seen for the ionic species as well. 
 These observations have similarity to some of the reported double-layer concepts. However, the calculated electric fields from the acceleration appear to be anomalously higher.
 Along with the acceleration observed for the ionic and neutral lines, significant asymmetry in spectral shape is observed for neutral nickel line (712.2 nm)  along with spectral broadening  as the background pressure increases. This large asymmetry is indicative of micro electric fields present inside the laser produced plasma plume which may result in a continuous acceleration of ions. Interestingly, the asymmetry in spectral broadening exhibits temporal and spatial dependence which indicates that electric field is present in the plasma plume even for longer duration and also  over a significant distance. 

\end{abstract}

\maketitle
\end{CJK*}

Interaction of nanosecond laser pulse  and the formation and evolution of plasma plume  has been extensively studied by a number of groups \cite{Farid_pr_dpendence,Amoruso_1999, RPSandTheraja, DLEBulgakova}.
Interaction of intense laser pulse with ultra thin films resulting in acceleration of protons and ions has been experimentally demonstrated and simulated effectively by different groups
\cite{ion_accela_film_pic, ion_accelaration_ps_foil, Capacitance_Accelaration, Ion_accelaration_Skar, Proton_Accelaration_film}. Various acceleration mechanisms e.g.  target normal sheath acceleration (TNSA) and radiation pressure acceleration (RPA) are currently attracting a substantial amount of experimental and theoretical attention.  Interaction of high power laser with a foil  resulting in collimated plasma ejection from the rear of thin foil target was reported by Kar et al \cite{plasmajet_kar}. All these acceleration mechanism of ions were reported for higher laser intensities ($ \geq 10^{19} w/cm^2$) for a few cycles of laser pulses. \par

 In this article we report studies on ns laser produced plasma evolution using optical emission and electrical probe in forward ablation geometry on nickel film.
 Acceleration is observed for the ionic species. However,  the observation of spectral asymmetry  at later times using a high resolution spectrometer with ICCD  suggests a continuous evolution of micro electric fields, which may be responsible for such an acceleration of plasma species. The dependence of background pressure and laser fluence on the observed acceleration is also reported in detail.
\section{Experimental setup}
\begin{figure}[ht]
\includegraphics[trim=4cm 3cm 7cm 6cm, clip=true, scale=0.6, angle=0]{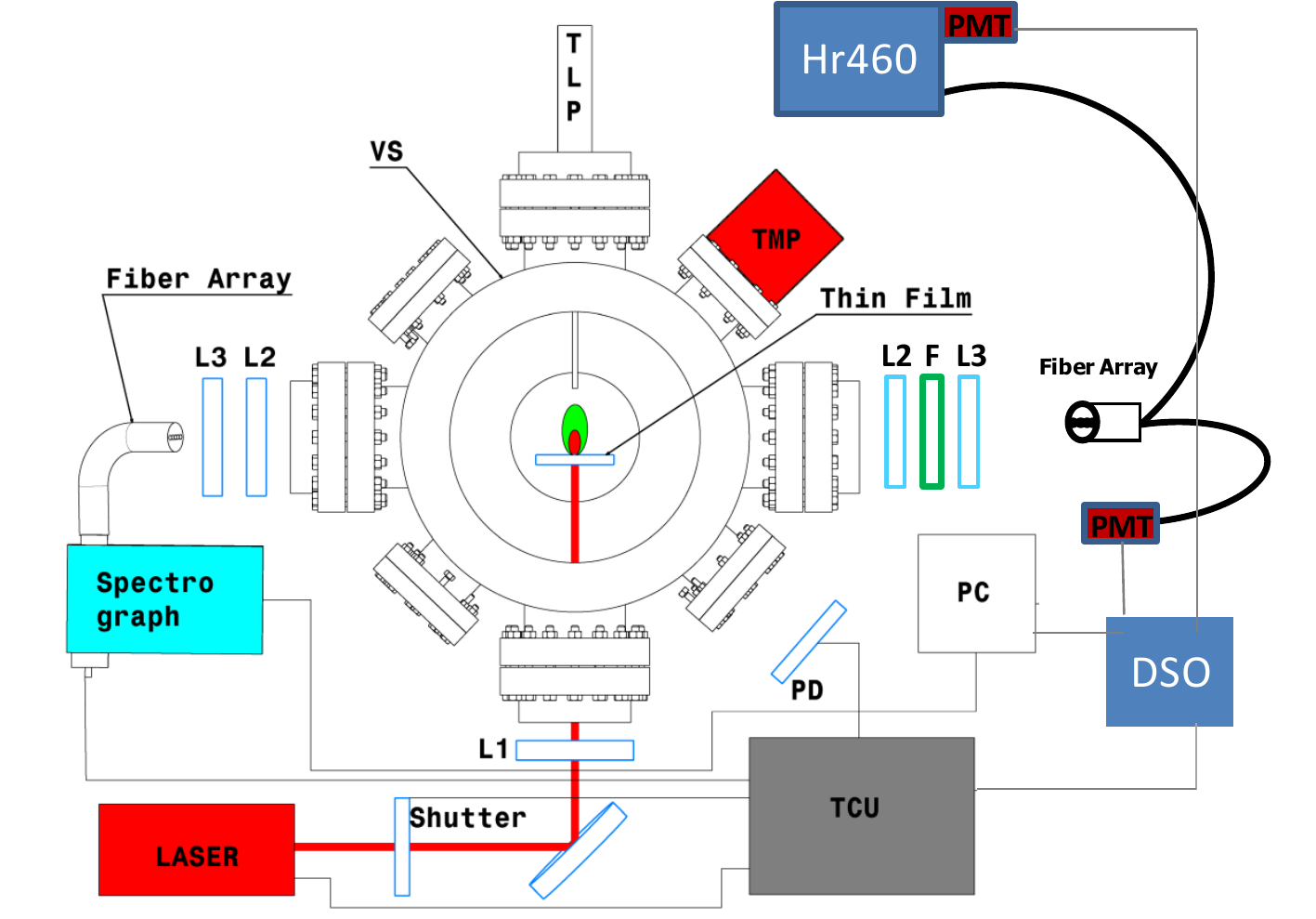}
\caption{\label{fig:expsetup} Schematic diagram of  experimental setup showing the arrangement of sample (thin film), triple langmuir probe (TLP) and laser system aligned to the vacuum vacuum system (VS). TMP is the turbo molecular pump, L1 is the focusing lens, L2 \& L3 are the lens system to image the plasma plume to fiber array. F is the band pass interference filter, Hr460 is the high resolution spectrometer, PMT is the fast photomultiplier Tube, DSO is the fast Digital Storage Oscilloscope, TCU is the trigger and control unit which synchronizes the instruments with laser pulses and the data are acquired on a personal computer (PC) }
\end{figure}

The experimental set up is similar to the one described in earlier work \cite{Jinto_POP2018} and is shown in figure~\ref{fig:expsetup}. The laser  used for this experiment is 1064 nm Nd:YAG laser with 10 ns pulse width. The laser beam, sample manipulator, vacuum system, gas feed, etc. are the same. In this experiment, an imaging system with magnification 1 is used to image the plasma plume into an optical fiber array with 10 branches of fibers having core diameter of 400 micron.
 The array is placed in a manner so that each individual fiber collects emission from various spatial locations from the sample along the axis of the laser beam. Anyone of the  optical fibers from this array can be coupled to a PMT so that recording of emission from different positions within the plasma is possible on a shot to shot basis.
 Narrow band pass interference filters (F as shown in figure) with pass band for prominent neutral nickel lines are used to allow the desired line emission to the PMT. PMT records the temporal emission of respective neutral lines from the respective locations using a fast digital storage oscilloscope(DSO).  Any of the remaining fibers from the fiber array can be coupled to another spectrometer(HR460) with PMT.
  This spectrometer and PMT are used to detect ionic spectral lines from the plasma. The spectral resolution of this spectrometer with PMT is better than 0.1 nm.
   The spectrometer is arranged with an F number matching optics for SMA fiber and hence the remaining fibers can be used to measure emission from the desired locations inside the plasma. \par
   
A 1 meter spectrometer is also coupled with another fiber array with 10 fibers using  F number matching optics to record the spectra of plasma. 
The fiber array is placed at the image plane of another imaging system aligned diametrically opposite to the earlier mentioned fiber array so that emission from plasma along the propagation axis can be imaged to the spectrometer. 
The slit of the spectrometer is fixed at 50 microns so that an effective spectral resolution of 0.12 nm is achieved. The spectrometer with fiber array and ICCD enables acquiring spectra from different locations of plasma plume simultaneously and for different time delays on a shot to shot basis.\par

In addition to the spectroscopic diagnostics, a triple Langmuir Probe (TLP) is also used for the measurement of floating potential and ion current. TLP records the ion saturation current and floating potential at a distance of 10 mm onwards. It can be noted here that the emission spectrum from ions and neutrals is recorded at the locations closer to the sample whereas ion saturation and floating potential from 10 mm onwards as it is not possible to put the probe closer to the target. The signals from PMT's and TLP are recorded using the DSO which is synchronised with the Timing control unit (TCU) and interfaced to the PC.

\begin{figure}
\centering
\begin{subfigure}{0.4\textwidth}
\includegraphics[width=\textwidth, trim=0cm 0cm 0cm 0cm, clip=true,angle=0]{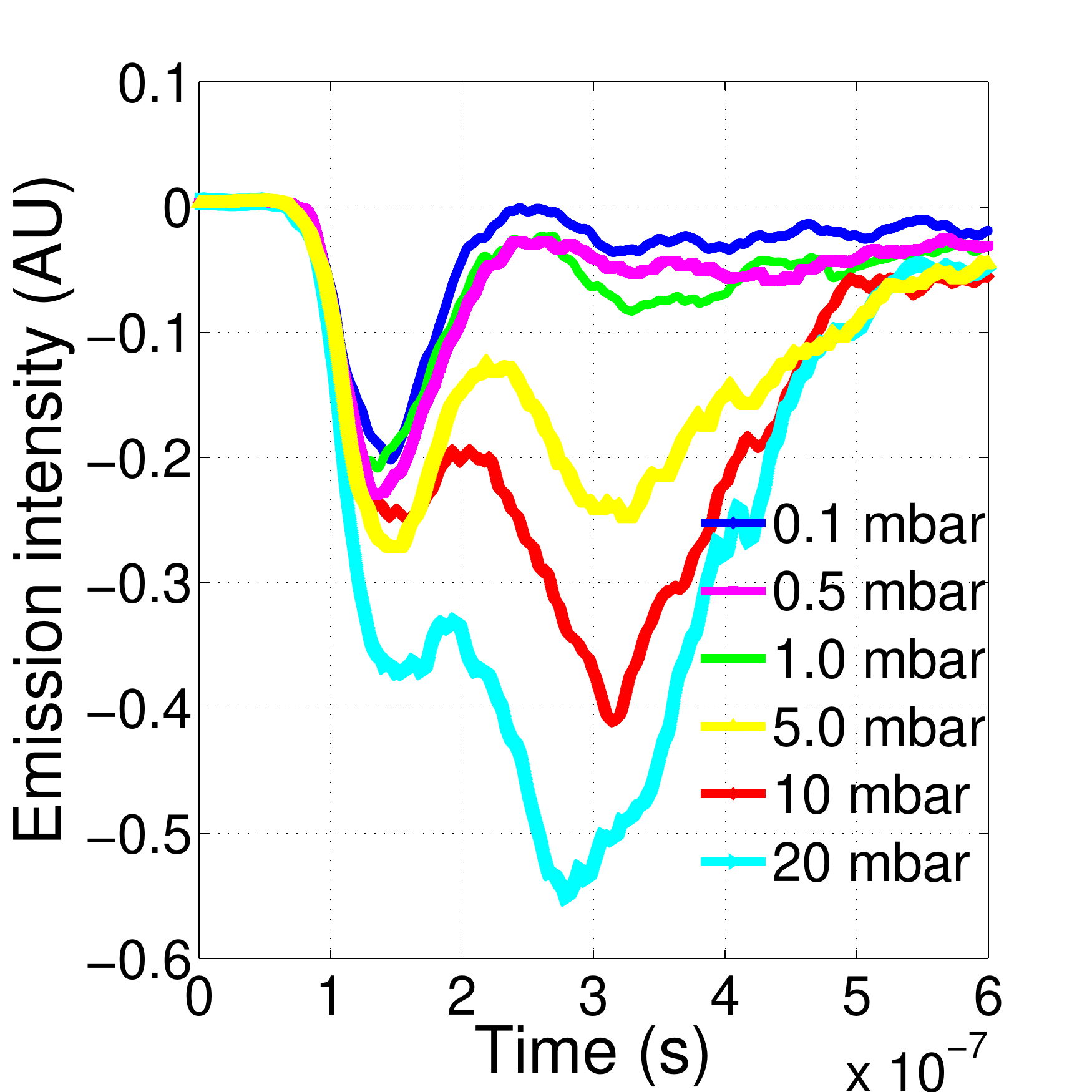}
\caption{\label{fig:DL_ion_Pr_var_1064nn_3mm_LE100mj}  Laser energy of 100 mJ.}
\end{subfigure}
\begin{subfigure}{0.4\textwidth}
\includegraphics[width=\textwidth, trim=0cm 0cm 0cm 0cm, clip=true,angle=0]{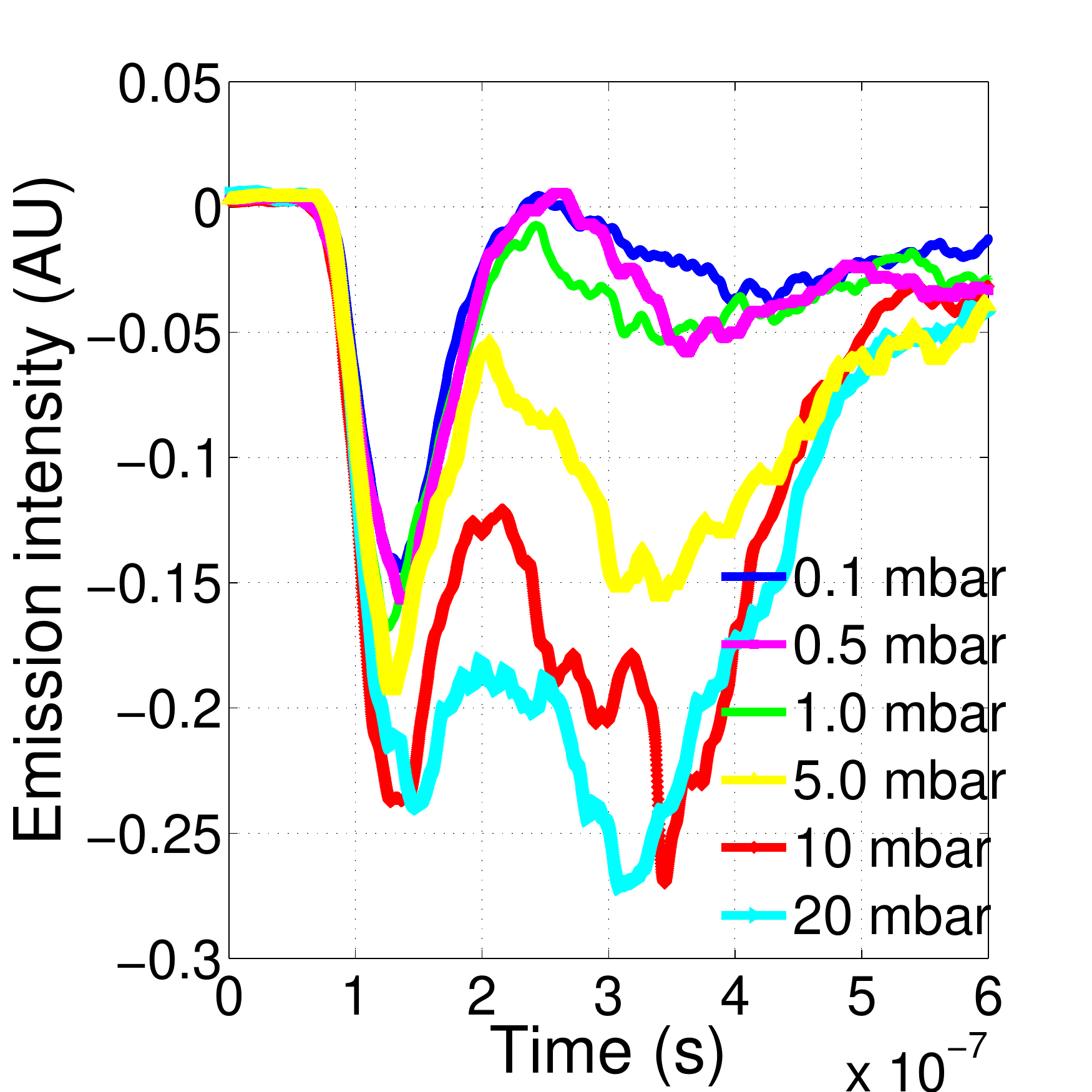}
\caption{\label{fig:DL_ion_Pr_var_1064nn_3mm_LE55mj}  Laser energy of 50 mJ.}
\end{subfigure}
\caption{\label{fig:DL_ion_Pr_var_1064nn_3mm_LE100_55mj}  Evolution of TOF spectrum of   ionic line (362.7 nm) at 3 mm from the sample for different background pressures for 10 ns, 1064 nm laser pulse at laser energies 100 mJ and  50 mJ.}
\end{figure}
\begin{figure}
\centering
\includegraphics[width=0.48\textwidth, trim=1cm 0.6cm 1cm 1cm, clip=true,angle=0]{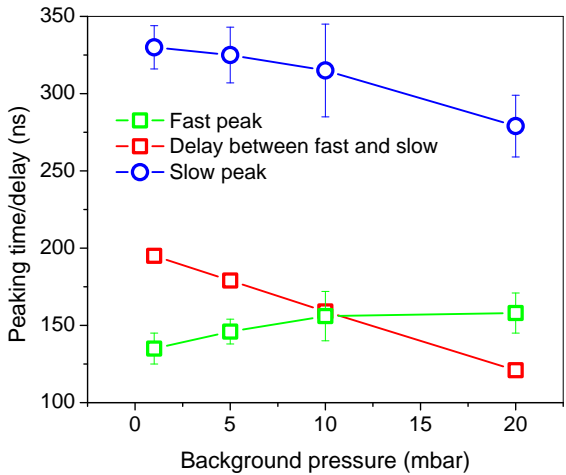}
\caption{\label{fig:DL_ionic_peaking_delay_3mm_100mj_1064nm}  Peaking time of fast and slow peaks of TOF spectra of  ionic species and the delay between fast and slow peaks with varying pressures for 1064 nm with 100 mJ of energy at 3 mm from sample.}
\end{figure}

\begin{figure*}
\centering
 \begin{subfigure}{0.32\textwidth}
\includegraphics[width=\textwidth, trim=0cm 0cm 0cm 0cm, clip=true,angle=0]{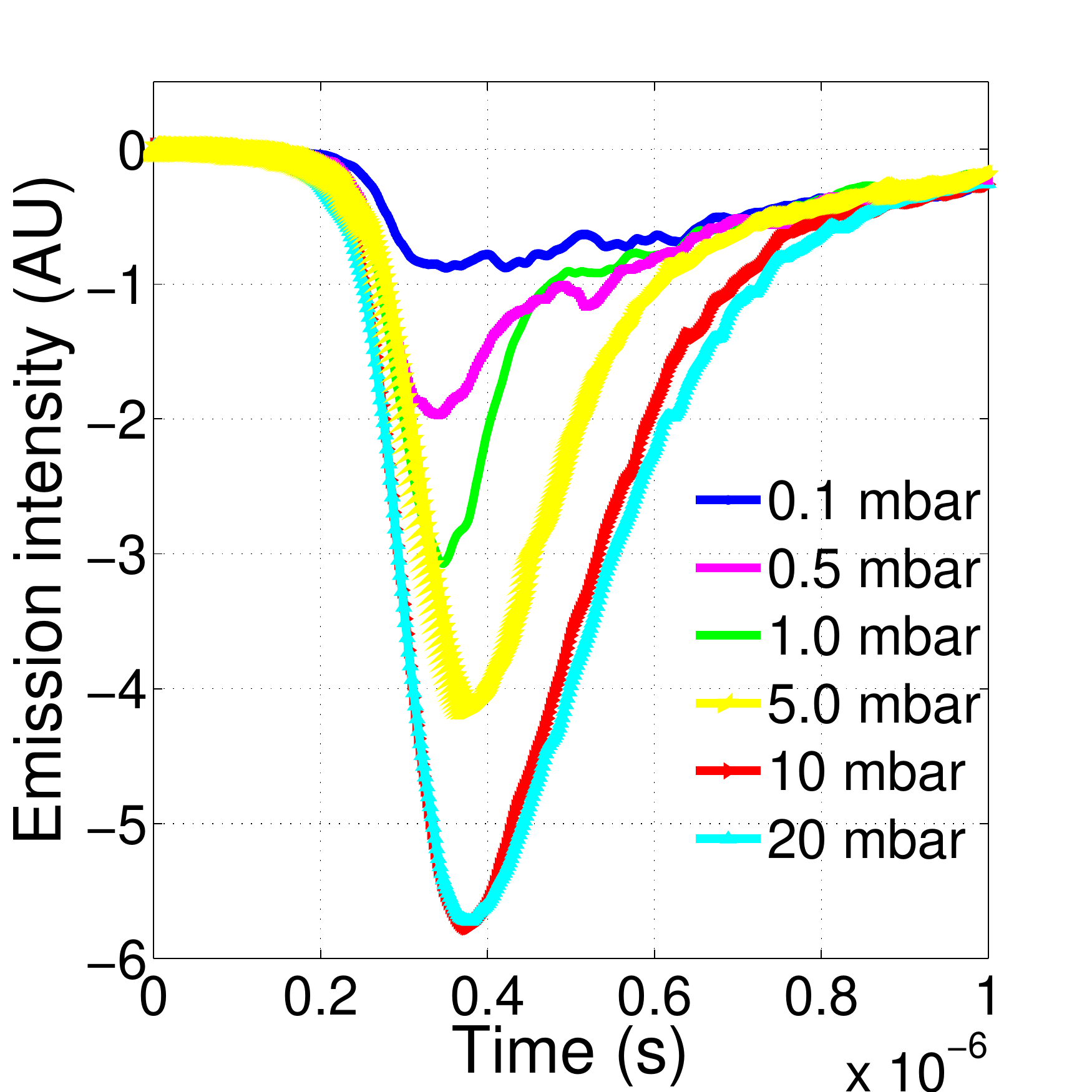}
\caption{\label{fig:DL_Neu_Pr_var_1064nn_3mm_LE100mj} 361.9 nm at 100 mJ ablation.}
\end{subfigure}
\begin{subfigure}{0.32\textwidth}
\includegraphics[width=\textwidth, trim=0cm 0cm 0cm 0cm, clip=true,angle=0]{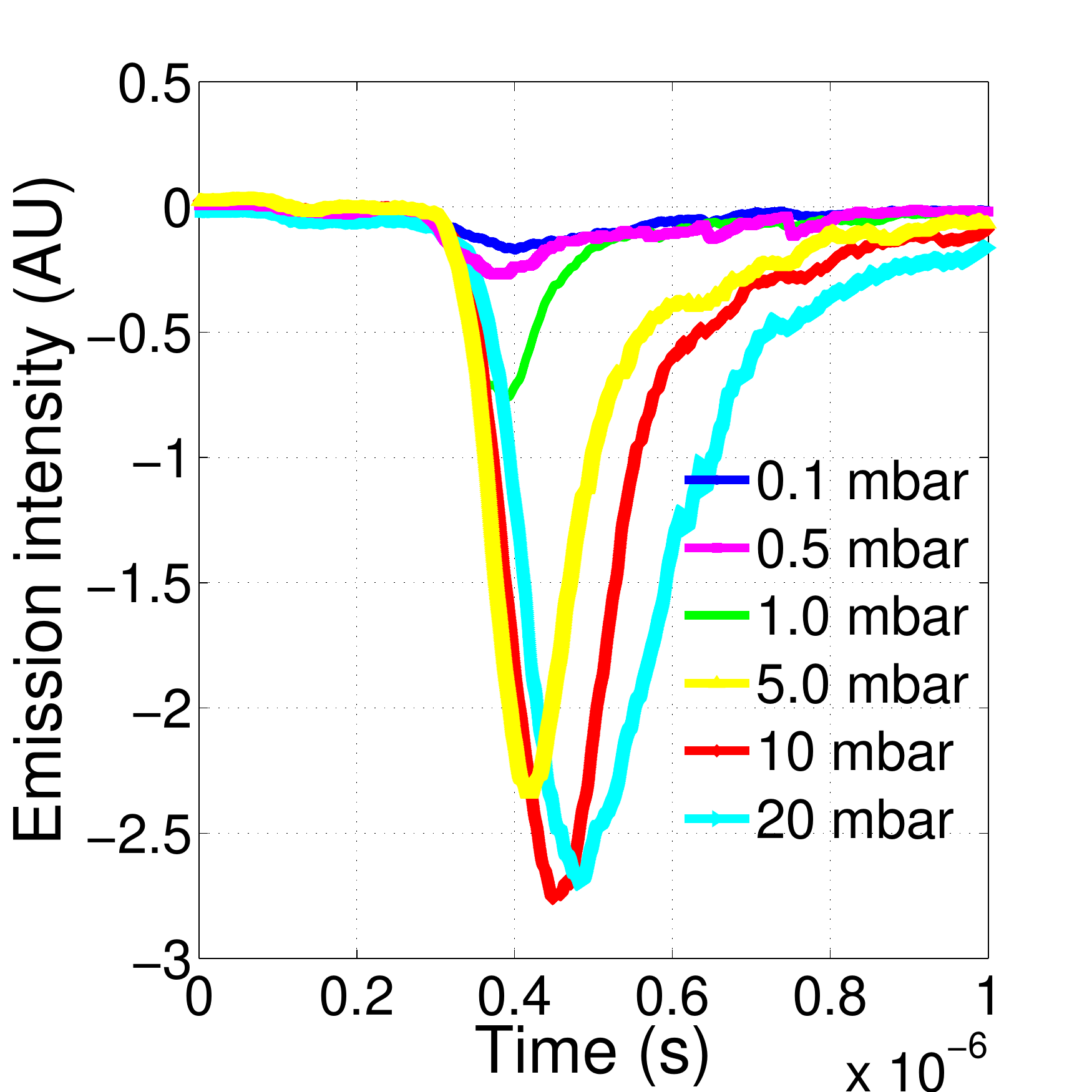}
\caption{\label{fig:DL_5080nm_Pr_var_1064nn_3mm_LE100mj} 508.0 nm at 100 mJ ablation.}
\end{subfigure}
\begin{subfigure}{0.32\textwidth}
\includegraphics[width=\textwidth, trim=0cm 0cm 0cm 0cm, clip=true,angle=0]{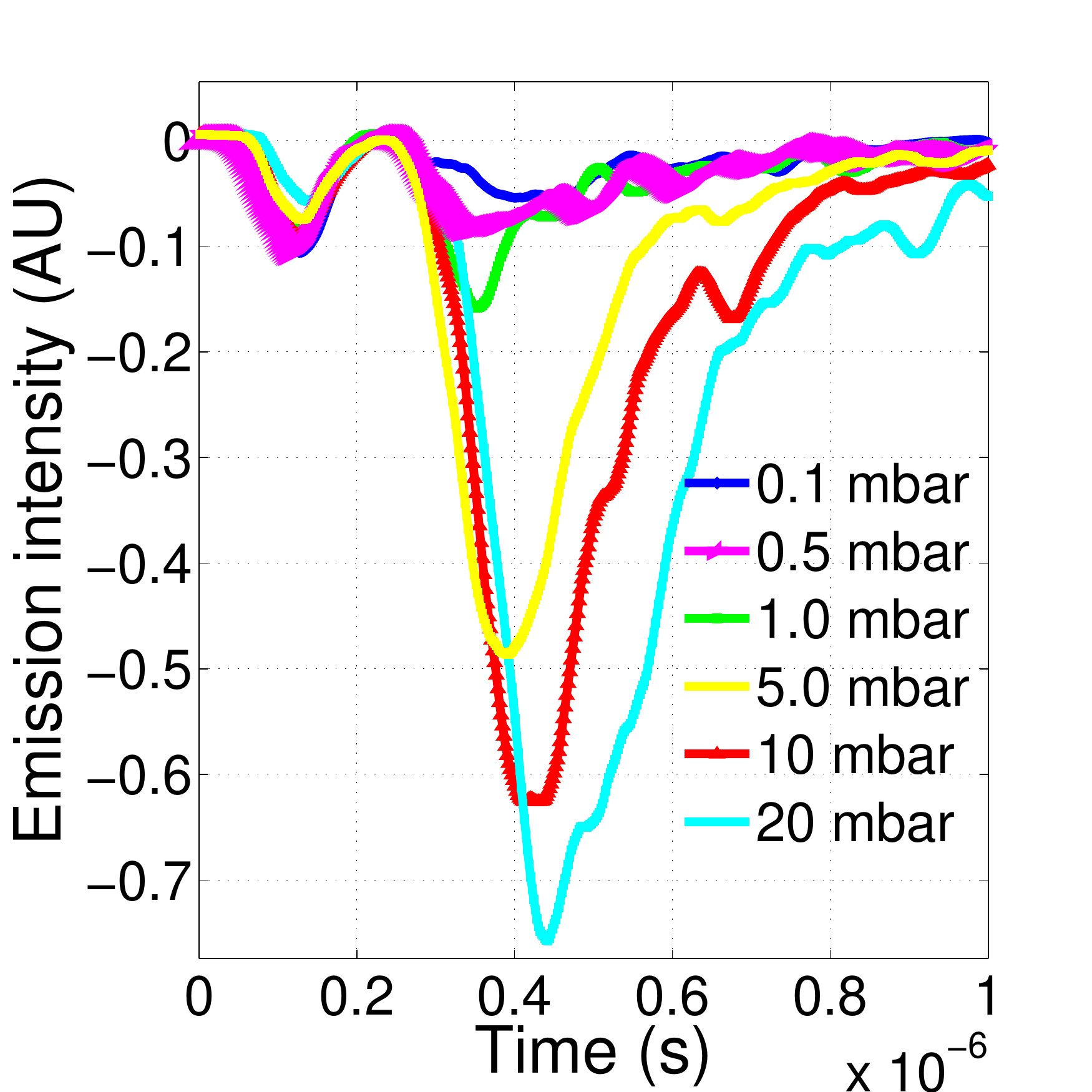}
\caption{\label{fig:DL_Neu_712_Pr_var_1064nn_3mm_LE100mj} 712.2 nm at 100 mJ ablation.}
\end{subfigure}
\begin{subfigure}{0.32\textwidth}
\includegraphics[width=\textwidth, trim=0cm 0cm 0cm 0cm, clip=true,angle=0]{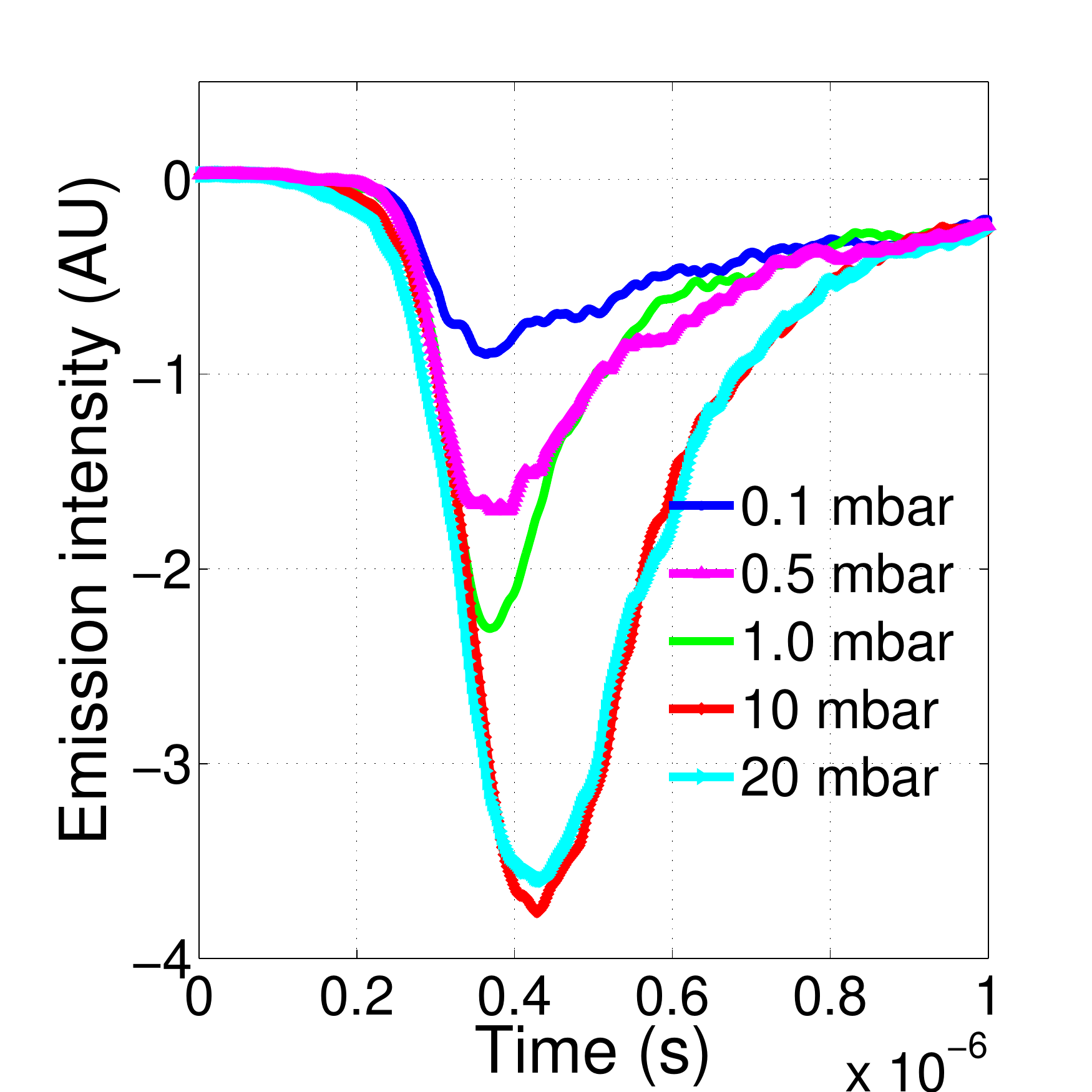}
\caption{\label{fig:DL_Neu_Pr_var_1064nn_3mm_LE55mj} 361.9 nm at 50 mJ ablation.}
\end{subfigure}
\begin{subfigure}{0.32\textwidth}
\includegraphics[width=\textwidth, trim=0cm 0cm 0cm 0cm, clip=true,angle=0]{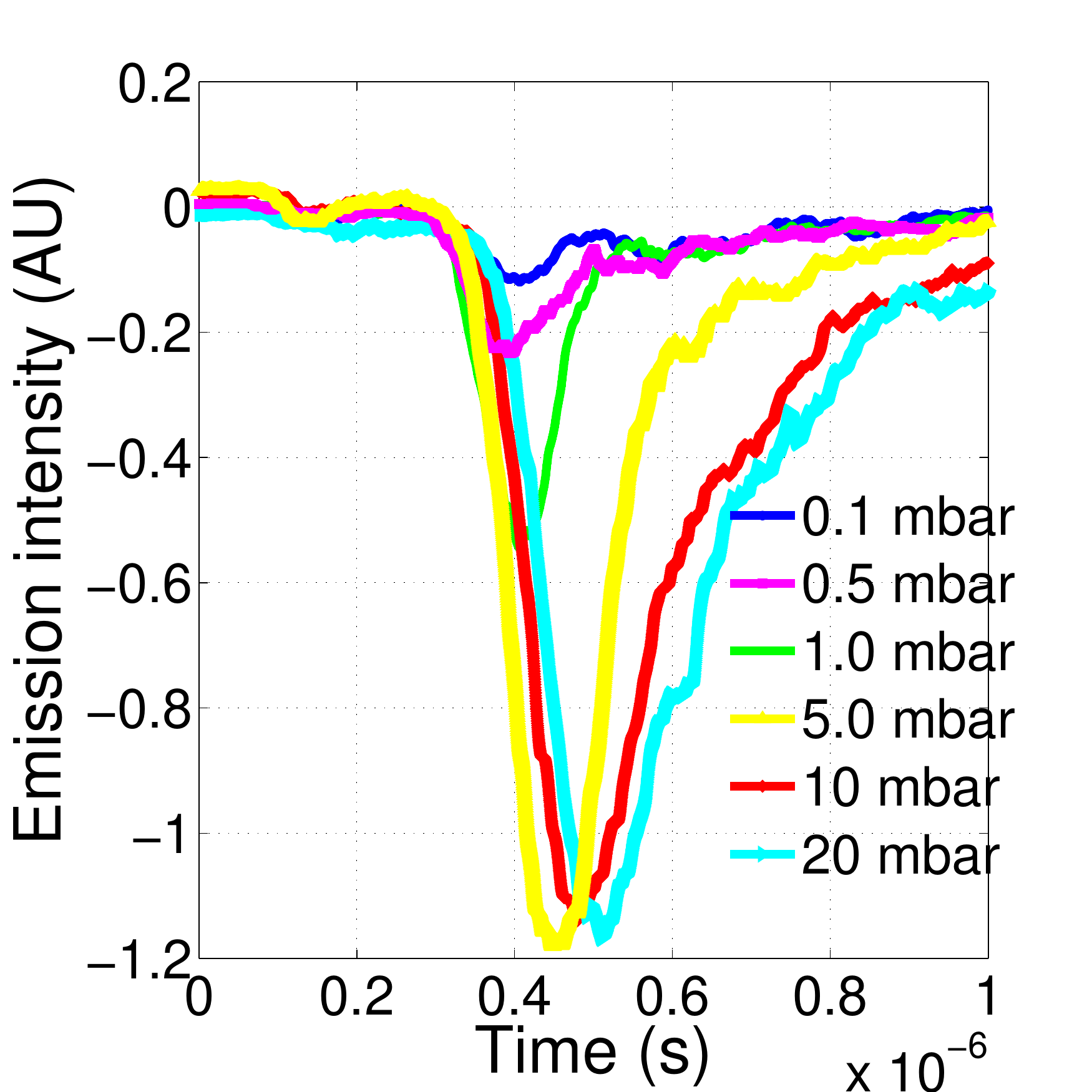}
\caption{\label{fig:DL_5080nm_Pr_var_1064nn_3mm_LE55mj} 508.0 nm at 50 mJ ablation.}
\end{subfigure}
\begin{subfigure}{0.32\textwidth}
\includegraphics[width=\textwidth, trim=0cm 0cm 0cm 0cm, clip=true,angle=0]{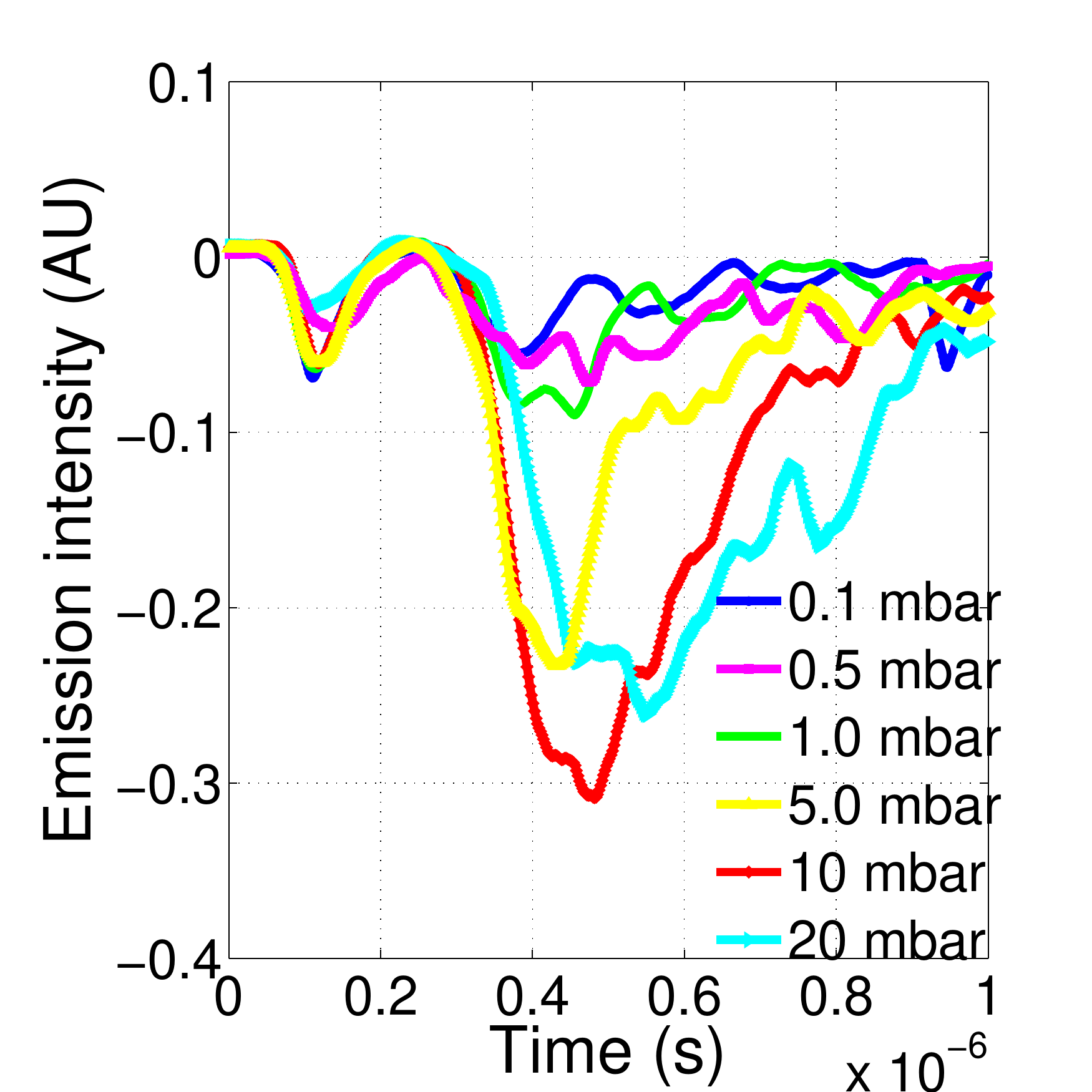}
\caption{\label{fig:DL_Neu_712_Pr_var_1064nn_3mm_LE55mj} 712.2 nm at 50 mJ ablation.}
\end{subfigure}
\caption{\label{fig:DL_Neu_Pr_var_1064nn_3mm_LE55_100mj}  Evolution of TOF spectrum of   neutral lines 3 mm away from the sample for different background pressures and two different laser energies (50 mJ and 100 mJ) of 10 ns, 1064 nm laser.}
\end{figure*}
\begin{figure}
\centering
\includegraphics[width=0.48\textwidth, trim=0cm 0cm 0cm 0cm, clip=true,angle=0]{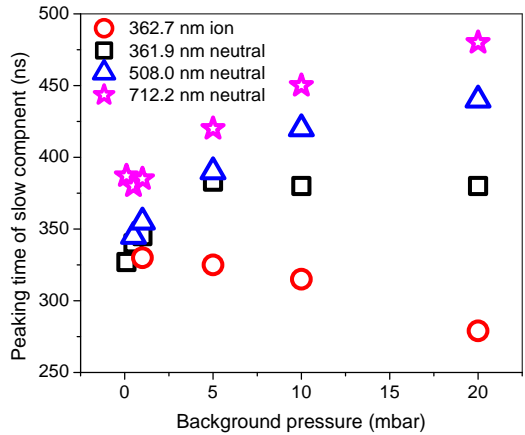}
\caption{\label{fig:DL_ionic_neutral_Slow_peaking_3mm_100mj_1064nm}  Comparison of peaking time of slow components of TOF spectra of  ionic and neutral species with varying background pressure for 1064 nm laser wavelength with 100 mJ of energy at 3 mm from the sample.}
\end{figure}
\begin{figure*} 
    \centering
    \begin{subfigure}{0.32\textwidth}
        \includegraphics[width=\textwidth,  trim=0cm 0cm 0cm 0cm, clip=true, angle=0]{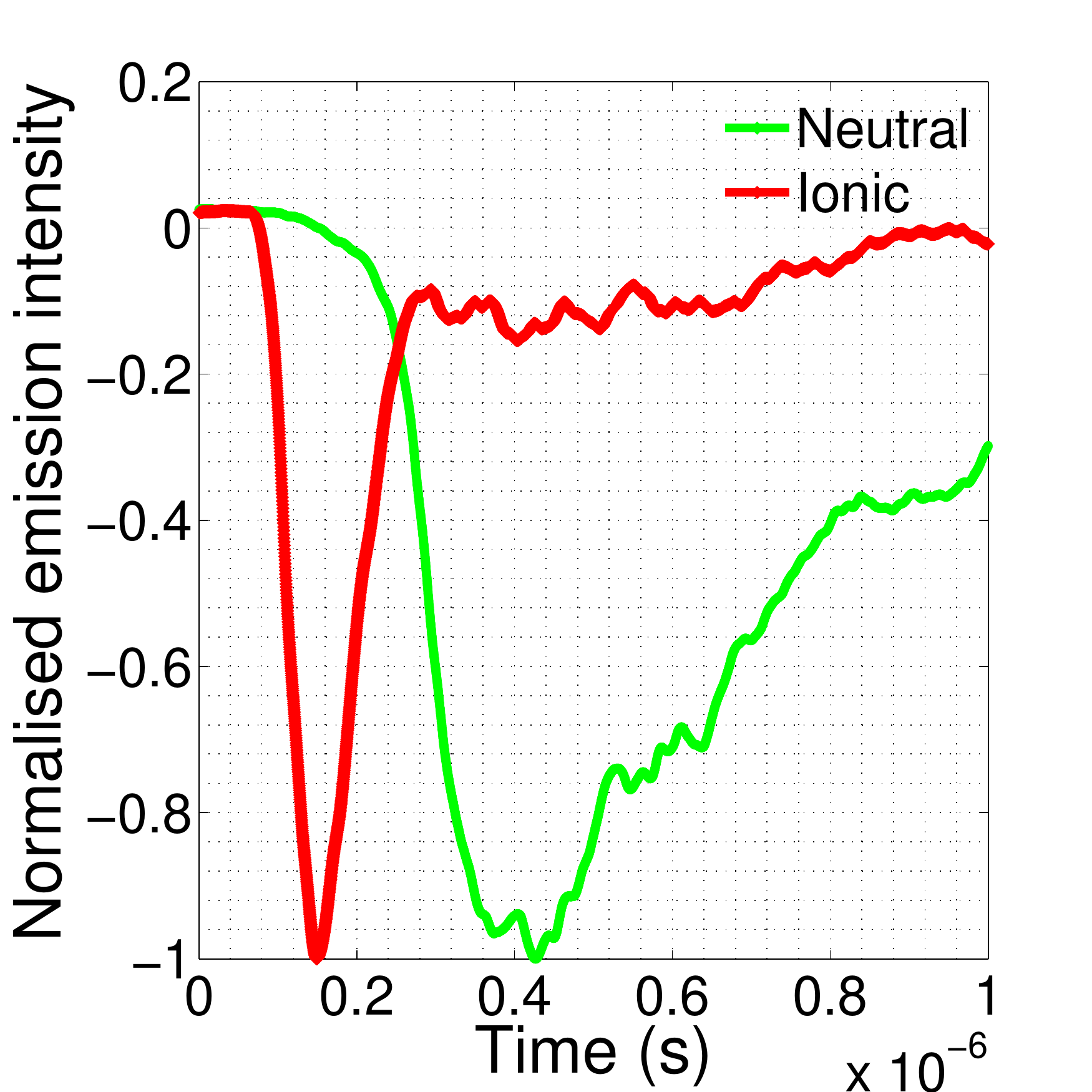}
          \caption{ 0.1 mbar 3 mm. }
        \label{fig:DL_Ion_Neu_3mm_Pr_1em1_1064nn_LE100mj}
    \end{subfigure}
    \begin{subfigure}{0.32\textwidth}
        \includegraphics[width=\textwidth,  trim=0cm 0cm 0cm 0cm, clip=true, angle=0]{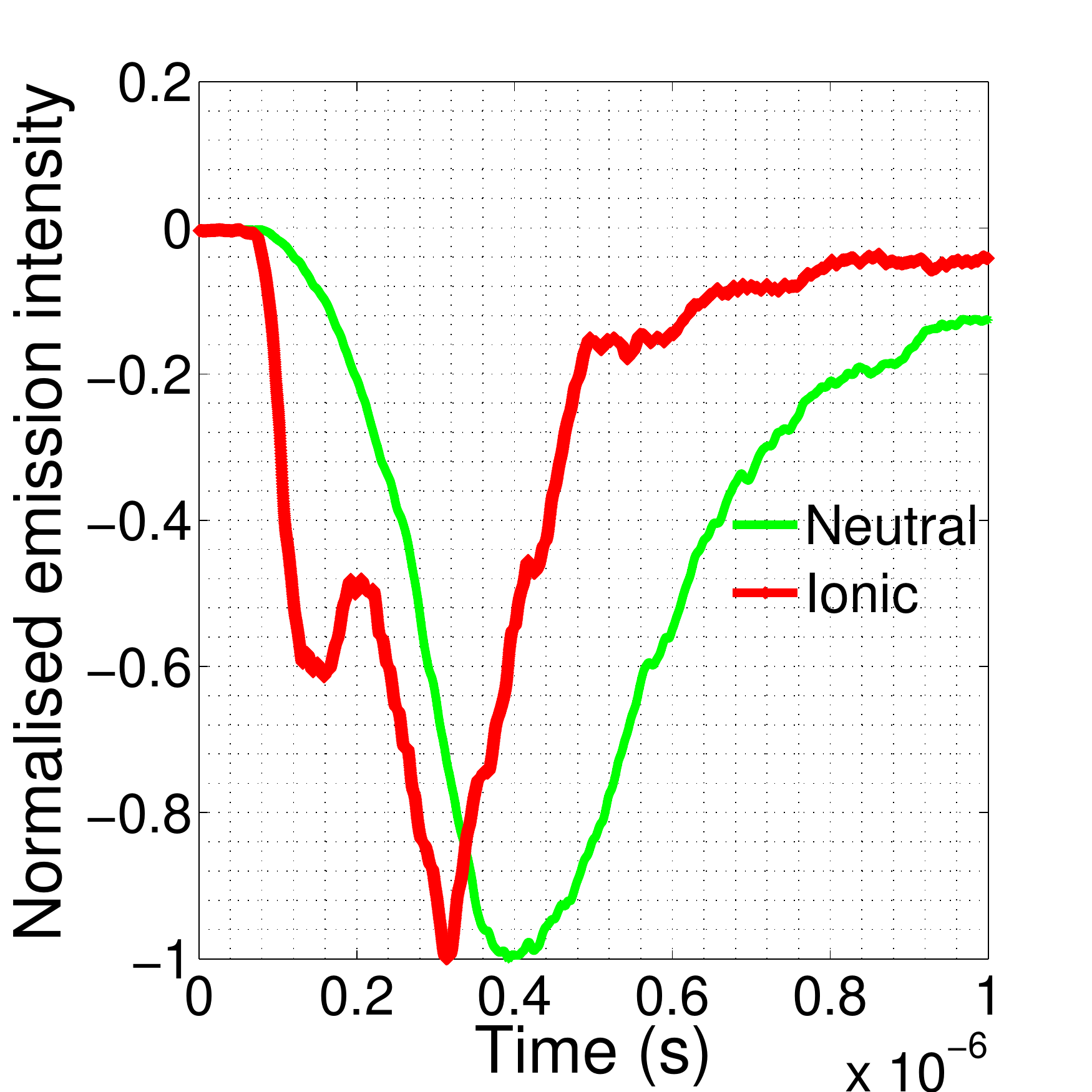}
          \caption{ 10 mbar 3 mm. }
        \label{fig:DL_Neu_ion_1064nn_3mm_1ep1mbar_LE100mj}
    \end{subfigure}
     \begin{subfigure}{0.32\textwidth}
        \includegraphics[width=\textwidth,  trim=0cm 0cm 0cm 0cm, clip=true, angle=0]{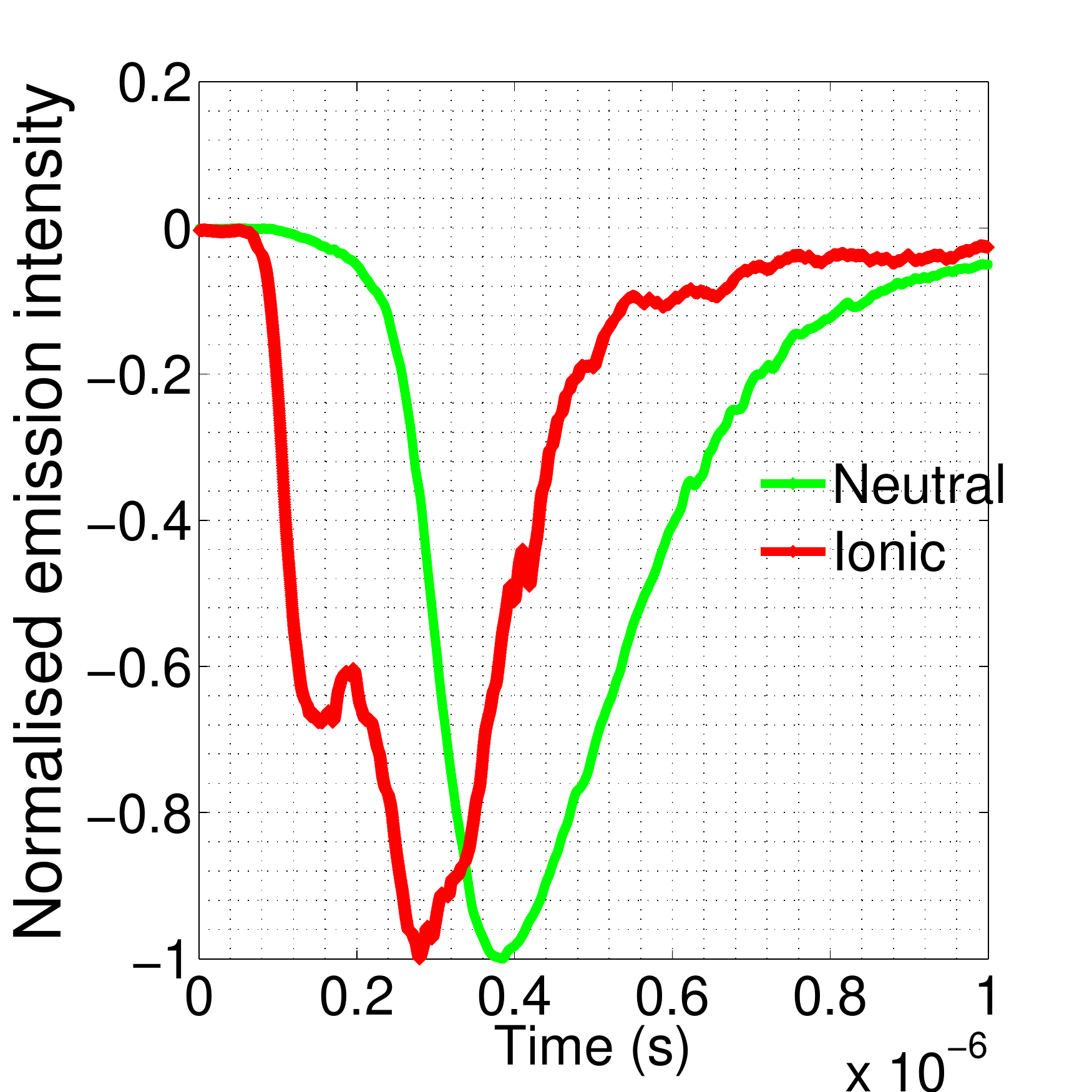}
          \caption{ 20 mbar 3 mm. }
        \label{fig:DL_Neu_ion_1064nn_3mm_2ep1mbar_LE100mj}
    \end{subfigure}
    \begin{subfigure}{0.32\textwidth}
        \includegraphics[width=\textwidth,  trim=0cm 0cm 0cm 0cm, clip=true, angle=0]{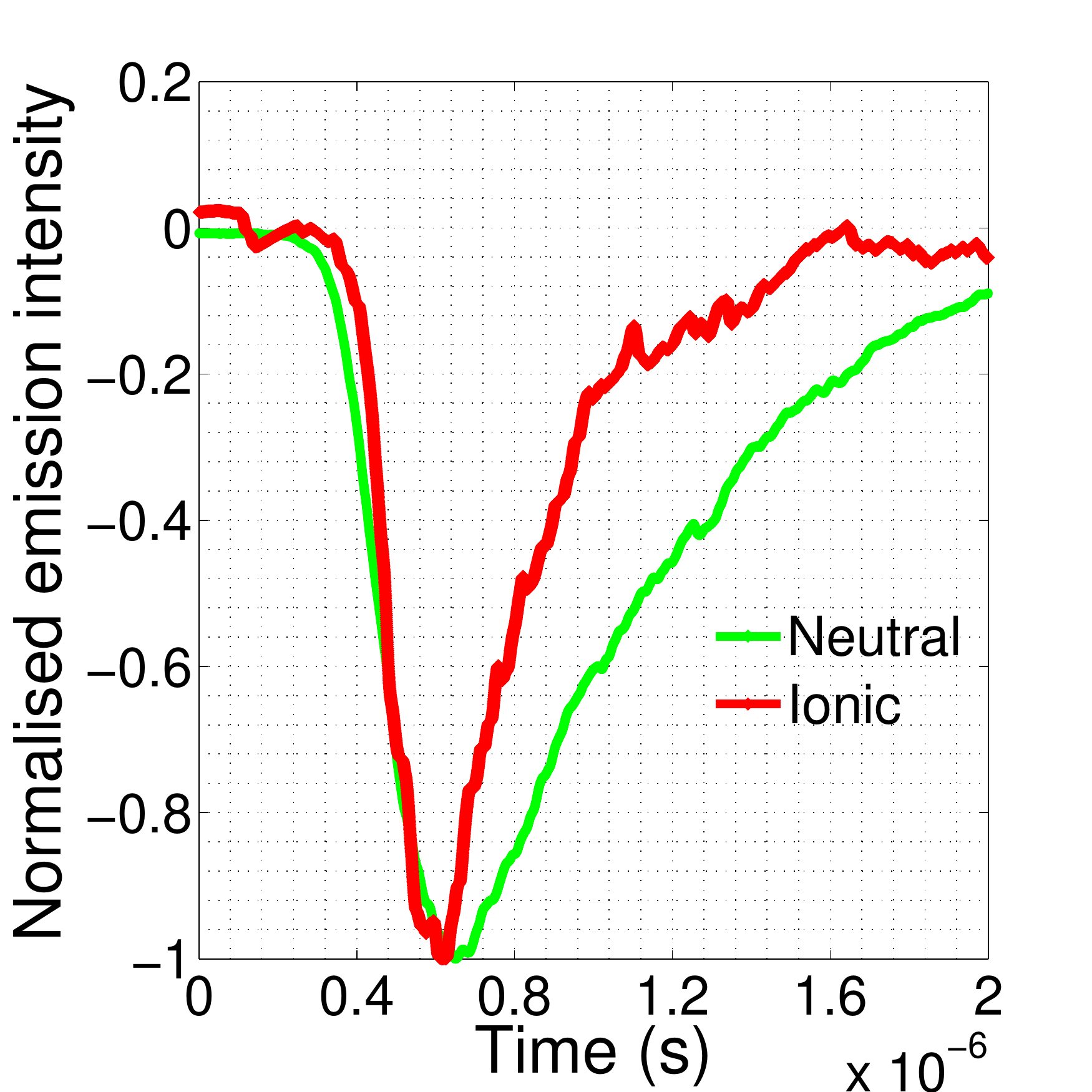}
        \caption{ 0.1 mbar 7 mm. }
        \label{fig:DL_Ion_Neu_7mm_Pr_1em1_1064nn_LE100mj}
    \end{subfigure} 
    \begin{subfigure}{0.32\textwidth}
        \includegraphics[width=\textwidth,  trim=0cm 0cm 0cm 0cm, clip=true, angle=0]{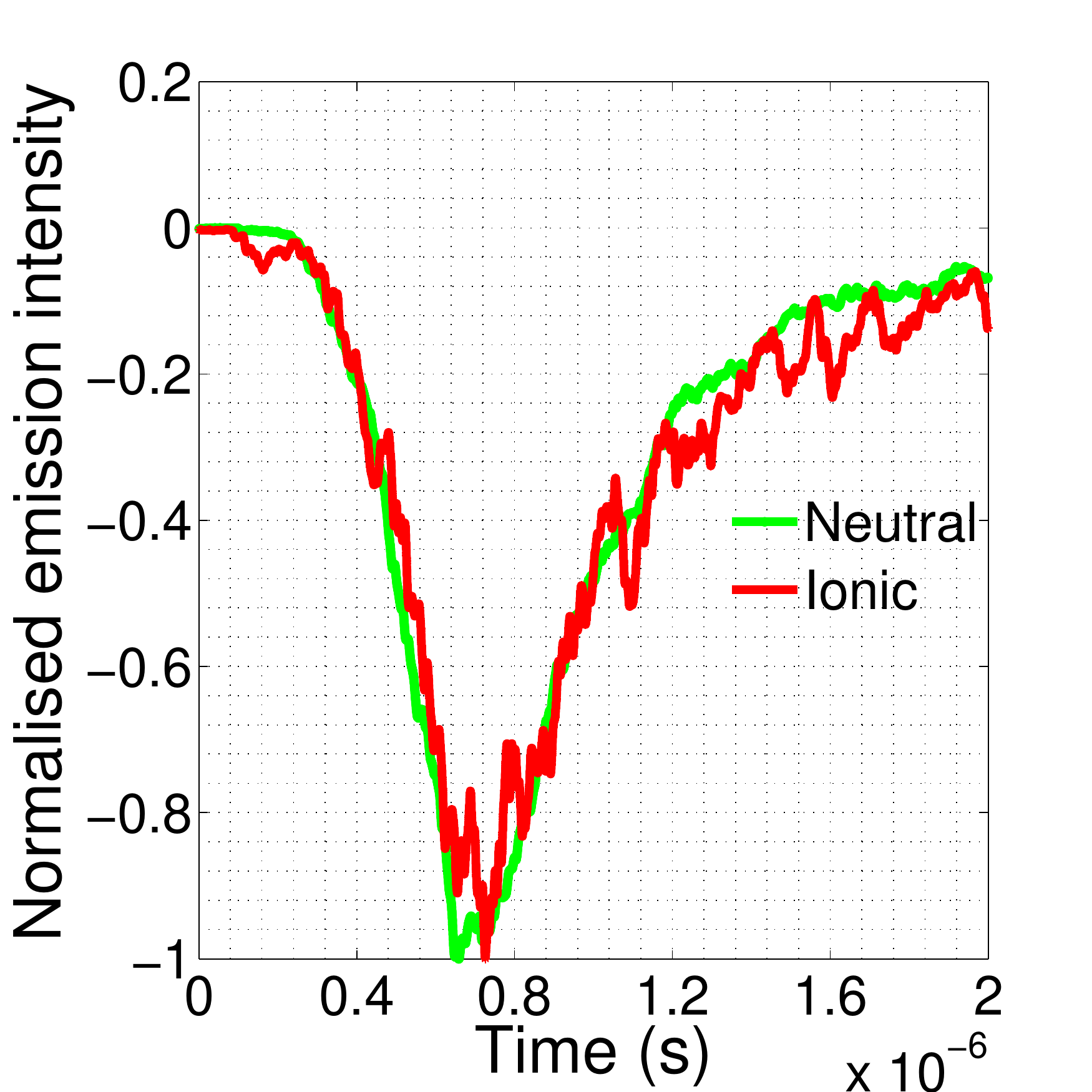}
        \caption{ 10 mbar 7 mm. }
        \label{fig:DL_Neu_ion_1064nn_7mm_1ep1mbar_LE100mj}
    \end{subfigure} 
    \begin{subfigure}{0.32\textwidth}
        \includegraphics[width=\textwidth,  trim=0cm 0cm 0cm 0cm, clip=true, angle=0]{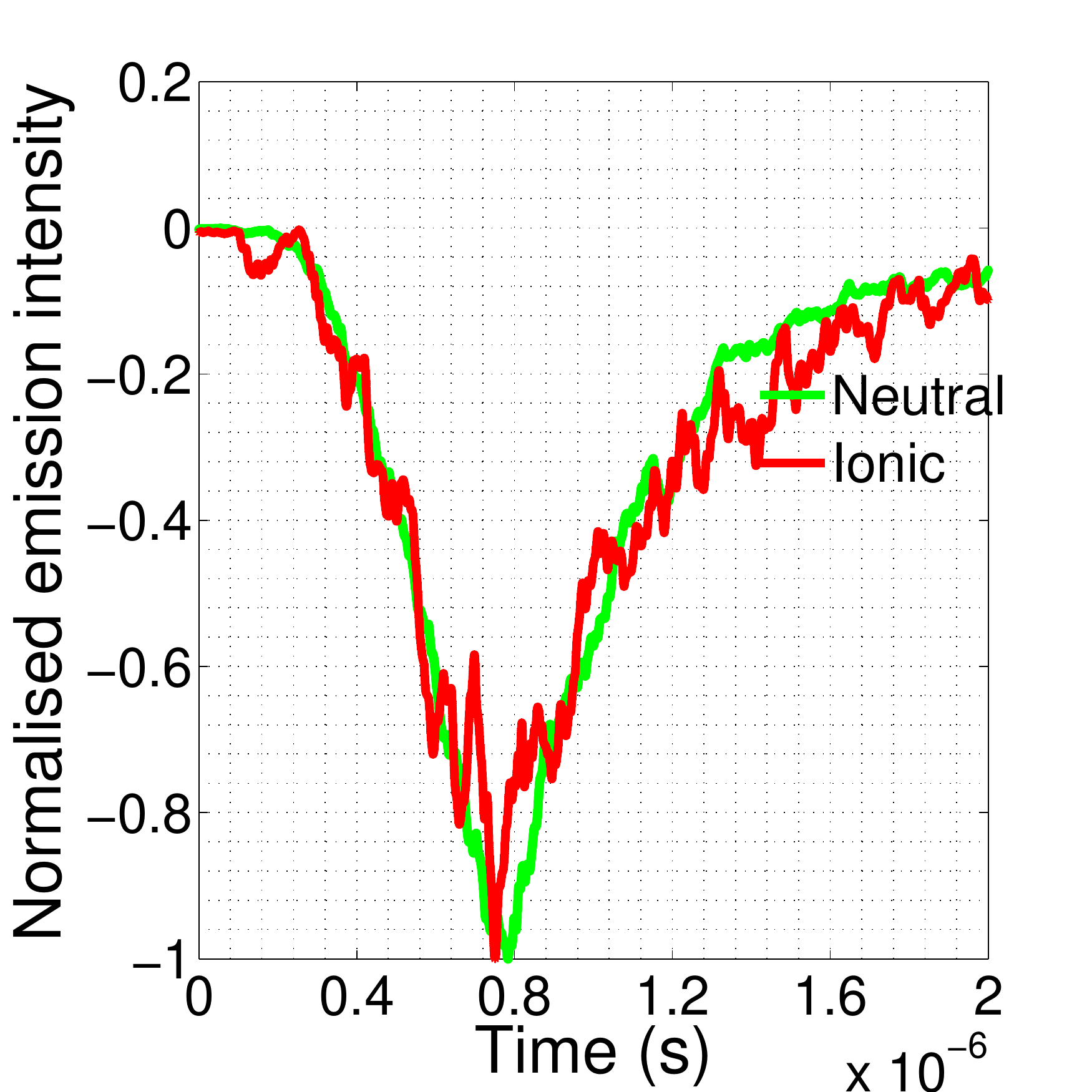}
        \caption{ 20 mbar 7 mm }
        \label{fig:DL_Neu_ion_1064nn_7mm_2ep1mbar_LE100mj}
    \end{subfigure} 
 \caption{\label{fig:DL_Neu_Ion_Prdependence_100mj} Temporal evolutions of neutral and ionic lines at 3 mm and 7 mm for different  background pressures  for laser energy of 100 mJ . }
\end{figure*}
\begin{figure}
\centering
\includegraphics[width=0.48\textwidth, trim=0cm 0cm 0cm 0cm, clip=true,angle=0]{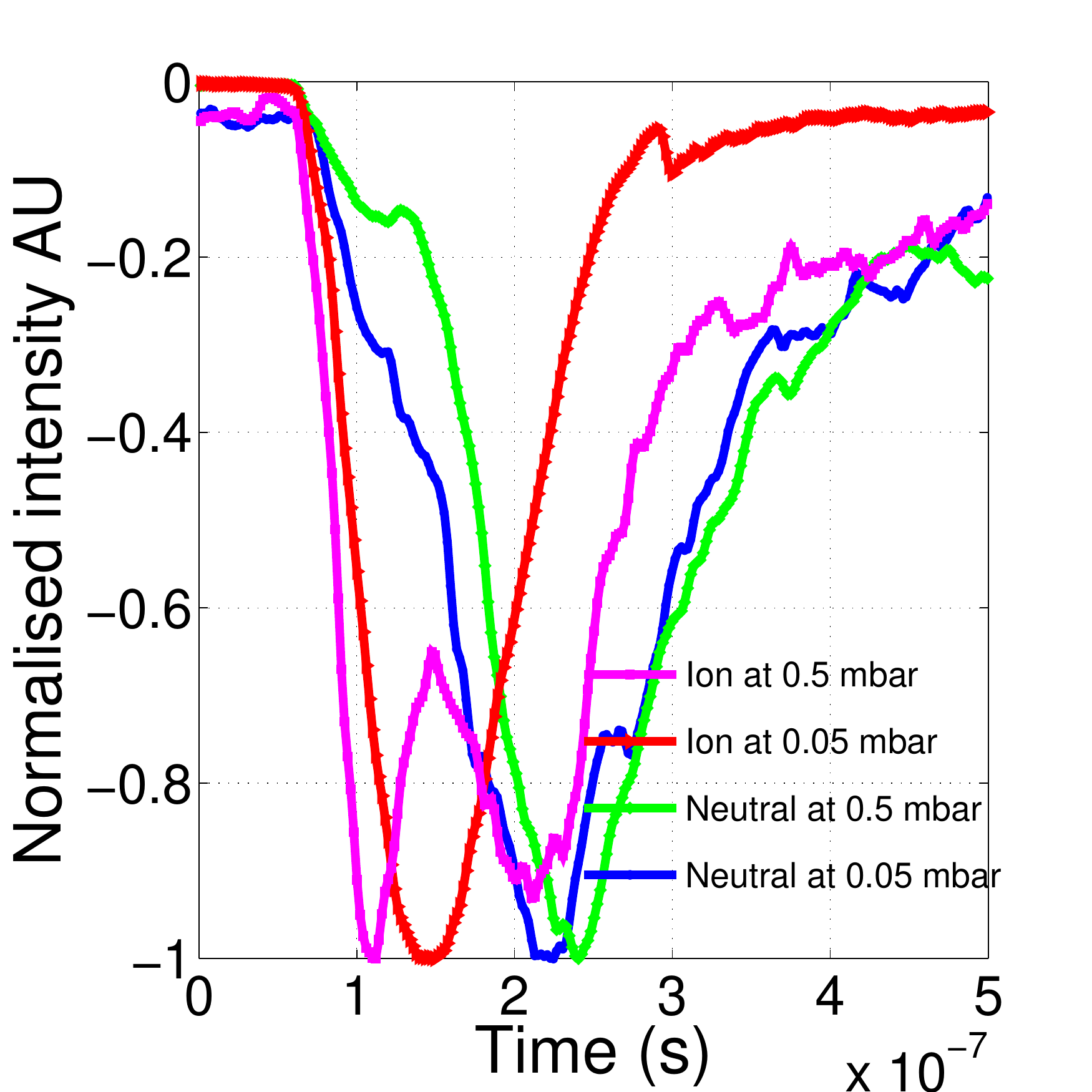}
\caption{\label{fig:DL_Neu_ion_1064nn_0mm_vr_pr_LELL25mj}  Temporal evolution of  TOF spectra of  ionic and neutral species for two background pressures for 1064 nm laser wavelength with 50 mJ of energy at 0.5 mm from the sample.}
\end{figure}
\begin{figure} 
    \centering
    \begin{subfigure}{0.48\textwidth}
        \includegraphics[width=\textwidth,  trim=0cm 0cm 0cm 0cm, clip=true, angle=0]{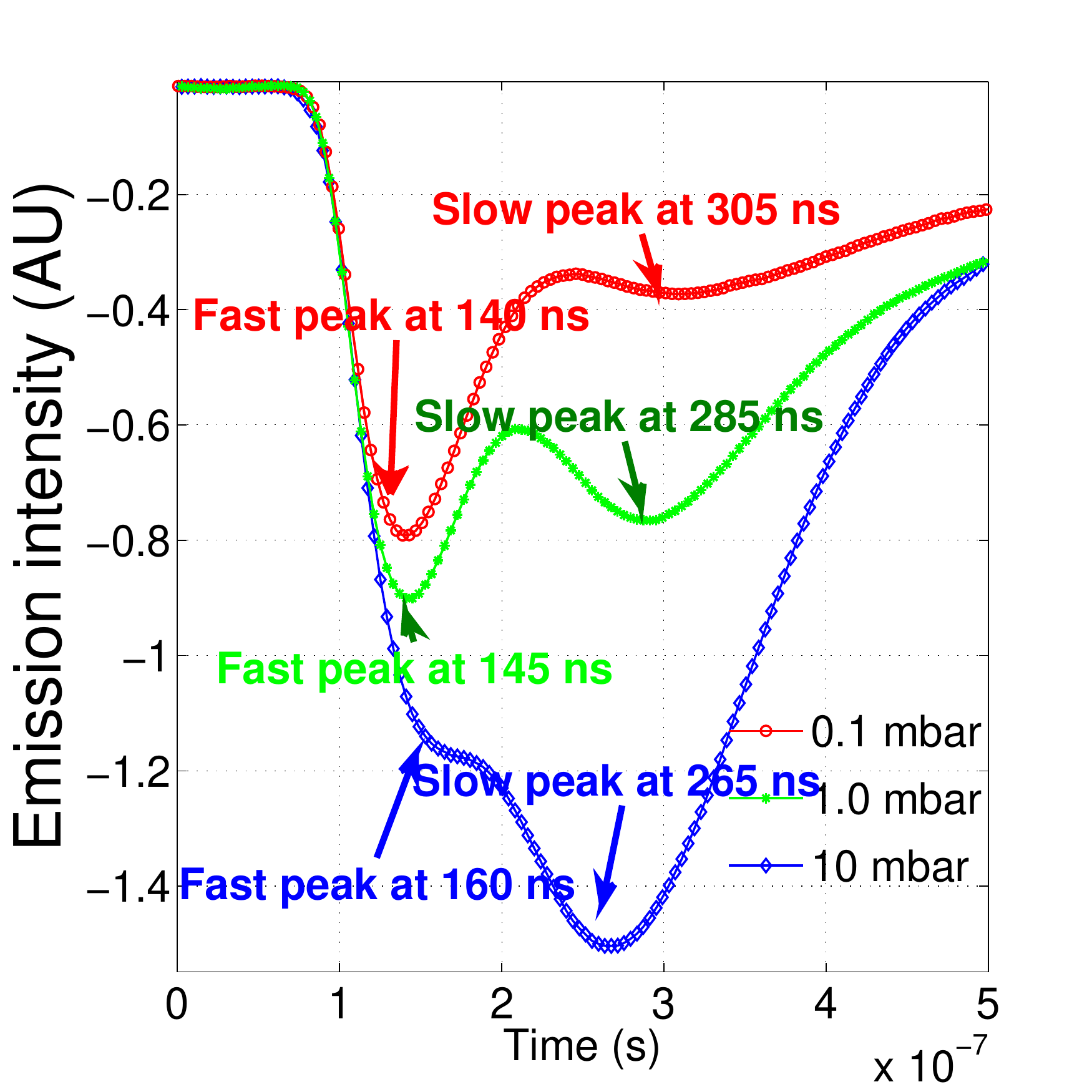}
          \caption{  361.9 nm neutral nickel line. }
        \label{fig:DL_Neu_362_1064nm_55mj_1mm_Var_Pr}
    \end{subfigure}   
    \begin{subfigure}{0.240\textwidth}
        \includegraphics[width=\textwidth,  trim=0cm 0cm 0cm 0cm, clip=true, angle=0]{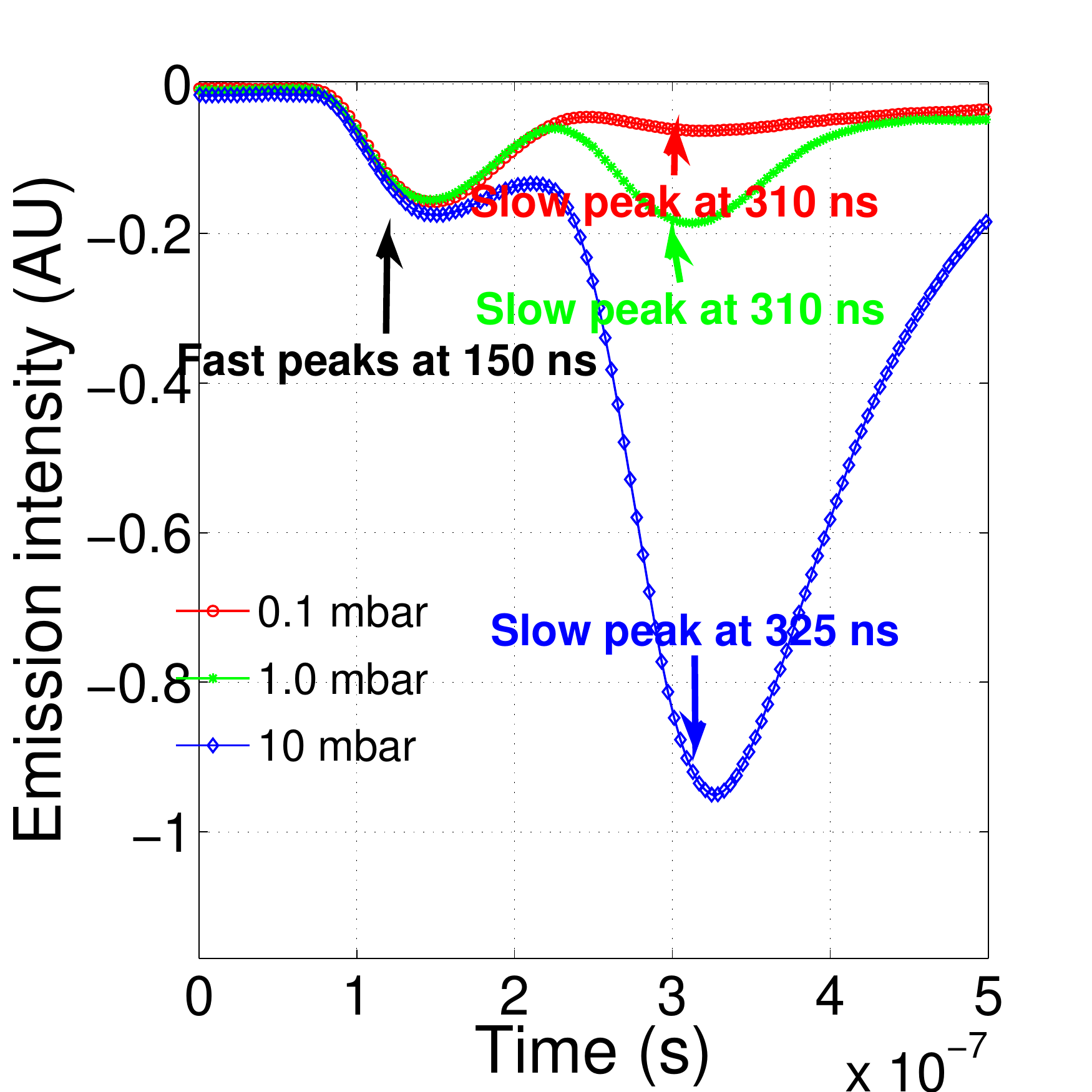}
          \caption{ 508.0 nm neutral nickel line. }
        \label{fig:DL_Neu_508_1064nm_55mj_1mm_Var_Pr}
    \end{subfigure}\begin{subfigure}{0.24\textwidth}
        \includegraphics[width=\textwidth,  trim=0cm 0cm 0cm 0cm, clip=true, angle=0]{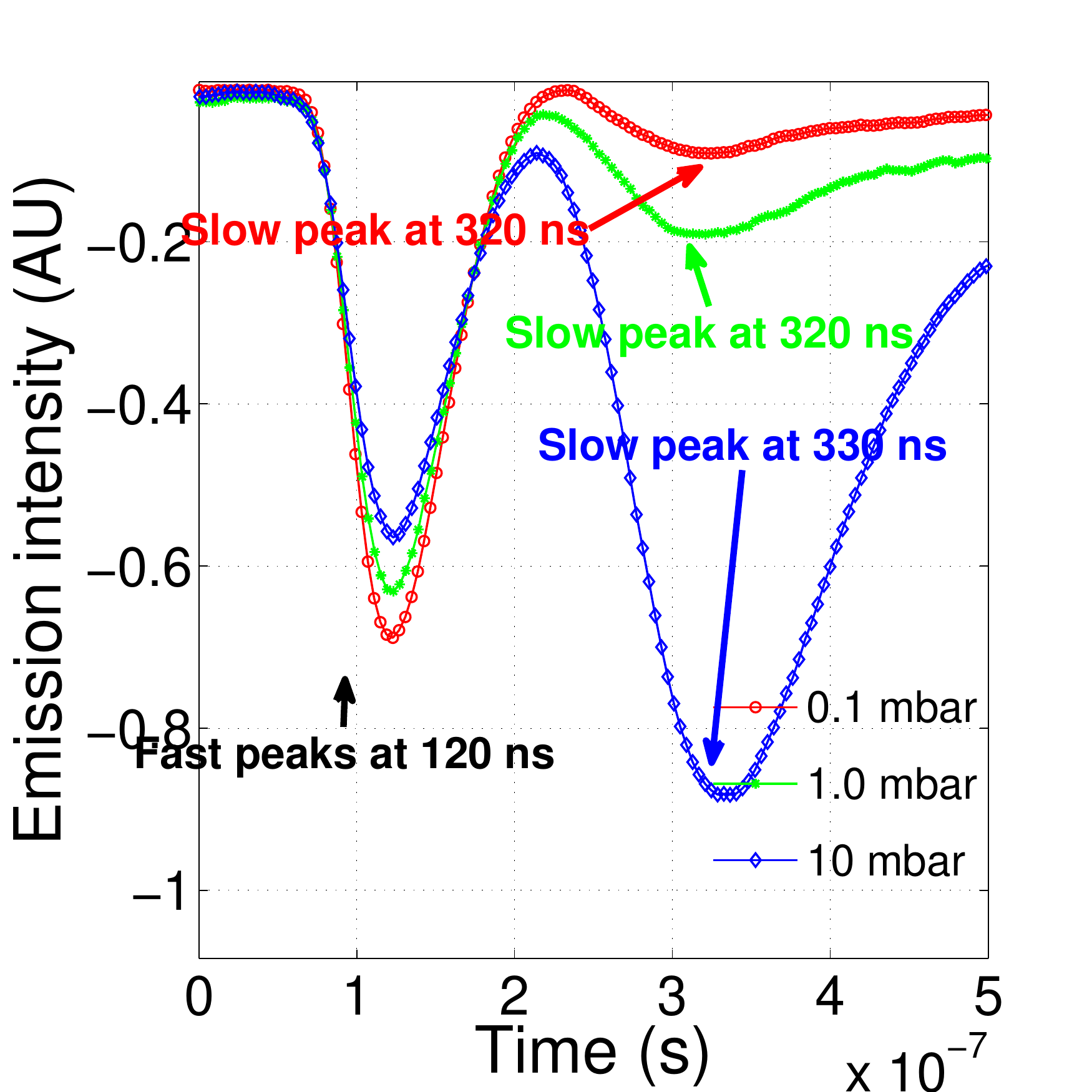}
        \caption{  712.2 nm neutral nickel line. }
        \label{fig:DL_Neu_712_1064nm_55mj_1mm_Var_Pr}
    \end{subfigure} 
 \caption{\label{fig:DL_Neu_3L_1064nm_55mj_1mm_Var_Pr} Temporal evolution of neutral lines of nickel at very close (1 mm) to the sample for 1064 nm laser ablation using 50 mJ energy. }
\end{figure}
\begin{figure} 
    \centering
    \begin{subfigure}{0.48\textwidth}
        \includegraphics[width=\textwidth,  trim=0cm 0cm 0cm 0cm, clip=true, angle=0]{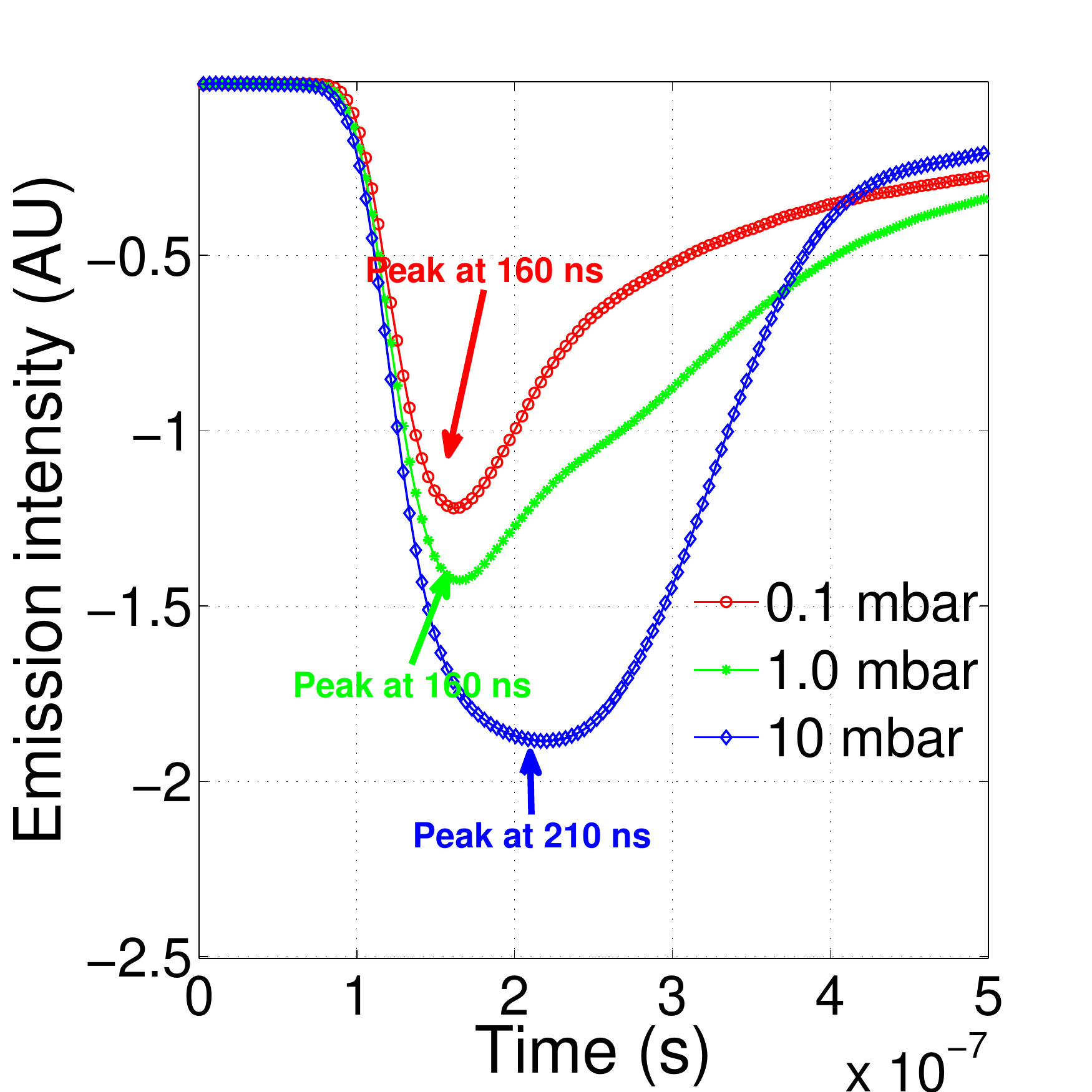}
          \caption{  361.9 nm neutral nickel line. }
        \label{fig:DL_Neu_362_1064nm_100mj_1mm_Var_Pr}
    \end{subfigure}
    \begin{subfigure}{0.2490\textwidth}
        \includegraphics[width=\textwidth,  trim=0cm 0cm 0cm 0cm, clip=true, angle=0]{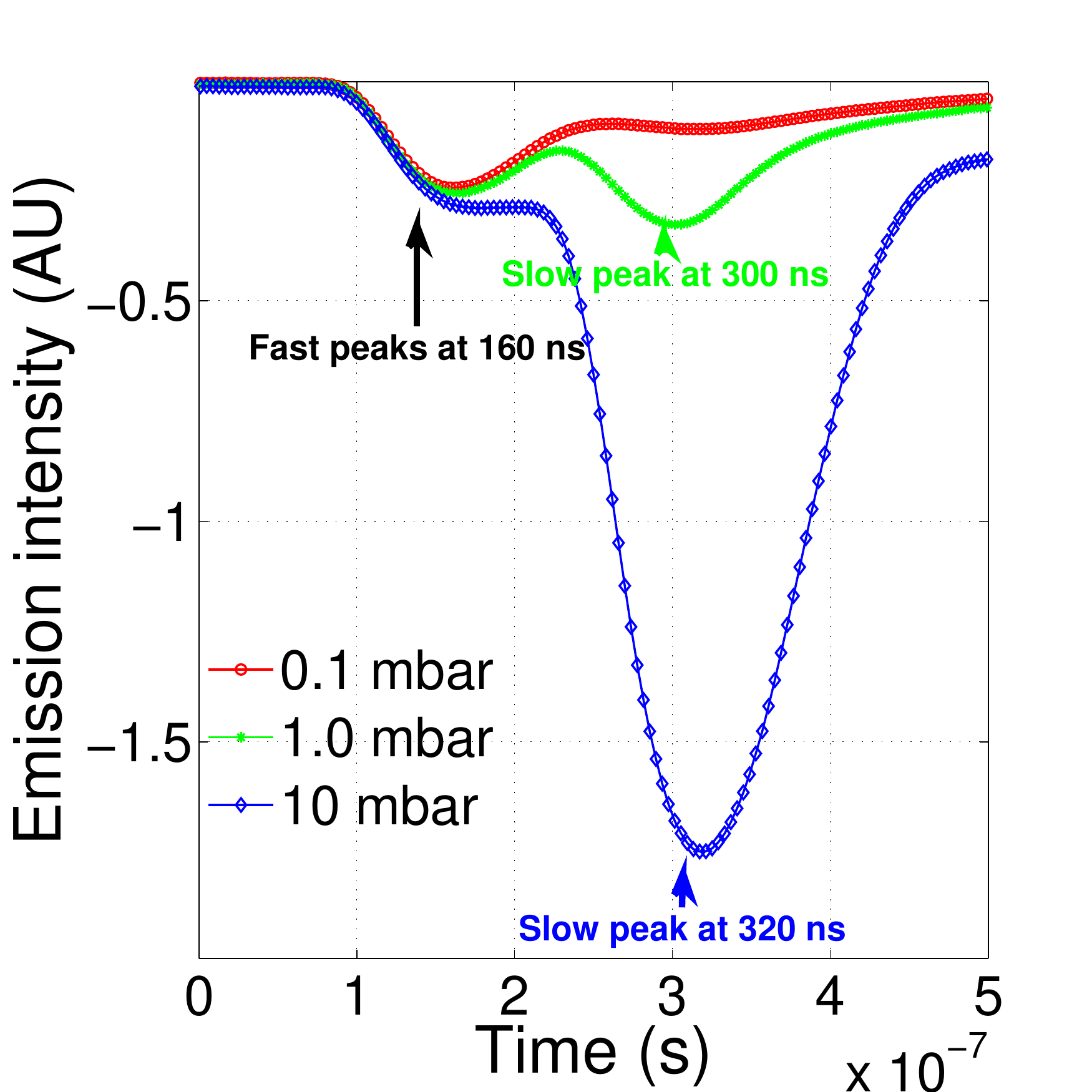}
          \caption{  508 nm neutral nickel line. }
        \label{fig:DL_Neu_508_1064nm_100mj_1mm_Var_Pr}
    \end{subfigure}\begin{subfigure}{0.249\textwidth}
        \includegraphics[width=\textwidth,  trim=0cm 0cm 0cm 0cm, clip=true, angle=0]{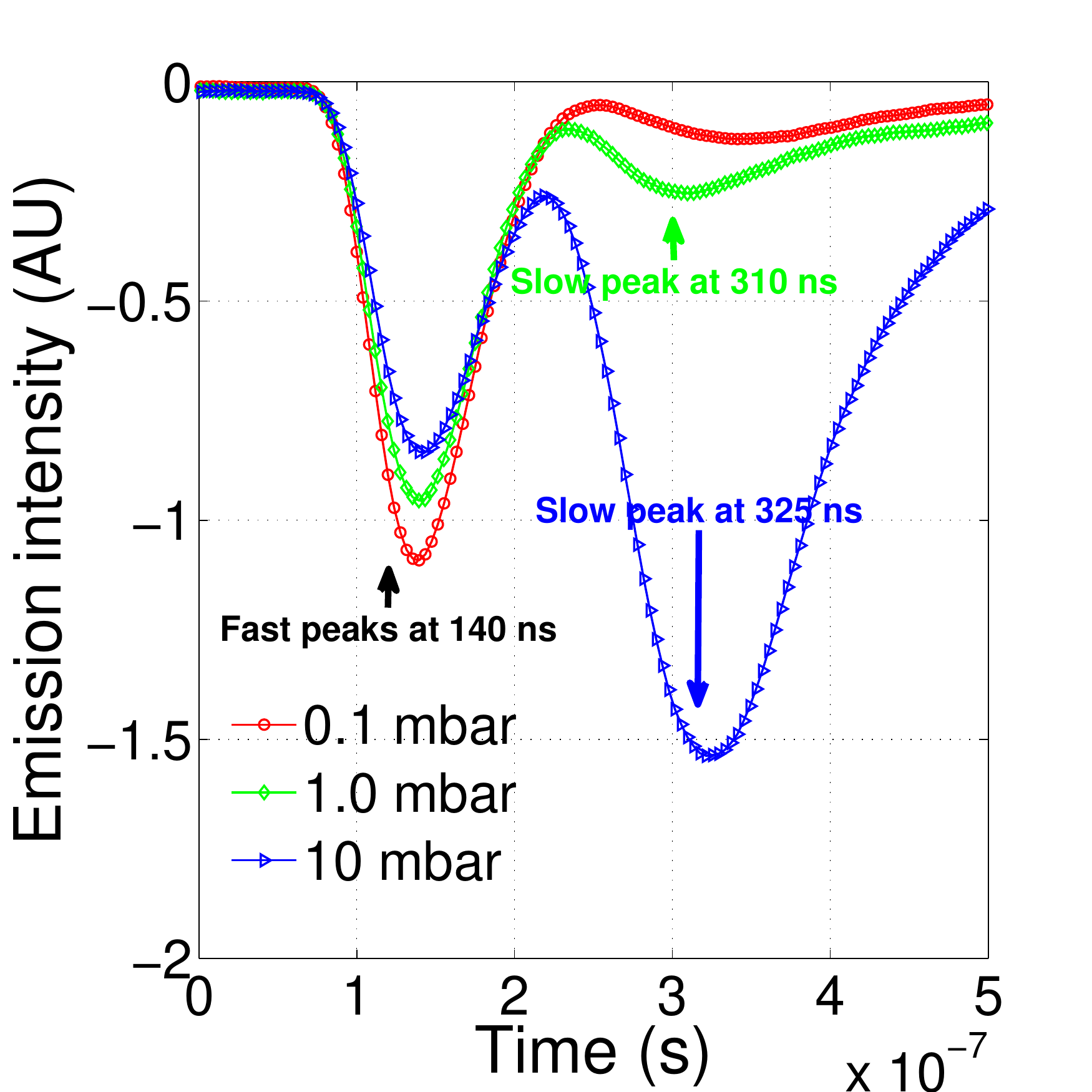}
        \caption{ 712 nm neutral nickel line. }
        \label{fig:DL_Neu_712_1064nm_100mj_1mm_Var_Pr}
    \end{subfigure} 
 \caption{\label{fig:DL_Neu_3L_1064nm_100mj_1mm_Var_Pr} Temporal evolution of neutral lines of nickel at very close (1 mm) to the sample for 1064 nm laser ablation using 100 mJ energy. }
\end{figure}
  \begin{table}[ht]
\caption{Spectroscopic details of lines  used in the present study \cite{NIST_ASD}.}
\centering
    \begin{tabular}{ | p{1.1cm} | p{1.4cm} | p{0.7cm}  |p{1.50cm} |p{2.50cm} |}
    \hline
    Lambda & Aij  & J & Ej &Transition  \\ 
   (nm) & ($10^7s^{-1}$)& -&($ cm^{-1}$) &   \\ \hline
361.94 Ni I& 6.6 & 3 & 31 037.02 & $ 3d^9(^2D)4p \rightarrow 3d^9(^2D)4s $ \\ \hline
362.68 Ni II& - & 5 & 29 070.93& $ 3p^6 3d^8 (4s) \rightarrow 3p^6 3D^9 $\\ \hline
508.11 Ni I& 5.70  &7 & 50 706.27& $3d^9(^2D_{3/2})4d \rightarrow 3d^9(^2D)4p $\\  \hline
712.22 Ni I& 2.10  &3 & 42 605.94 & $3d^9(^2D_{5/2})5s \rightarrow 3d^9(^2D)4p $\\  \hline
    \end{tabular}
    \label{table:Ni_ion_neutral_details}
\end{table}

\section{Results and Discussion}
Figure~\ref{fig:DL_ion_Pr_var_1064nn_3mm_LE100mj} shows the TOF emission recorded for the ionic line 362.7 nm of nickel at 3 mm from the sample for different background pressures at a laser energy of 100 mJ. The figure shows that the evolution of the slower peak becomes significant as the background pressure increases from 0.1 mbar to 20 mbar. At 0.1 mbar the slower peak is not evident as can be seen from the figure, however, it appears with same intensity as that of the faster peak as the background pressure reaches 5 mbar, which further increases substantially  with increase in background pressure.
 Further, it can be noted that the slower peak gradually advances in time as the background pressure increases. A  closer look at the faster peak shows that it slows down marginally as the pressure increases (figure~\ref{fig:DL_ionic_peaking_delay_3mm_100mj_1064nm}). Similar trend is seen in the TOF spectra of ionic species of the plume generated with 50 mJ laser energy as shown in figure~\ref{fig:DL_ion_Pr_var_1064nn_3mm_LE55mj}. 
 It is interesting to see that the temporal separation between the faster and slower peaks decreases from 195 ns at 1 mbar of background pressure to 120 ns at 20 mbar of background pressure when ablated with 100 mJ of laser energy (figure~\ref{fig:DL_ionic_peaking_delay_3mm_100mj_1064nm}).   In case of ablation with 50 mJ laser energy, the delay changes from 220 ns to 150 ns in the same range of background pressure.\par
 
It is clear from the figure that the slow ionic peak, which is not prominent at lower pressure slowly marks its presence at higher pressures. 
At lower energy, the structure of peaks is well defined even at higher pressure. From TOF spectrum, it can be noted that the fast peak is not much affected (due to the background pressure) and the slow peak has enhanced intensity and also indicates increased velocity. 
This observation can be used to rule out the possibility of splitting of the fast peak due to multiple collisions with background gas as reported in some of the earlier studies\cite{Harilal_plume_spliting,HARILAL_spliting}. 
The fast peak is apparently due to the ions that penetrate the ambient whereas slow peak due to the ions that are confined by it \cite{Harilal_plume_spliting,Hari_spliting_sn}. In these studies it is clearly observed that the fast peak is not much affected by the increase of background pressure but the slower pulse further slows down significantly as the background pressure increases \cite{Hari_spliting_sn,Plume_spliting_Fe_Al}. The striking difference between the the present and earlier reported works is probably configuration of ablation (forward vs backward ablation geometries).
It is interesting to note that at lower laser energy (50 mJ), the slower peak of ionic emission is more prominent at lower background pressure (0.1 mbar) in comparison  to higher laser energy (100 mJ)   there is no significant intensity of the slow peak .The intensity of ionic line is higher at higher laser energy. On plume spliting earlier works \cite{Hari_spliting_sn} have observed that the fast peak is not significantly retarded with increase in background pressure, however,the the emission intensity decreases significantly with increase in pressure. On the other hand, the slower pulse is significantly retarded with increase in background pressure. In our observation also the faster peak is not retarded much with increase in background pressure. However, a contradicting behavior is seen for the intensity of fast peak, which increases with background pressure.  
Another important observation from the TOF spectra of ionic species for various background pressures and laser energies is that the velocity of the slower component gets enhanced with background pressure and laser energy.
 For instance, the faster peak gets slowed down whereas the slower peak becomes faster with pressure as can be seen in figure~\ref{fig:DL_ionic_peaking_delay_3mm_100mj_1064nm}. Though it is not clear why slow peak becomes faster, in case of fast peak it can be attributed to colliding with the ambient and subsequently losing part of its momentum. Further, it is interesting to see that the faster peak has relatively higher intensity for 50 mJ energy. It is expected that more charged species should be present at higher energy. This anomalous behavior may be due to screening of the ions and hence decreased acceleration.\par

Akin to the ionic spectra, emission spectra of neutrals are also recorded using the same configuration by changing the central wavelength of the spectrometer to three prominent neutral nickel lines as listed in table~\ref{table:Ni_ion_neutral_details}.
Figure~\ref{fig:DL_Neu_Pr_var_1064nn_3mm_LE55_100mj} shows the TOF emission spectra of all the three neutral lines recorded at 3 mm from the sample for both the laser energies.  
As can be seen from the figure for particular laser energy and pressure, the neutral line (361.9 nm) peaks faster than the other two neutral lines (slower peak).
 508.0 nm and 712.2 nm neutral lines show similar trend in peaking time for the slower peak. 
Both lines have lower intensity. However, faster peak for 712.2 nm line (figure~\ref{fig:DL_Neu_712_Pr_var_1064nn_3mm_LE100mj} and \ref{fig:DL_Neu_712_Pr_var_1064nn_3mm_LE55mj}) has relatively higher intensity. As the background pressure increases, the intensity of slower peak increases significantly whereas the intensity o faster peak decreases slightly. 
However, 361.9 nm neutral line does not show any trace of faster peak for both the laser energies at this distance.  
The neutral lines at 3 mm show that the peaks get slightly delayed with increase in the background pressure.
However, for the 361.9 nm emission, the drag offered by the background gas is not significant as can be seen in figure~\ref{fig:DL_Neu_Pr_var_1064nn_3mm_LE100mj} in the respect that the peak position does not change as the pressure increases from 5 mbar to 20 mbar (in fact in this range the peak advances a little in this range). However, for the other two emission lines the peak position delays gradually with the increase of background pressure (figure~\ref{fig:DL_5080nm_Pr_var_1064nn_3mm_LE100mj} and \ref{fig:DL_Neu_712_Pr_var_1064nn_3mm_LE100mj}). \par

From the observation of the TOF spectra of ionic and neutral lines of nickel at 3 mm in this experimental configuration, it is clear that the ionic species within the plasma have significantly higher velocities, in comparison to neutral species. 
Also, the velocity of ionic species increases with increase in background pressure which is worthwhile to note because instead of getting slowed down due to resistive force of the medium, ions get accelerated. This may be probably increase in electron density and subsequent increase in electric field as will be discussed latter.
Figure~\ref{fig:DL_ionic_neutral_Slow_peaking_3mm_100mj_1064nm} shows the peaking time of the slower component for  ionic emission and for the neutral lines for varying background pressures at 100 mJ of laser energy at 3 mm distance. 
The peaking time and its variation with background pressure are distinctly different for the ions and neutrals. For ions, as the pressure increases, the slower peak gets faster. However, this is not the case for neutral lines.   
The 361.9 nm line shows that the pressure does not affect the peaking time as background pressures increases( from 5 mbar to 20 mbar the peaking time remains almost constant as can be seen in the figure ~\ref{fig:DL_ionic_neutral_Slow_peaking_3mm_100mj_1064nm}). However, the other two neutral emission lines(508.0 nm 712.2 nm) show that the peaks slow down further as the pressure increases.  As the peaking time for ions as well as the neutrals is almost the same at lower pressures (figure~\ref{fig:DL_ionic_neutral_Slow_peaking_3mm_100mj_1064nm}), it can be assumed that the neutrals are initially formed from the ionic species  closer to the sample. As the background pressure increases, neutrals are affected to a very small amount, whereas, the velocity of the ionic species is increased with pressure.
However, as the role of background gas is same for neutrals and ions, it has to be assumed that the ions get some additional acceleration as the pressure increases, so that they gains more velocity. As mentioned earlier, it can be assumed that part of neutrals may be formed from the recombination process in ionic species and may continue with initial energy they acquire as an ion until the gas pressure drags the components and slows it down. 
Earlier studies have demonstrated  the effect of background pressure, laser energy and dependence of spatial location on plasma plume splitting in LPP (backward ablation) geometry. Some studies show a completely different behavior as expected for plume splitting and propagation. 
For instance, the dynamics of carbon plume in helium background \cite{HARILAL_C_He} has shown that the fast peak gets slowed down as the laser irradiance increased.
 It is also mentioned that the velocity of fast peak increases and of the slow peak decreases as the helium pressure increases. The observation of decrease in faster peak velocity with the increase in laser irradiance has been explained as
a selective depletion of high velocity $C_2$ species and the effect of background pressure on the velocity of faster peak is explained using the inverse relation of pressure and energy in the adiabatic expansion model\cite{HARILAL_C_He}. We would like to mention that the present results also show a totally unexpected nature of the behavior of the slow peak.\par

The behavior of the ionic (362.7 nm ) and the neutral (361.9 nm) lines at 7 mm distance were also recorded similarly by using the optical fiber at 7 mm distance. For an easy comparison of the differences in TOF of neutral and ionic lines at 3 mm and 7 mm the TOF spectrum with sufficient visibility, it is shown in 
figure~\ref{fig:DL_Neu_Ion_Prdependence_100mj} for 100 mJ of energy . The figure shows a single peak for neutral emission, but distinctly two or more peaks for the ionic line (specifically at 3 mm). Hereafter the peaks will be denoted as: i) fast ion peak ii) slow ion peak and iii) neutral peak for further discussion. 
The maximum intensity of each emission line is normalized to 1 for ease in comparison.  Though ionic and neutral lines are separated by 0.8 nm, the resolution of the spectrometer is sufficient enough to resolve them.  At 3 mm, as can be seen from the figures, the neutral peak lags significantly with respect to the fast ionic peak. However, at lower pressure, the slow ion peak (very broad and shallow) coincides with the neutral peak. 
As the background pressure increases to 10 mbar, the peaking time of slow ion peak advances but the neutral peak slightly lags to the neutral peak at 0.1 mbar. On further increasing the background pressure to 20 mbar, both the slow ion and the neutral peaks slightly advance further in comparison to 10 mbar pressure. 
In contrast to this behavior at 3 mm, the peaks of ionic and neutral species coincide at 7 mm distance which indicates that equilibrium between neutrals and ions is established at this distance. It also shows that the neutral peak is broader than the ionic peaks, especially at lower background pressure. It is likely that neutrals may originate from recombination process from ion as well directly formed in the ablation process, This can explain rather broad distribution of neutral TOF spectrum.  It can be noted that at lower fluence also similar behavior of spectral emission of ionic and neutral species is present. Figure ~\ref{fig:DL_Neu_Ion_Prdependence_100mj} shows that the ionic peaks get accelerated as the background pressure increases. 

The striking differences between ionic and neutral emission and its spatial dependence are quite interesting and prompt to the analysis of data collected from even closer distances from the film $\approx 1 $ mm.
 However, very close to the sample the emission intensity of the ionic component (corresponding to 362.7nm) is less and it becomes unobservable as the background pressure increases. Using the same spectrometer and PMT set up, the ionic and neutral line behavior is recorded for $5\times 10 ^{-2} mbar$ and  $5\times 10 ^{-1} mbar$. 
 The behavior of neutral transitions listed in table~\ref{table:Ni_ion_neutral_details} at a closer distance(two adjacent fibers separated by less than 1 mm are used, with the one closer to the sample is used with spectrometer  and the other one is used with filter and PMT. The fiber connected to spectrometer is mapped closer to the sample by less than 1 mm. ) is recorded using band pass filters of respective neutral lines at different background pressure and laser fluence is recorded using fast PMT.
  The plasma plume is  imaged using an imaging system and fiber array as described earlier. Figure~\ref{fig:DL_Neu_ion_1064nn_0mm_vr_pr_LELL25mj} shows the effect of background pressure on ionic(362.7 nm) and neutrals line (361.9 nm) for background pressures of $5\times 10 ^{-2} mbar$ and  $5\times 10 ^{-1} mbar$. 
  The intensity of ionic emission is low and hence higher gain for recording the signal is used. 
  The TOF spectrum is normalized with the maximum intensity for each spectrum. As can be seen from the figure, the ionic emission for either of the pressures is faster than the neutral emission. 
  As the background pressure increases slightly the neutral lines show slight drag whereas, the ionic line shows a double peak structure and the fast peak of ion advances ahead of the ionic peak at low pressure.
   It may also be noted from the figure that even the neutral spectra at 0.5 mbar show an onset of a faster peak at around 120 ns which is prominent for higher pressures.
     These observations clearly underline the possibility of neutral formation through recombination.\par

TOF spectra of neutral transitions (table~\ref{table:Ni_ion_neutral_details}) are also recorded using band pass filters for respective neutral lines and fast PMT for different background pressure and laser fluences closer to the target. Figure~\ref{fig:DL_Neu_3L_1064nm_55mj_1mm_Var_Pr} shows the TOF spectra for three different neutral lines (361.9 nm, 508.0 nm and 712.2 nm) using the same optical fiber and PMT for different background pressures. 
As can be seen from figure~\ref{fig:DL_Neu_362_1064nm_55mj_1mm_Var_Pr},for all the three pressures, the TOF spectra show two well-resolved peaks for 50 mJ energy. It can be noted that the neutral emission spectra recorded by spectrometer(figure~\ref{fig:DL_Neu_ion_1064nn_0mm_vr_pr_LELL25mj}) do not show well resolved peaks as seen here at the same energy, which may be due to smaller distance from the target(less than 1 mm). 
It may be noted that interestingly, the slower peak becomes faster (observed at shorter time) as marked in the figure and the faster peak gets slightly slowed down as the pressure increases from 0.1 mbar to 10 mbar which shows similar behavior as the ionic line.  
 The slower peak at 305 ns for 0.1 mbar advances to 265 ns by increasing the background pressure to 10 mbar. At the same time, the faster peak trails from 140 ns to nearly 160 ns. Although 20 ns deviation is rather small and may be argued as uncertainty, it should be noted here that triggering of acquisition is done by a fast photo diode of 1 ns rise time and every component used in this experiment like PMT, filter, cables and triggering mechanism are exactly the same.
  Moreover, the emission earlier to the fast peak due to laser line leak matches well and hence timing is not a matter of concern. Figure~\ref{fig:DL_Neu_508_1064nm_55mj_1mm_Var_Pr} and figure~\ref{fig:DL_Neu_712_1064nm_55mj_1mm_Var_Pr} are the evolutions of TOF spectra of 508.0 nm and 712.2 nm respectively, which show distinctly different temporal behavior, as compared to 361.9 nm. 
  The fast peak for 361.9 nm emission shows enhancement with increase in pressure, for 508.0 nm it does not show any signification enhancement  whereas for 712.2 nm there is appreciable decrease in intensity. 
  However, for the slower peak there is enhancement in the intensity for all the three lines. The 508.0 nm emission  has significantly low intensity for the fast peak even at higher pressures in comparison to the slow peak. However, the TOF spectrum of 712.2 nm shows significant intensity for the fast peak for all the three background pressures. 
  From figure~\ref{fig:DL_Neu_3L_1064nm_55mj_1mm_Var_Pr} it is evident that the slower peak of 361.9 nm advances significantly with pressure but the other two neutral lines do not show as much variation as that in case of 361.9 nm.
   it is likely that the governing atomic processes e.g. recombination process for these three transitions get modified distinctly for the different transitions resulting in different behavior for different characteristics corresponding different neutral lines. It can be noted that transition probability (Table~\ref{table:Ni_ion_neutral_details}) for 712.2 nm line is lowest but still it is prominent. This clearly indicates that recombination process has highest contribution for this line. As a result the faster peak which is created by recombination process in ion shows up with highest intensity in case of 712.2 nm line. this is also supported from the results of fig~\ref{fig:DL_Neu_Pr_var_1064nn_3mm_LE55_100mj}.\par
 
 We also recorded the TOF spectra for these lines at higher energy 100mJ and are shown in Figure~\ref{fig:DL_Neu_3L_1064nm_100mj_1mm_Var_Pr}. For this laser energy, in the case of 361.9 nm emission, the slow and fast peaks  appear merged together as can be seen in figure~\ref{fig:DL_Neu_362_1064nm_100mj_1mm_Var_Pr}. The peak position advances significantly to 210 ns for 10.0 mbar background pressure. This again indicates that the fast peak gets slowed down and the slower peak gets accelerated. 
 As a result the peaks appear merged.  From a closer look at the figure, it can be inferred, though distinctly, that for 1.0 mbar a slower peak appears at later time nearly at 260 ns. Figure~\ref{fig:DL_Neu_508_1064nm_100mj_1mm_Var_Pr} and figure~\ref{fig:DL_Neu_712_1064nm_100mj_1mm_Var_Pr} show TOF spectra of 508.0 nm and 712.2 nm respectively, where no significant deviation in evolution pattern or peak position is observed in comparison to the TOF spectra observed at 50 mJ (figure~\ref{fig:DL_Neu_3L_1064nm_55mj_1mm_Var_Pr}). However, the emission intensity is increased significantly for both these lines as compared to the 50 mJ excitation. \par
 
As mentioned earlier the dynamics of emission spectra of different transitions are slightly different, but it shows a consistent double peak structure for all the transitions. The 361.9 nm neutral emission shows a trend similar to the ionic spectra recorded for different pressures, i.e. as the background pressure increases, the second peak advances. Hence the assumption of evolution of neutrals from the ions appears to be reasonable. These observations also highlight the role of particular atomic transition for the observations using optical time of flight measurements.  

\begin{figure} 
    \centering
    \begin{subfigure}[b]{0.45\textwidth}
        \includegraphics[width=\textwidth,  trim=0cm 0cm 0cm 0cm, clip=true, angle=0]{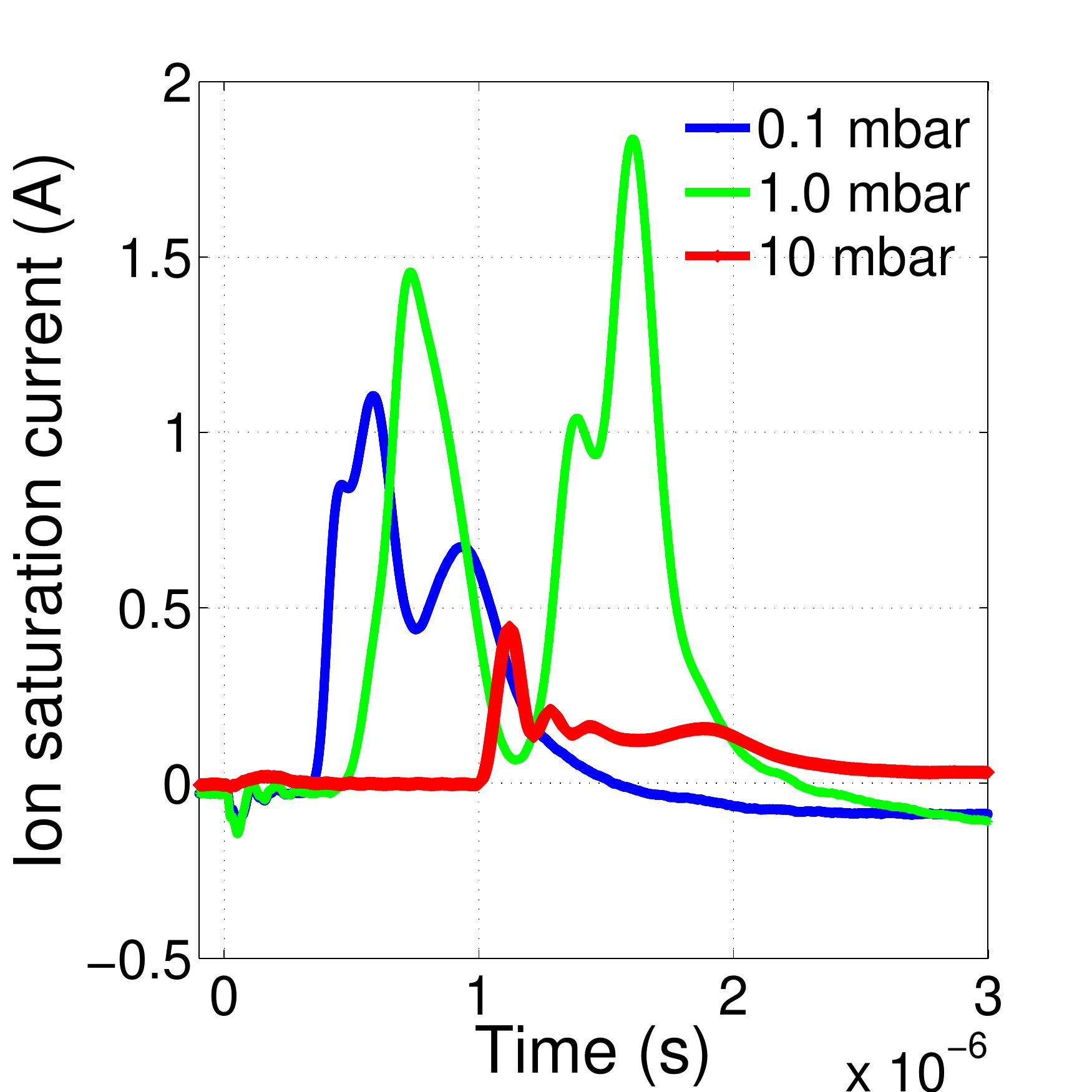}
        \caption{ Ion saturation current for 100 mJ laser energy }
        \label{fig:DL_IC_1064nm_100mj_10mm_Var_Pr}
    \end{subfigure} 
    \begin{subfigure}[b]{0.45\textwidth}
        \includegraphics[width=\textwidth,  trim=0cm 0cm 0cm 0cm, clip=true, angle=0]{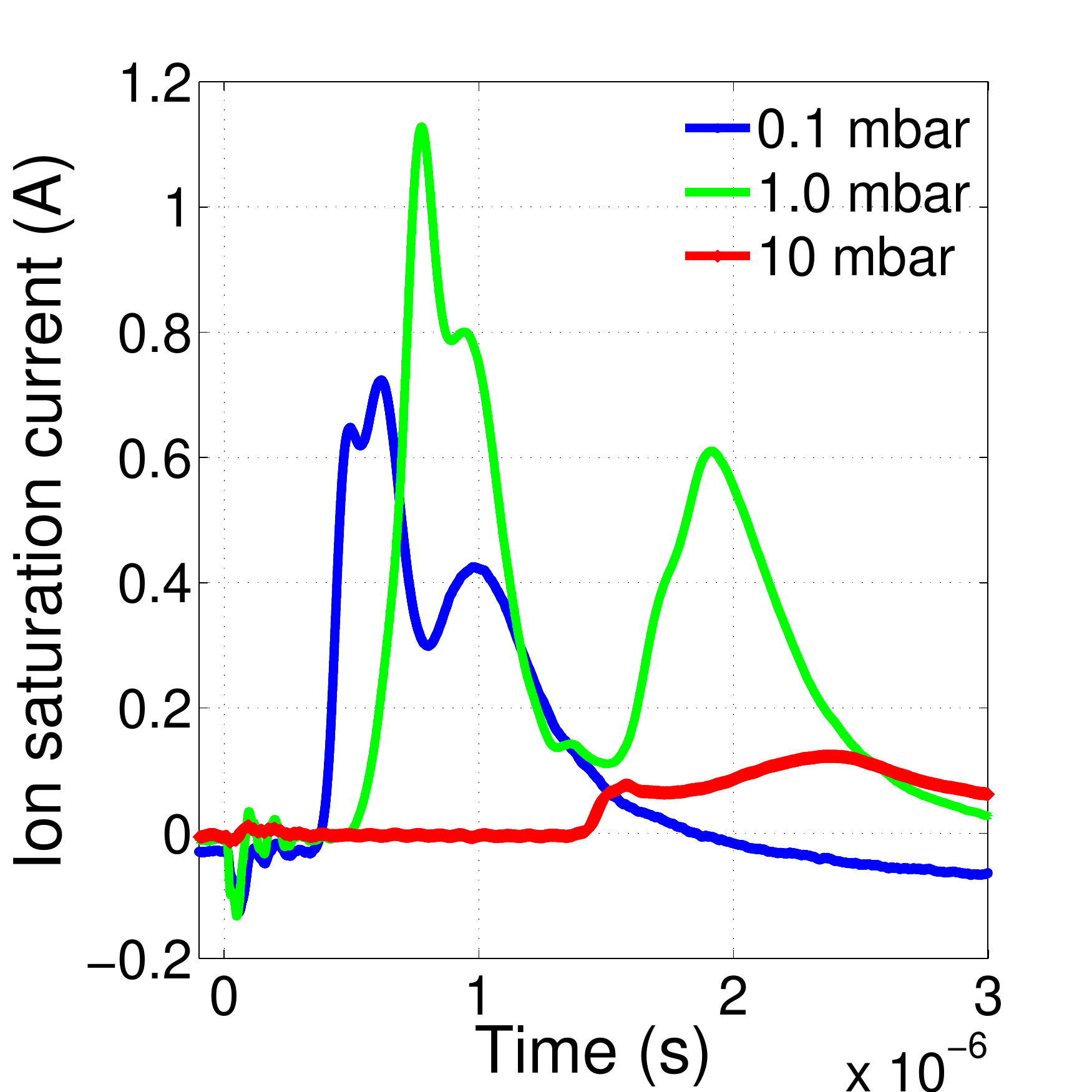}
          \caption{ Ion saturation current at 50 mJ of energy. }
        \label{fig:DL_IC_1064nm_55mj_10mm_Var_Pr}
    \end{subfigure}
 \caption{\label{fig:DL_IC_1064nm_100and55mj_10mm_Var_Pr} Temporal evolution of ion saturation current at 10 mm from the sample for different background pressures and two different laser energies (50 mJ and 100 mJ). }
\end{figure}

\begin{figure} 
    \centering
    \begin{subfigure}[b]{0.45\textwidth}
        \includegraphics[width=\textwidth,  trim=0cm 0cm 0cm 0cm, clip=true, angle=0]{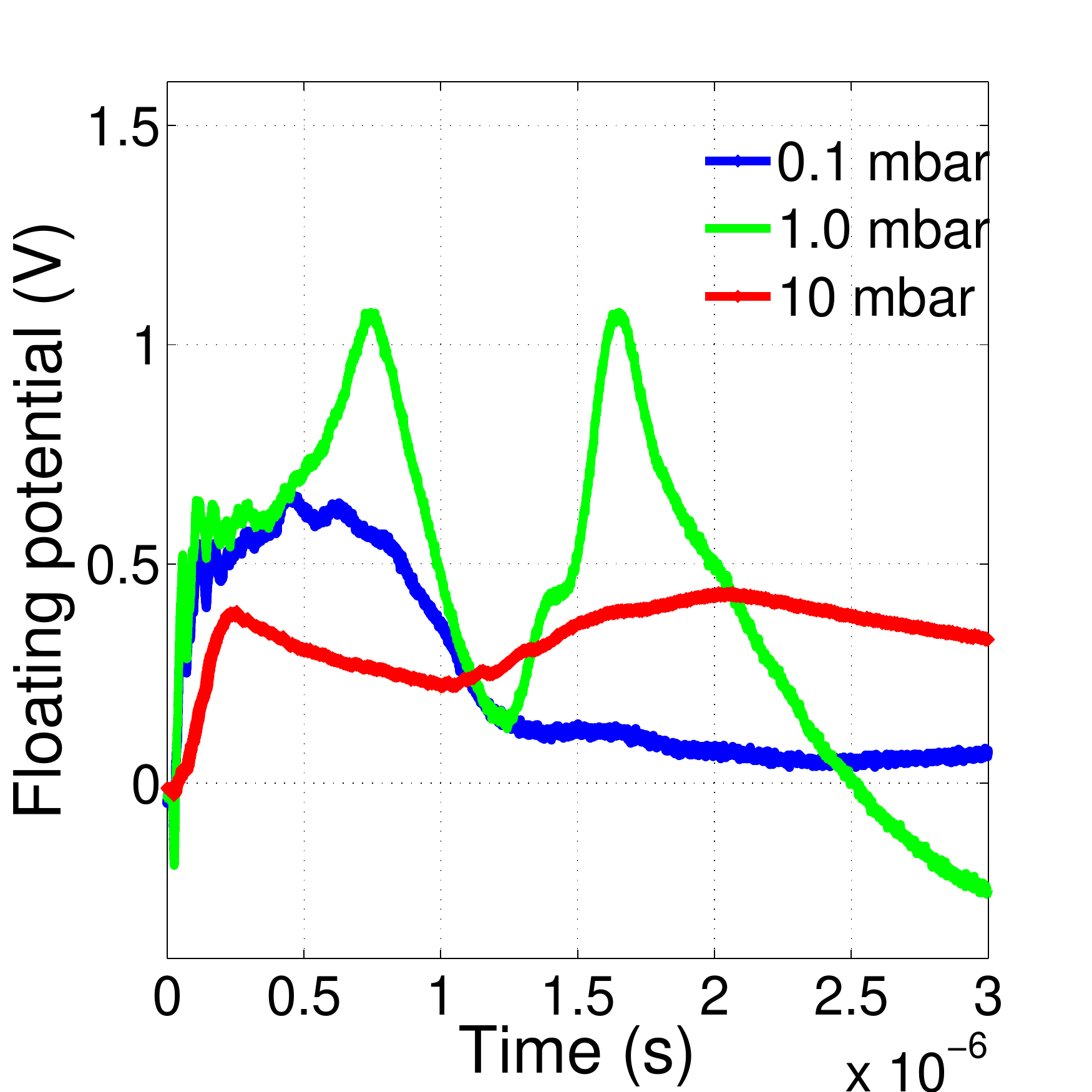}
        \caption{ Floating potential for 100 mJ laser energy }
        \label{fig:DL_FP_Pr_var_1064nn_10mm_LE100mj}
    \end{subfigure} 
    \begin{subfigure}[b]{0.45\textwidth}
        \includegraphics[width=\textwidth,  trim=0cm 0cm 0cm 0cm, clip=true, angle=0]{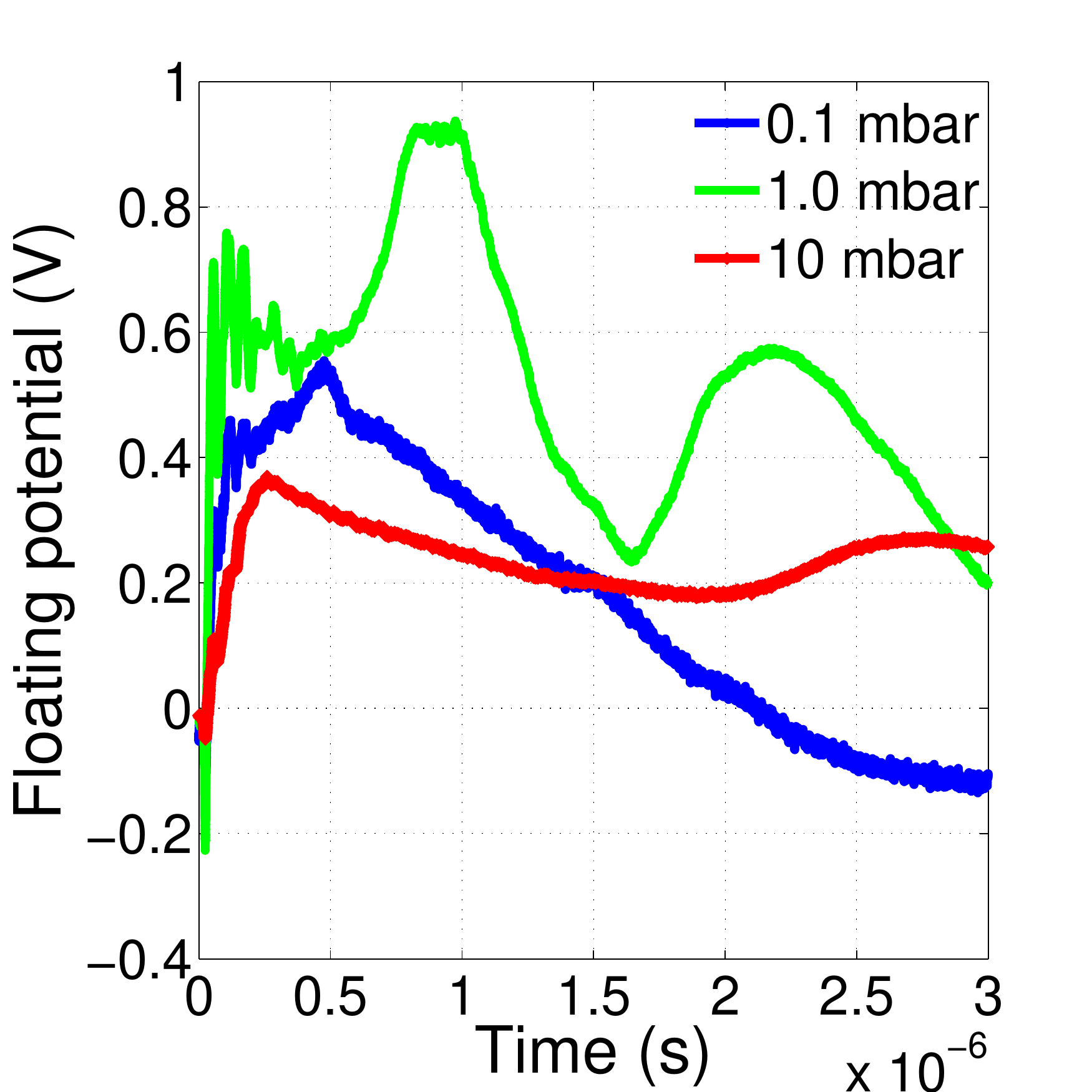}
          \caption{ Floating potential at 50 mJ of Laser energy. }
        \label{fig:DL_FP_Pr_var_1064nn_10mm_LE55mj}
    \end{subfigure}
 \caption{\label{fig:DL_FP_Pr_var_1064nn_10mm_LE55_100mj} Temporal evolution of floating potential at 10 mm from sample for different background pressures and two different laser energies (50 mJ and 100 mJ) . }
\end{figure}

\begin{figure} 
    \centering
    \begin{subfigure}[b]{0.45\textwidth}
        \includegraphics[width=\textwidth,  trim=0cm 0cm 0cm 0cm, clip=true, angle=0]{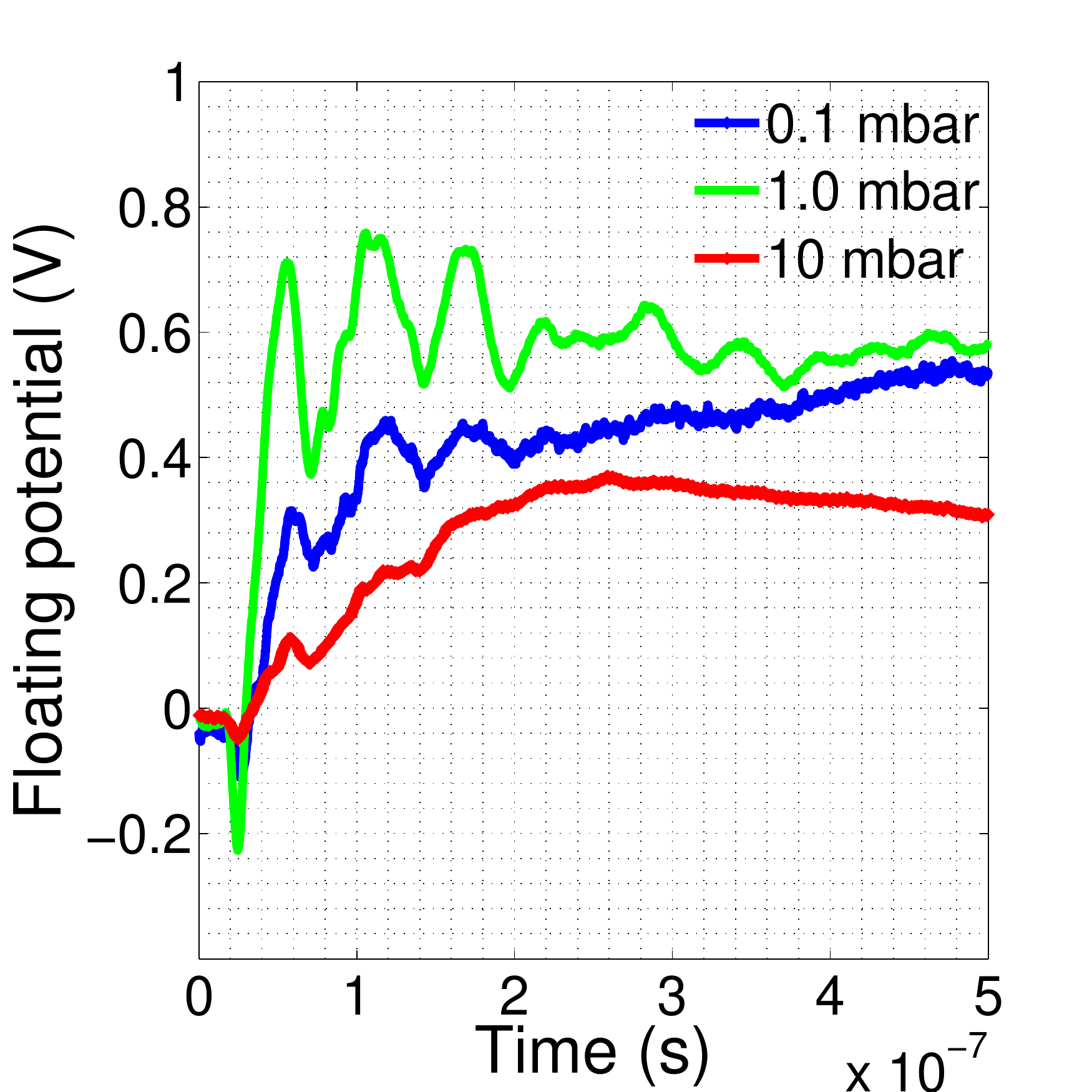}
        \caption{ Floating potential. }
        \label{fig:DL_FP_Pr_var_1064nn_10mm_LE55mj_Zoom}
    \end{subfigure} 
    \begin{subfigure}[b]{0.45\textwidth}
        \includegraphics[width=\textwidth,  trim=0cm 0cm 0cm 0cm, clip=true, angle=0]{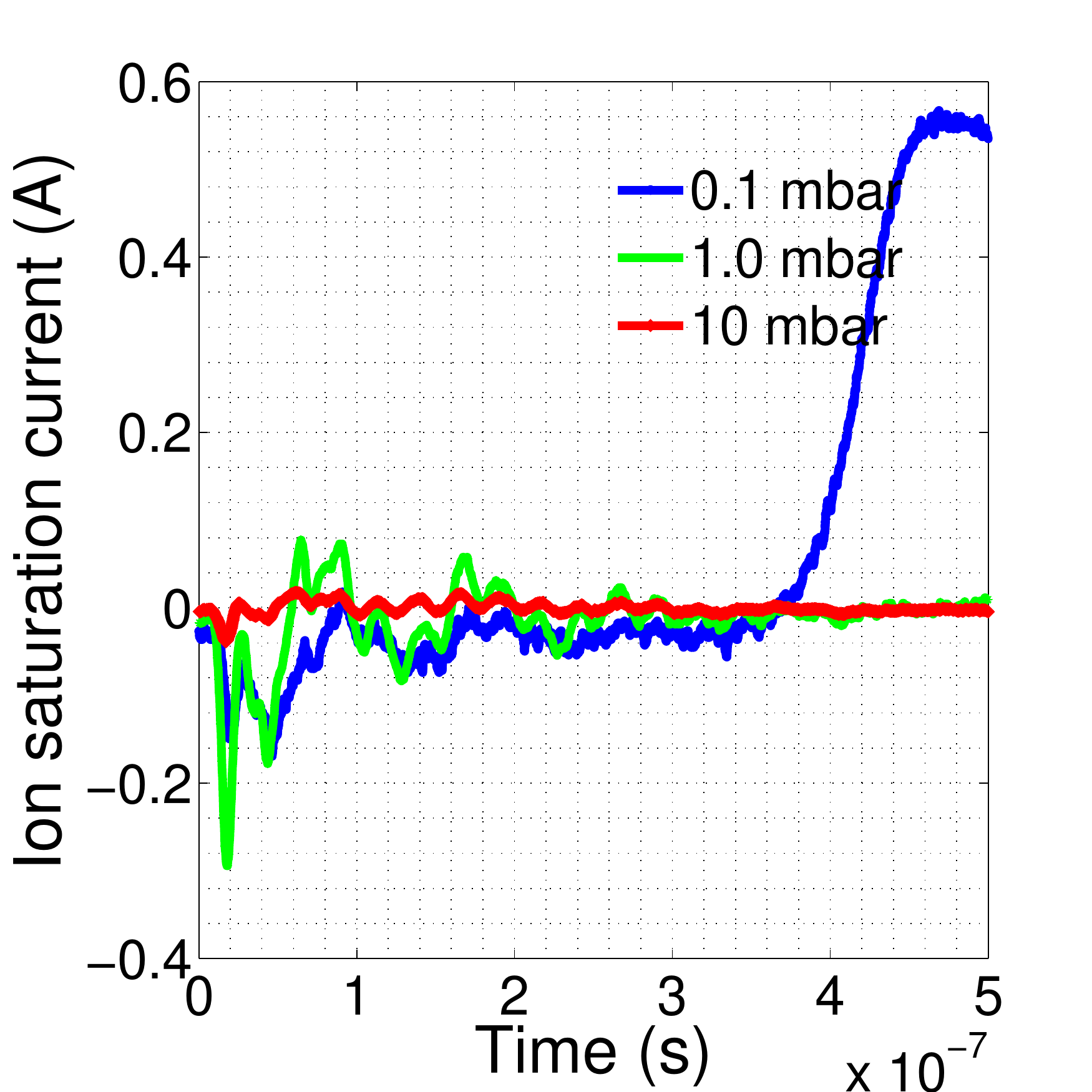}
          \caption{ Ion saturation current.}
        \label{fig:DL_IC_Pr_var_1064nn_10mm_LE55mj_Zoom}
    \end{subfigure}
 \caption{\label{fig:DL_IC_FP_Pr_var_1064nn_10mm_LE55mj_Zoom} Zoomed section of the floating potential and ion current recorded at 10 mm away from sample at early stages of plasma plume for 50 mJ energy, showing the traces of oscillations. }
\end{figure} 

The neutral and ionic emissions at different locations clearly show an enhancement in the velocity of slower component of ions and some specific neutral species up to certain distance. To extract more information we have used triple Langmuir probe to observe behavior of ions as well as the evolution of potential inside the plasma. However, it is not feasible to make any measurements close to the sample considering the rather high plasma density and temperature \cite{TLP_3,AjaiTLPJAP}.
 Hence, for ascertaining the dynamics of ions Langmuir probe measurements have been used after 10 mm from the target.

Figure~\ref{fig:DL_IC_1064nm_100and55mj_10mm_Var_Pr} shows the evolution of ion saturation current recorded using TLP at 10 mm from the sample for different background pressures for both the laser energies. It is interesting to monitor the evolution of ion saturation current considering the observation of TOF spectra of ionic species, where the slower component advances in time with background pressure. However, no such observation is seen in the evolution of the ion saturation current at 10 mm from the sample. Instead, it slows down with background pressure. 
 The ion saturation current shows that there may be multiple components of ionic species present as distinct peaks. 

It can be noted that the first peak position of ion saturation current for low background pressure (0.1 mbar) is not largely different for both the energies.  For instance, at 0.1 mbar the ion saturation current peaks at 452 ns and 490 ns for 100 mJ and 50 mJ respectively. The second peak of ion saturation currents is observed at 584 and 616 ns respectively for the 0.1 mbar background pressure. At 1.0 mbar the first peak for both the laser energies is observed at 730 and 770 ns for 100 mJ and 50 mJ of laser energy respectively. However, at 10 mbar the first peak of 100 mJ ablation is significantly faster (1120 ns) compared to the peak (1510) at 50 mJ of ablation. 
 A straightforward comparison among  other peaks is difficult as the peaks are not distinctly visible for many instances. At this stage, it is interesting to point out that at low pressure (0.1 mbar), the velocity of the fast peak (at 135 ns figure~\ref{fig:DL_ionic_peaking_delay_3mm_100mj_1064nm}) for ionic line at 3 mm can be correlated  with that of the fast peak (first peak which is not well separated at 435 ns, figure~\ref{fig:DL_IC_1064nm_100mj_10mm_Var_Pr}) observed for ion saturation current at 10 mm. However, the velocity corresponding to the ionic emission at 7 mm (at 600 ns) does not match these velocities. Instead, it is comparable to the slowest peak (at 950 ns)observed for ion saturation current.  Hence at lower pressure the optical TOF measurement and the TLP measurements are reasonably correlated.
  The first and last peaks of ion saturation current at 10 mm from the sample at a background pressure of 0.1 mbar and laser energy of 100 mJ (figure~\ref{fig:DL_IC_1064nm_100mj_10mm_Var_Pr})matches with the TOF recorded for ionic species at 3 mm and 7 mm. This correlation between two measurements is interesting and confirms that ionic species do not undergo any significant drag at lower background pressures.
\par
Figure~\ref{fig:DL_FP_Pr_var_1064nn_10mm_LE55_100mj} shows the temporal evolution of floating potential recorded at 10 mm from the sample along with the ion saturation current using the TLP. Here, the floating potential shows immediate response to the laser interaction with the sample (within 10-20 ns of laser incidence) which shows a fast rise in floating potential at a time where the fast negative current is observed in ion saturation current.
The floating potential in plasma can provide insight into the electron temperature  at the particular location. 
 However, it should be noticed that in this case the TLP is at a distance of 10 mm from the sample.  Hence it is possible to rule out the plasma generated on the sample (thin film) by the laser pulse reaching the probe location within 10 ns. Hence, the potential recorded on the TLP appears  to be due to some other mechanism. 
  Zoomed views of floating potential and ion current at the early stages of plasma are given in figure~\ref{fig:DL_IC_FP_Pr_var_1064nn_10mm_LE55mj_Zoom}. The figure shows well defined damping oscillations in floating potential as well as some non periodic negative peaks in ion saturation current at early stages (10 to 25 ns) of the plume evolution.  Negative value of ion saturation current shows that the contribution to the current is due to electrons. Moreover, the peaking time of the negative peak is not affected by the background gas pressure unlike ion saturation peak which significantly gets delayed as the background pressure increases.
 Here it has to be mentioned that the probes are biased over -20 V for measuring the ion saturation current and electrons with energy higher than 20 eV is reaching the probe to record the negative current.  These electrons interact with the background gas and ionizes as seen in earlier studies \cite{Hari_spliting_sn}   Hence, the negative values of ion saturation current and the detection of potential by TLP shows the ejection of electrons from the initial plasma formed from the sample. The possible cause of generation of  hot/fast electrons will be discussed later. 
  However, at this instant, the reason for oscillations observed on the floating potential does not have a clear cut explanation.   Oscillations in ion saturation current  for laser plasma were also reported earlier \cite{gurlui_osc_PRE} , which are usually explained by plasma double-layer effects. Gurlui et. al \cite{gurlui_osc_PRE} further analyzed these oscillations and concluded that it is due to the self structuring of the interface of double layers couple with temperature fluctuations.

For some other instances, the occurrence of such fast response on probes is attributed to prompt electrons due to photo ionization by the stray laser photons (reflections from the sample) on the probes. 
 However, in this case, the photon energy is significantly low for photo ionization. In addition to this, due to the configuration (rear ablation geometry) of the experiment the laser photons can not reach the probe as the film is opaque. 
  Moreover, the traces show variation in oscillation amplitude, negative peak increases at 1 mbar pressure. At 10 mbar oscillation amplitude is rather small, which is not expected in case of photoionisation with stray photons. Hence, the onset of negative peaks and oscillations in the ion saturation current and floating potential may be assumed due to the interaction of fast electrons generated from the laser interaction with nickel film.  
  The decrease in the influx of electrons at higher pressures may be due to increased number of collisions with the background gas.  It is well known that the electron neutral collision rate  is a function of number density of background gas molecules and temperature (velocity) of the electrons. For instance, for electron with velocity equivalent to 3 eV at a background 
gas pressure of 0.1 mbar, the collision rate is estimated of the order of $10^{12} s^{-1}$, which increases by two orders of magnitude when background pressure increases to 10.0 mbar.

 The observation regarding a large number of electrons (resulting in a current of few hundred of mA) is very important due to the fact that plasma plume at its initial stages loses a large number of electrons, which create a charge imbalance within the plasma plume. The spectroscopic observation of acceleration of ions near the sample can be explained in the light of this observation. 
 However, prior to attempting an explanation, we shall characterize the plasma in terms of density and temperature near the sample where the acceleration of ions as well as neutrals is observed. \par

As described in earlier article\cite{Jinto_POP2018}, estimation of density of plasma plume at early times is performed using a high resolution spectrograph (1 m with 0.07 nm resolution) and an ICCD. The spectra of neutral nickel lines are recorded and the Stark broadening parameters for these lines were estimated using the width of $H_{\alpha}$ recorded for the same laser energy and background pressure\cite{Jinto_POP2018} .The estimation of density and  temperature of the plasma parameter at early stage of plume evolution (200 ns) are $N_e=1\times 10^{18} /cc$ and temperature of 2-3 eV as reported earlier\cite{Jinto_POP2018}. 
 
 As commonly adopted, the observation of enhancement in the speed of ionic species as well as neutrals can be explained using the concept of double-layer (DL) or multi-layer formed in plasma as observed in a few earlier experimental and theoretical works\cite{DLEBulgakova,DL_S_Eliezier}. 
 It is noticed that due to the presence of relativistic electrons produced in laser plasma interaction, significant charge separation occurs within the plasma. Due to the property of shielding, this charge separation is possible within a few Debye lengths, which is known as the thickness of the double layer. It is assumed that within the plasma, at a very small distance (few times the Debye length)\cite{DL_S_Eliezier}, a large electric field is generated. This electric field can accelerate/decelerate the ions/electrons based on the polarity. 
 Plasma will try to restore the charge neutrality by re-organizing the charged species \cite{DLEBulgakova}. Due to the escape of fast electrons, a region is formed within the plasma where quasi neutrality is broken as the ions trail the electrons. In this scenario  the ions from the core region of plasma can be accelerated to establish quasi neutrality. 
 Due to the large mass of accelerated ions, the core region can have less number of ions and an inverse process of deceleration is also possible
 \par

The formation of DL in plasma is explained with the concept of formation of hot  electrons, by getting additional energy from the plasma plume. The major processes responsible for this  can be three body recombination\cite{Hotel_Threebody} and inverse bremstrahlung absorption\cite{Hotel_InvB}.
In three-body recombination, the electron is recombined to an ion to some excited level and can transfer 
the excess energy to another electron. Harilal et. al. \cite{WavelengthHarilal} observed that three body recombination (TBR) dominates radiative recombination for higher density and lower temperatures.
 In our case the temperature is not very high. It is likely that TBR increases the electron energy at a longer distance.  TBR (which has a rate proportional to $ T^{-9/2}$) is more prominent at lower temperature and higher densities. The radiative recombination (which has a rate proportional $T^{-3/4}$) is expected to be dominating at higher temperature\cite{PhysRev.Recombinations,Hotel_Threebody,WavelengthHarilal}. As estimated in reference\cite{WavelengthHarilal} the three body recombination dominates if the density $N_e \geq \frac{3\times 10^{19} T_e^{15/4}}{Z} m^{-3}$ where $T_e$ expressed in eV and Z is the charge state. 
In the present scenario, where the density is measured as $2\times 10^{18} cm^{-3}$ (for higher pressure  close to the sample and at a delay of 200 -300 ns)  and the temperature estimated is in the range of 2-3 eV, three body recombination will be the dominant mechanism for recombination. 
The absorption of energy from the incident laser through inverse bremsstrahlung is also expected to have hot electrons generated. 
In fact, laser of 1064 nm with longer pulse width is a favorable case for inverse bremsstrahlung heating.

For a plasma with density of the order of $10^{18}/cc$ and temperature of 2-3 eV, Debye length is estimated around 10 nm, and the possible thickness for DL can be assumed to of the order of few Debye lengths (say 100 nm to 1000 nm).
  From the negative peak observed for ion saturation current and the oscillations on floating potential, it can be assumed that the electrons escape from the plasma with moderate to high velocity ( minimum $10^8$ cm/s, it is expected that certain minimum time is required to get the hot electron generation as the laser pulse is of 10 ns duration and the hot electron generated has to get through the film which is in the process of melting). In addition to the hot electrons, we also have to consider the posibility of thermal electrons generated from the ablation of the film which escape. At the same time, it can also be noted that the laser intensity may not be sufficient to generate relativistic electrons as reported \cite{ion_accelaration_ps_foil,ion_accelaration2,ion_accela_film_pic} with the experiments using ultra fast lasers. 
In earlier experiments with comparable laser fluences at ns pulse width, the acceleration of ions is observed at longer distances using charge collector diagnostics at relatively higher pressures($6.8\times 10^{-2}$ mbar) \cite{DLEBulgakova}. In our experiment the maximum fluence used is $12 Jcm^{-2}$ (100 mJ ) and a background pressure of 20 mbar and have observed acceleration of ionic species as well as neutral even at closer distance (within 7 mm).  
  The present experimental configuration (rear ablation geometry of thin film) is expected to be helpful in generating higher plasma density and also the generation of fast electrons from the plasma as observed in the TLP. 
  The rigid thin film at the front side and glass substrate on the rear side confines the initial nickel plumes until the duration of complete melting of the film (few 10s of nanosecond). During this time, the high density plasma plume absorbs energy from laser light through IB, which heats up the plume and positively contributes to the generation of fast electrons by TBR and IB. \par
 
 The DL formation mechanism by the loss of fast electrons was reported by Bulgakova et. al.\cite{DLEBulgakova}. At this stage pictorial explanation (Figure 8 of reference \cite{DLEBulgakova} ) of formation of DL or multi-layer appears to be a plausible model to explain the observation of fast ions and neutrals we observed. 
Figure~\ref{fig:DL_Neu_Ion_Prdependence_100mj} show the acceleration of ionic peaks as the background pressure increases from 0.1 mbar to 10 mbar and 20 mbar for 100 mJ energy. 
Table~\ref{tab:PrVar_peaks_Le55mJ} and Table~\ref{tab:PrVar_peaks_Le100mJ} list the peaking time of these peaks for different background pressure for 50 mJ and 100 mJ excitation. 
\begin{table*}
\caption{Peaking time and respective velocities of various peaks at 3 mm from the sample at different background pressure ans laser energy of 50 mJ}
\begin{tabular}{|r|r|r|r|}\hline 
Ion or Neutral & Pressure (0.1 mbar) & Pressure (10 mbar)  & Pressure (20 mbar)  \\ \hline 
    Fast ion peak time &     135 ns &   145 ns&    150 ns\\ 
    Fast ion velocity&      $2.2\times10^{6} cms^{-1}$ & $2.1\times10^{6} cms^{-1}$&   $2.0\times10^{6} cms^{-1}$\\ \hline
     Slow ion peak time&    400 ns&    350 ns&    315 ns  \\
     Slow ion velocity&   $7.5\times10^{5} cms^{-1}$&  $8.6\times10^{5} cms^{-1}$& $9.5\times10^{5} cms^{-1}$ \\ \hline 
      Neutral peak time &  375 ns&   450 ns &    405 ns \\ 
       Neutral velocity &  $8\times10^{5} cms^{-1}$ & $6.6\times10^{5} cms^{-1}$ &$7.4\times10^{5} cms^{-1}$\\ \hline 
\end{tabular}
\label{tab:PrVar_peaks_Le55mJ}
\end{table*}

\begin{table*}
\caption{Peaking time and respective velocities of various peaks at 3 mm from sample at different background pressure ans laser energy of 100 mJ}
\begin{tabular}{|r|r|r|r|}\hline 
Ion or neutral & Pressure (0.1 mbar) & Pressure (10 mbar)  & Pressure (20 mbar)  \\ \hline 
    Fast ion peak time &    145 ns &   145 ns &    155 ns \\ 
    Fast ion velocity&      $2.1\times10^{6} cms^{-1}$ & $2.1\times10^{6} cms^{-1}$&   $1.9\times10^{6} cms^{-1}$\\ \hline
     Slow ion peak time&    410 ns &    315 ns &    280 ns \\
     Slow ion velocity&   $7.3\times10^{5} cms^{-1}$&  $9.5\times10^{5} cms^{-1}$& $10.7\times10^{5} cms^{-1}$ \\ \hline 
      Neutral peak time &  410 ns &    390 ns &    380 ns \\ 
       Neutral velocity &  $7.3\times10^{5} cms^{-1}$ & $7.7\times10^{5} cms^{-1}$ &$7.9\times10^{5} cms^{-1}$\\ \hline 
\end{tabular}
\label{tab:PrVar_peaks_Le100mJ}
\end{table*}

\begin{figure*} 
    \centering
    \begin{subfigure}{\textwidth}
        \includegraphics[width=\textwidth,  trim=0cm 0cm 0cm 0cm, clip=true, angle=0]{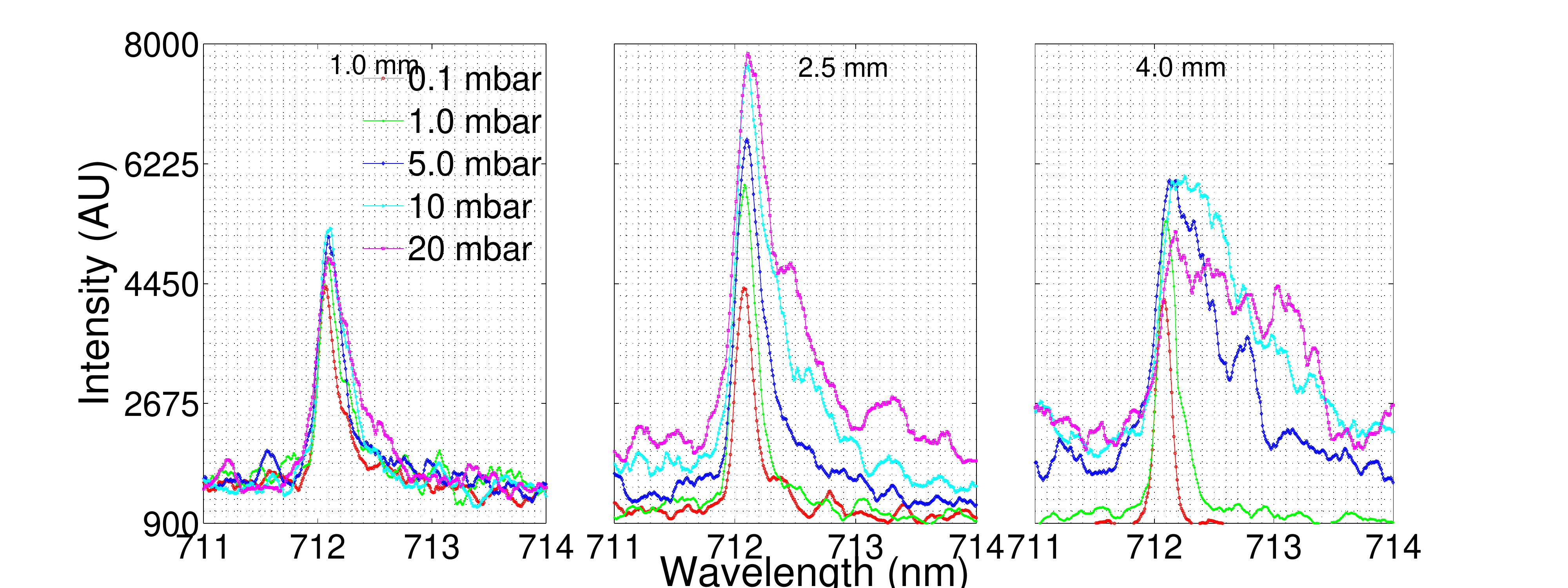}
          \caption{At 200 ns. }
        \label{fig:Assy_Neu_712_D200ns_1064nn_Varpos_Varpr_LE100mj}
    \end{subfigure}
    \begin{subfigure}{\textwidth}
        \includegraphics[width=\textwidth,  trim=0cm 0cm 0cm 0cm, clip=true, angle=0]{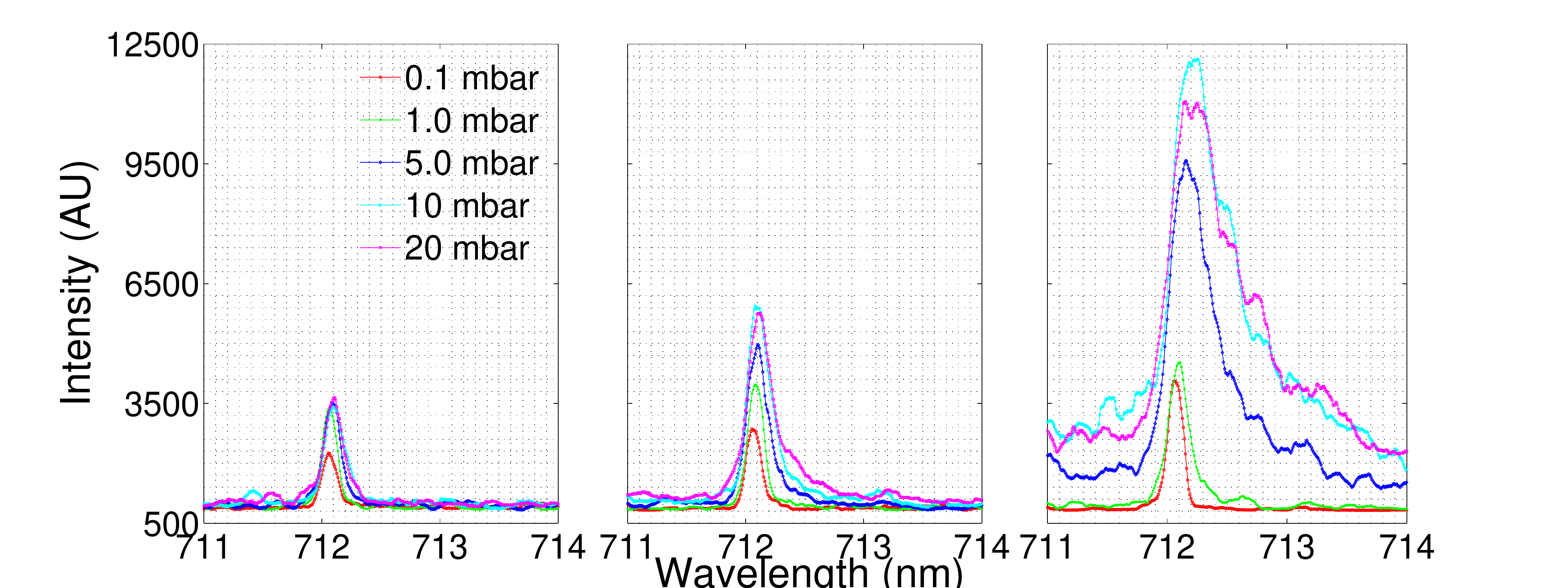}
        \caption{ At 300 ns.}
        \label{fig:Assy_Neu_712_D300ns_1064nn_Varpos_Varpr_LE100mj}
    \end{subfigure} 
    \begin{subfigure}{\textwidth}
        \includegraphics[width=\textwidth,  trim=0cm 0cm 0cm 0cm, clip=true, angle=0]{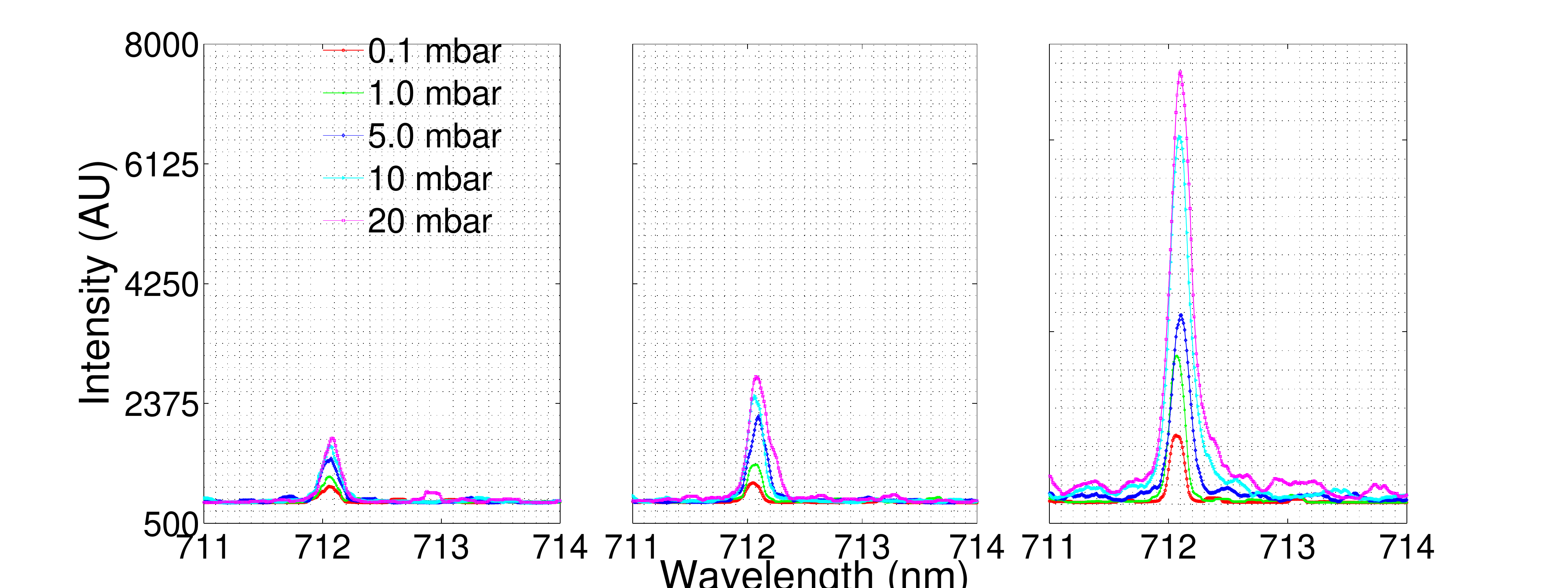}
          \caption{ at 500 ns. }
        \label{fig:Assy_Neu_712_D500ns_1064nn_Varpos_Varpr_LE100mj}
    \end{subfigure}
   
 \caption{\label{fig:Assy_Neu_712_VarD_1064nn_Varpos_Varpr_LE100mj} Asymmetric broadening of 712.2 nm neutral line of nickel at nearer points and different delay times for different  background pressures  for laser energy of 100 mJ . }
\end{figure*}

As mentioned, the density close to the sample at a delay of 200 ns is measured as $2\times 10^{18} cm^{-3}$ using Stark width and the temperature at 10 mm is measured using TLP as 1.5 eV. 
If interpolating these values with existing understanding and the observations elsewhere, it can be assumed that the plasma at an early stage will have higher density and temperature than the measured values at later times and longer distances. 
Hence, with the help of NIST data base\cite{NIST_ASD} we can estimate the extent of ionization. At a density of $10^{18} cm^{-3}$ and a temperature of 3 eV, as in the case of initial state of the plume, the plasma will be completely ionized and will have negligible neutral contribution ($3.5 \times 10^{-3} \%$). However, as the temperature and density fall down to $10^{17} cm^{-3}$ and 1 eV respectively as happens the case of expanding plasma plume, there will be significant amount of neutrals (6.5 \%)  present in the plasma.
 This information is helpful in assuming that the observed neutrals at early stages of plasma evolution are formed by the recombination of ions and hence have almost the same velocity profile as that of ions. For the recombination process, the density and temperature range of initial plasma plume will be favorable for TBR, fast electrons will be generated which may leave the plasma plume and generate a charge imbalance. 
 
 The peak values for slow ionic and neutral at 0.1 mbar and 100 mJ  are almost same (410 ns table~\ref{tab:PrVar_peaks_Le100mJ}) which correspond to a velocity of $7.3\times 10^{5} cms^{-1}$ . Assuming there is no significant acceleration/deceleration for this slow peak, the velocity of ions/neutrals can be assumed as the initial velocity. 
At 20 mbar, the slow ion peak advances to 280 ns corresponding to a velocity of $10.7\times 10^{5} cms^{-1}$.   Here it should be noted from the table~\ref{tab:PrVar_peaks_Le100mJ} that  the velocity of neutrals also increases with pressure for 100 mJ excitation. 
However, the drag of background gas can be seen for lower energy excitation where the velocity of neutrals decreases.  
The change in velocity of neutrals or ions can be attributed to the combined effect of acceleration attained from the DL and the drag force of the background medium throughout the distance it travels. 
For the ease of estimation of acceleration gained by the nickel ions, the drag force is ignored. The estimation of the electric field produced within the DL and the acceleration it generates is estimated as described by Eliezer et.al \cite{DL_accelaration_calc}. 
Considering the velocity of slow neutrals at 0.1 mbar as the initial velocity (not accelerated), and 1000 nm ($10-100\times \lambda_D$) as the DL thickness, the acceleration of ions responsible for the slow peak to increase the velocity to $10.7\times 10^{5} cms^{-1}$ at 20 mbar where the maximum of acceleration is seen  is estimated  as  $3\times 10^{15} cm s^{-2}$. To attain this acceleration for singly charged nickel ion, the electric field required is estimated as  $1.8\times 10^{9} Vcm^{-1}$. A similar estimation performed for other energies and background pressures also gets a value of electric field within one order of magnitude lower than the highest observed field. In this work, the observed values of electric field generated within the plume are rather high considering the fluence range we are using. The earlier reported \cite{ELIEZER_DL,ion_accelaration2} values of DL fields are also of this order, however, with a much higher laser power. \par
\section{Asymmetry in spectral emission}
\begin{figure} 
    \centering
    \begin{subfigure}{0.48\textwidth}
        \includegraphics[width=\textwidth,  trim=0cm 0cm 0cm 0cm, clip=true, angle=0]{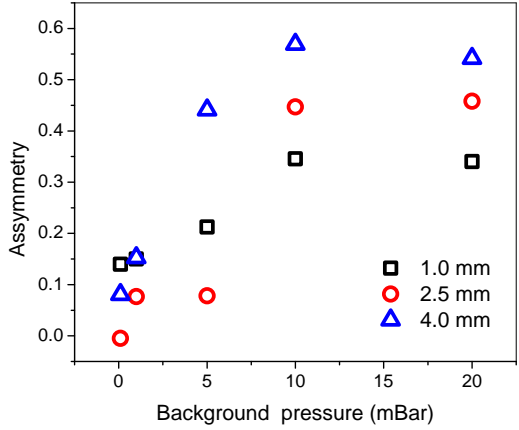}
          \caption{Asymmetry Parameter estimated. }
        \label{fig:Assy_Par_Var_Pr_200ns_712nm}
    \end{subfigure}
    \begin{subfigure}{0.48\textwidth}
        \includegraphics[width=\textwidth,  trim=0cm 0cm 0cm 0cm, clip=true, angle=0]{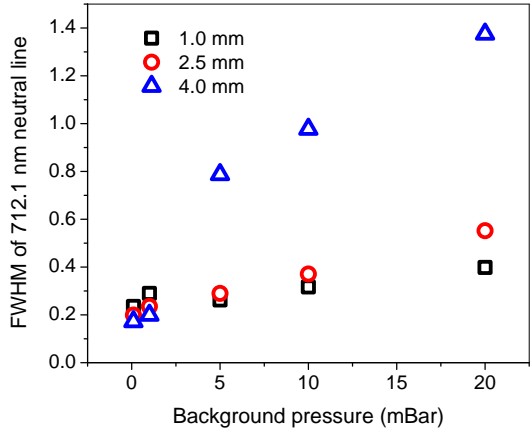}
        \caption{FWHM of spectral emission.}
        \label{fig:Assy_FWHM_Var_Pr_200ns_712nm}
    \end{subfigure}

 \caption{\label{fig:Assy_param_FWHM_Neu_712_VarD_1064nn_Varpos_Varpr_LE100mj} Asymmetry of 712.2 nm neutral line of nickel estimated as per equation~\ref{eq:assy} and the FWHM of the same line for 200 ns delay for different  background pressures and distance up to 4.5 mm  for laser energy of 100 mJ . }
 \end{figure}
 
The spectral shape of 712.2 nm line ($ 3d^9(^2D_{5/2})4p \rightarrow 3d^9(^2D)4p $ ) of neutral nickel recorded using high-resolution spectrometer in the range of few 10ns appears to show a large asymmetric broadening as the background pressure increases. Figure~\ref{fig:Assy_Neu_712_VarD_1064nn_Varpos_Varpr_LE100mj} shows the spectral shape of this line at different background pressures and for three spatial positions for three different delay times (200 ns ,300 ns and 500 ns).  Figure shows that the spectral line is broadened asymmetrically towards the red  as the background gas pressure increase. 
The NIST database shows that there is no strong emission line present for nickel (even for ions) near the red region with significant transition probability. Hence, the asymmetric broadening can be interpreted considering some perturbation in atomic levels for this specific transition. 
Interestingly other neutral lines recorded in the present work do not exhibit such asymmetry. 
As can be seen from the figure, asymmetry is strongly correlated with the spatial position and plume evolution time. 
For instance, at 200 ns for almost all the background pressures, asymmetry appears identical at a distance very close to the sample 1.0 mm. However, as the distance increases a significant variation in asymmetry is observed with increase in background pressure. Figure~\ref{fig:Assy_Neu_712_D200ns_1064nn_Varpos_Varpr_LE100mj}  shows evolution of asymmetry with background pressure at 2.5 mm, where no significant asymmetry in spectral shape is seen for 0.1 mbar of background pressure. However, a large asymmetrical broadening can be seen at 20 mbar of background pressure. As  the time of recording of emission spectrum increases to 500 ns, the spectral broadening is smallest and no significant asymmetry is observed up to 4.5 mm from the sample. \par

The asymmetry ($A_s$ )of spectral broadening at a width of $\Delta \lambda$ can be expressed using a simple formula \cite{Assymmetry_broadening}
\begin{eqnarray}
A_s(\Delta\lambda)=\frac{I_R-I_B}{I_R+I_B}
\label{eq:assy}
\end{eqnarray}
Where $I_R$ and $I_B$ are the line intensities  at wavelength separation of $\Delta \lambda$ and $- \Delta \lambda$ from the line center respectively.  A positive value of $A_s$ is indicative of an asymmetry towards the red wavelength and a negative value corresponds to the asymmetry towards blue region.   Figure~\ref{fig:Assy_param_FWHM_Neu_712_VarD_1064nn_Varpos_Varpr_LE100mj} shows the calculated  $A_s$ (figure~\ref{fig:Assy_Par_Var_Pr_200ns_712nm}) and the respective FWHM (figure~\ref{fig:Assy_FWHM_Var_Pr_200ns_712nm}) for three distances at delay time of 200 ns for a laser energy of 100 mJ (spectra shown in figure~\ref{fig:Assy_Neu_712_D200ns_1064nn_Varpos_Varpr_LE100mj}).
 As can be seen from figure~\ref{fig:Assy_Par_Var_Pr_200ns_712nm} the asymmetry is increases with background pressure and maximum asymmetry is observed at 4.0 mm from the sample. As seen in figure~\ref{fig:Assy_FWHM_Var_Pr_200ns_712nm}, at 4 mm the FWHM of the spectral line also higher as compared to other locations, an observation seen in earlier reports \cite{Jinto_POP2018} for the emission of other nickel lines as well.\par

The experimental observations indicate the role of plasma density for spectral asymmetry as the rest of all parameters like laser fluence, pulse width and wavelength, etc. remain the same in this experiment. 
The variation of background pressure in the range of few mbar is also not expected significantly influence the spectral shape due to pressure broadening.
 As already reported\cite{Jinto_POP2018}, with increase in background pressure the plasma density also increases substantially due to the confinement of plasma plume considering the drag nature of background medium, which peaks around 4-6 mm. 
 Hence, the asymmetry observed in this experiment can be correlated with the enhanced density of plasma and the spectroscopic properties of the particular line which showed the asymmetry. \par
Such asymmetry in spectral line shape has been described in literature\cite{Griem,Asymetry_sherbini,Asymmetry_review}  with very limited experimental data and the cause of the asymmetry has attributed to the \enquote{generated micro electromagnetic fields} (ion micro-field). 
In the present scenario where we observe a significantly large acceleration for the ionic species and resulting neutral species generated due to recombination, observation of asymmetry of this neutral line is further underlining the possibility of the presence of significant electric field.
Assuming the ion micro-field (possibly due to density and temperature gradients)  is the origin of asymmetry of the neutral line profile, the observed asymmetry at  later times as well as at longer distances from the sample suggests the existence of such fields for longer distances as well as longer times.
 Hence, the acceleration observed for the plasma species (ions and neutrals) with increase in background pressure, appears to be a continuous process, and hence the assumption of acceleration occurring only in the DL (of the dimension of few times $\lambda_D$) may be not valid assumption which can lead to extremely large electric fields for acceleration. 
 This simultaneous observation of ion acceleration and asymmetric broadening of neutral line is interesting and requires further detailed study with more experiments.

\section{Summery}

 The study of ionic species with varying background pressures shows a consistent evolution of the slower peak and  advancement in peaking time with an increase in background pressure.  As the background pressure increases, the temporal separation between fast and slow peaks of the ionic line decreases considerably. 
 This indicates, that there is a significant acceleration  in the case of the slower peak. The acceleration seen for the neutrals at closer distances eventually vanishes at moderate distances. 
 However, for ionic species, the acceleration can be observed for some what longer distances. 
 An interplay between the drag of background medium and acceleration is appears to be present for   different atomic species. In the case of neutrals at medium distance, the drag force dominates, however, for ionic species the acceleration is dominant  .
  As the distance further increases, the drag of background medium  for neutrals and ionic almost varies in a similar fashion. At longer distances and for low background pressure, the TLP measurements show a matching between temporal responses for the fast peak observed for ionic TOF emission. \par

The significant acceleration observed for ionic species at a closer distance can be attributed to the double layer formation within the plasma plume. The escape of the hot electrons from the plasma plume is recorded on TLP.  The acceleration shown by the ionic species is modeled to a double layer concept with an assumption of layer thickness and shows a very large electric field. The reason for such a large field for  such a low laser fluence is can be attributed to the present experimental configuration of rear ablation. 

However, the observed asymmetry of neutral line and its possible connection with the presence of micro fields indicates the possibility of continuous acceleration of ionic species within the plasma plume over longer distances and extended time as the asymmetry itself shows such a trend. 
Hence, the large acceleration seen for ionic species appears to be a continuous process in contrast to the DL concept, where the acceleration can take place within few Debye lengths. 
We believe further detailed experiments and modeling may shed more light on the observed acceleration of ionic species in this configuration. Nonetheless, the present study brings out some interesting features of ionic and neutral dynamics in forward laser ablation of nickel plasma plume in the forward ablation geometry.
\section{Bibliography}


\begin{thebibliography}{32}%
\makeatletter
\providecommand \@ifxundefined [1]{%
 \@ifx{#1\undefined}
}%
\providecommand \@ifnum [1]{%
 \ifnum #1\expandafter \@firstoftwo
 \else \expandafter \@secondoftwo
 \fi
}%
\providecommand \@ifx [1]{%
 \ifx #1\expandafter \@firstoftwo
 \else \expandafter \@secondoftwo
 \fi
}%
\providecommand \natexlab [1]{#1}%
\providecommand \enquote  [1]{``#1''}%
\providecommand \bibnamefont  [1]{#1}%
\providecommand \bibfnamefont [1]{#1}%
\providecommand \citenamefont [1]{#1}%
\providecommand \href@noop [0]{\@secondoftwo}%
\providecommand \href [0]{\begingroup \@sanitize@url \@href}%
\providecommand \@href[1]{\@@startlink{#1}\@@href}%
\providecommand \@@href[1]{\endgroup#1\@@endlink}%
\providecommand \@sanitize@url [0]{\catcode `\\12\catcode `\$12\catcode
  `\&12\catcode `\#12\catcode `\^12\catcode `\_12\catcode `\%12\relax}%
\providecommand \@@startlink[1]{}%
\providecommand \@@endlink[0]{}%
\providecommand \url  [0]{\begingroup\@sanitize@url \@url }%
\providecommand \@url [1]{\endgroup\@href {#1}{\urlprefix }}%
\providecommand \urlprefix  [0]{URL }%
\providecommand \Eprint [0]{\href }%
\providecommand \doibase [0]{http://dx.doi.org/}%
\providecommand \selectlanguage [0]{\@gobble}%
\providecommand \bibinfo  [0]{\@secondoftwo}%
\providecommand \bibfield  [0]{\@secondoftwo}%
\providecommand \translation [1]{[#1]}%
\providecommand \BibitemOpen [0]{}%
\providecommand \bibitemStop [0]{}%
\providecommand \bibitemNoStop [0]{.\EOS\space}%
\providecommand \EOS [0]{\spacefactor3000\relax}%
\providecommand \BibitemShut  [1]{\csname bibitem#1\endcsname}%
\let\auto@bib@innerbib\@empty
\bibitem [{\citenamefont {Farid}\ \emph {et~al.}(2014)\citenamefont {Farid},
  \citenamefont {Harilal}, \citenamefont {Ding},\ and\ \citenamefont
  {Hassanein}}]{Farid_pr_dpendence}%
  \BibitemOpen
  \bibfield  {author} {\bibinfo {author} {\bibfnamefont {N.}~\bibnamefont
  {Farid}}, \bibinfo {author} {\bibfnamefont {S.~S.}\ \bibnamefont {Harilal}},
  \bibinfo {author} {\bibfnamefont {H.}~\bibnamefont {Ding}}, \ and\ \bibinfo
  {author} {\bibfnamefont {A.}~\bibnamefont {Hassanein}},\ }\href {\doibase
  10.1063/1.4862167} {\bibfield  {journal} {\bibinfo  {journal} {Journal of
  Applied Physics}\ }\textbf {\bibinfo {volume} {115}},\ \bibinfo {pages}
  {033107} (\bibinfo {year} {2014})},\ \Eprint
  {http://arxiv.org/abs/https://doi.org/10.1063/1.4862167}
  {https://doi.org/10.1063/1.4862167} \BibitemShut {NoStop}%
\bibitem [{\citenamefont {Amoruso}\ \emph {et~al.}(1999)\citenamefont
  {Amoruso}, \citenamefont {Bruzzese}, \citenamefont {Spinelli},\ and\
  \citenamefont {Velotta}}]{Amoruso_1999}%
  \BibitemOpen
  \bibfield  {author} {\bibinfo {author} {\bibfnamefont {S.}~\bibnamefont
  {Amoruso}}, \bibinfo {author} {\bibfnamefont {R.}~\bibnamefont {Bruzzese}},
  \bibinfo {author} {\bibfnamefont {N.}~\bibnamefont {Spinelli}}, \ and\
  \bibinfo {author} {\bibfnamefont {R.}~\bibnamefont {Velotta}},\ }\href
  {\doibase 10.1088/0953-4075/32/14/201} {\bibfield  {journal} {\bibinfo
  {journal} {Journal of Physics B: Atomic, Molecular and Optical Physics}\
  }\textbf {\bibinfo {volume} {32}},\ \bibinfo {pages} {R131} (\bibinfo {year}
  {1999})}\BibitemShut {NoStop}%
\bibitem [{\citenamefont {Singh}\ \emph {et~al.}(2013)\citenamefont {Singh},
  \citenamefont {Gupta},\ and\ \citenamefont {Thareja}}]{RPSandTheraja}%
  \BibitemOpen
  \bibfield  {author} {\bibinfo {author} {\bibfnamefont {R.~P.}\ \bibnamefont
  {Singh}}, \bibinfo {author} {\bibfnamefont {S.~L.}\ \bibnamefont {Gupta}}, \
  and\ \bibinfo {author} {\bibfnamefont {R.~K.}\ \bibnamefont {Thareja}},\
  }\href {\doibase 10.1063/1.4846897} {\bibfield  {journal} {\bibinfo
  {journal} {Physics of Plasmas}\ }\textbf {\bibinfo {volume} {20}},\ \bibinfo
  {pages} {123509} (\bibinfo {year} {2013})},\ \Eprint
  {http://arxiv.org/abs/https://doi.org/10.1063/1.4846897}
  {https://doi.org/10.1063/1.4846897} \BibitemShut {NoStop}%
\bibitem [{\citenamefont {Bulgakova}\ \emph {et~al.}(2000)\citenamefont
  {Bulgakova}, \citenamefont {Bulgakov},\ and\ \citenamefont
  {Bobrenok}}]{DLEBulgakova}%
  \BibitemOpen
  \bibfield  {author} {\bibinfo {author} {\bibfnamefont {N.~M.}\ \bibnamefont
  {Bulgakova}}, \bibinfo {author} {\bibfnamefont {A.~V.}\ \bibnamefont
  {Bulgakov}}, \ and\ \bibinfo {author} {\bibfnamefont {O.~F.}\ \bibnamefont
  {Bobrenok}},\ }\href {www.scopus.com} {\bibfield  {journal} {\bibinfo
  {journal} {Physical Review E - Statistical Physics, Plasmas, Fluids, and
  Related Interdisciplinary Topics}\ }\textbf {\bibinfo {volume} {62}},\
  \bibinfo {pages} {5624} (\bibinfo {year} {2000})},\ \bibinfo {note} {cited By
  :144}\BibitemShut {NoStop}%
\bibitem [{\citenamefont {Chen}\ \emph {et~al.}(2010)\citenamefont {Chen},
  \citenamefont {Pukhov}, \citenamefont {Yu},\ and\ \citenamefont
  {Sheng}}]{ion_accela_film_pic}%
  \BibitemOpen
  \bibfield  {author} {\bibinfo {author} {\bibfnamefont {M.}~\bibnamefont
  {Chen}}, \bibinfo {author} {\bibfnamefont {A.}~\bibnamefont {Pukhov}},
  \bibinfo {author} {\bibfnamefont {T.-P.}\ \bibnamefont {Yu}}, \ and\ \bibinfo
  {author} {\bibfnamefont {Z.-M.}\ \bibnamefont {Sheng}},\ }\href {\doibase
  10.1088/0741-3335/53/1/014004} {\bibfield  {journal} {\bibinfo  {journal}
  {Plasma Physics and Controlled Fusion}\ }\textbf {\bibinfo {volume} {53}},\
  \bibinfo {pages} {014004} (\bibinfo {year} {2010})}\BibitemShut {NoStop}%
\bibitem [{\citenamefont {Iwata}\ \emph {et~al.}(2017)\citenamefont {Iwata},
  \citenamefont {Mima}, \citenamefont {Sentoku}, \citenamefont {Yogo},
  \citenamefont {Nagatomo}, \citenamefont {Nishimura},\ and\ \citenamefont
  {Azechi}}]{ion_accelaration_ps_foil}%
  \BibitemOpen
  \bibfield  {author} {\bibinfo {author} {\bibfnamefont {N.}~\bibnamefont
  {Iwata}}, \bibinfo {author} {\bibfnamefont {K.}~\bibnamefont {Mima}},
  \bibinfo {author} {\bibfnamefont {Y.}~\bibnamefont {Sentoku}}, \bibinfo
  {author} {\bibfnamefont {A.}~\bibnamefont {Yogo}}, \bibinfo {author}
  {\bibfnamefont {H.}~\bibnamefont {Nagatomo}}, \bibinfo {author}
  {\bibfnamefont {H.}~\bibnamefont {Nishimura}}, \ and\ \bibinfo {author}
  {\bibfnamefont {H.}~\bibnamefont {Azechi}},\ }\href {\doibase
  10.1063/1.4990703} {\bibfield  {journal} {\bibinfo  {journal} {Physics of
  Plasmas}\ }\textbf {\bibinfo {volume} {24}},\ \bibinfo {pages} {073111}
  (\bibinfo {year} {2017})},\ \Eprint
  {http://arxiv.org/abs/https://doi.org/10.1063/1.4990703}
  {https://doi.org/10.1063/1.4990703} \BibitemShut {NoStop}%
\bibitem [{\citenamefont {Shen}\ \emph {et~al.}(2019)\citenamefont {Shen},
  \citenamefont {Qiao}, \citenamefont {Zhang}, \citenamefont {Xie},
  \citenamefont {Kar}, \citenamefont {Borghesi}, \citenamefont {Zepf},
  \citenamefont {Zhou}, \citenamefont {Zhu},\ and\ \citenamefont
  {He}}]{Capacitance_Accelaration}%
  \BibitemOpen
  \bibfield  {author} {\bibinfo {author} {\bibfnamefont {X.~F.}\ \bibnamefont
  {Shen}}, \bibinfo {author} {\bibfnamefont {B.}~\bibnamefont {Qiao}}, \bibinfo
  {author} {\bibfnamefont {H.}~\bibnamefont {Zhang}}, \bibinfo {author}
  {\bibfnamefont {Y.}~\bibnamefont {Xie}}, \bibinfo {author} {\bibfnamefont
  {S.}~\bibnamefont {Kar}}, \bibinfo {author} {\bibfnamefont {M.}~\bibnamefont
  {Borghesi}}, \bibinfo {author} {\bibfnamefont {M.}~\bibnamefont {Zepf}},
  \bibinfo {author} {\bibfnamefont {C.~T.}\ \bibnamefont {Zhou}}, \bibinfo
  {author} {\bibfnamefont {S.~P.}\ \bibnamefont {Zhu}}, \ and\ \bibinfo
  {author} {\bibfnamefont {X.~T.}\ \bibnamefont {He}},\ }\href {\doibase
  10.1063/1.5088340} {\bibfield  {journal} {\bibinfo  {journal} {Applied
  Physics Letters}\ }\textbf {\bibinfo {volume} {114}},\ \bibinfo {pages}
  {144102} (\bibinfo {year} {2019})},\ \Eprint
  {http://arxiv.org/abs/https://doi.org/10.1063/1.5088340}
  {https://doi.org/10.1063/1.5088340} \BibitemShut {NoStop}%
\bibitem [{\citenamefont {Scullion}\ \emph {et~al.}(2017)\citenamefont
  {Scullion}, \citenamefont {Doria}, \citenamefont {Romagnani}, \citenamefont
  {Sgattoni}, \citenamefont {Naughton}, \citenamefont {Symes}, \citenamefont
  {McKenna}, \citenamefont {Macchi}, \citenamefont {Zepf}, \citenamefont
  {Kar},\ and\ \citenamefont {Borghesi}}]{Ion_accelaration_Skar}%
  \BibitemOpen
  \bibfield  {author} {\bibinfo {author} {\bibfnamefont {C.}~\bibnamefont
  {Scullion}}, \bibinfo {author} {\bibfnamefont {D.}~\bibnamefont {Doria}},
  \bibinfo {author} {\bibfnamefont {L.}~\bibnamefont {Romagnani}}, \bibinfo
  {author} {\bibfnamefont {A.}~\bibnamefont {Sgattoni}}, \bibinfo {author}
  {\bibfnamefont {K.}~\bibnamefont {Naughton}}, \bibinfo {author}
  {\bibfnamefont {D.~R.}\ \bibnamefont {Symes}}, \bibinfo {author}
  {\bibfnamefont {P.}~\bibnamefont {McKenna}}, \bibinfo {author} {\bibfnamefont
  {A.}~\bibnamefont {Macchi}}, \bibinfo {author} {\bibfnamefont
  {M.}~\bibnamefont {Zepf}}, \bibinfo {author} {\bibfnamefont {S.}~\bibnamefont
  {Kar}}, \ and\ \bibinfo {author} {\bibfnamefont {M.}~\bibnamefont
  {Borghesi}},\ }\href {\doibase 10.1103/PhysRevLett.119.054801} {\bibfield
  {journal} {\bibinfo  {journal} {Phys. Rev. Lett.}\ }\textbf {\bibinfo
  {volume} {119}},\ \bibinfo {pages} {054801} (\bibinfo {year}
  {2017})}\BibitemShut {NoStop}%
\bibitem [{\citenamefont {Powell}\ \emph {et~al.}(2015)\citenamefont {Powell},
  \citenamefont {King}, \citenamefont {Gray}, \citenamefont {MacLellan},
  \citenamefont {Gonzalez-Izquierdo}, \citenamefont {Stockhausen},
  \citenamefont {Hicks}, \citenamefont {Dover}, \citenamefont {Rusby},
  \citenamefont {Carroll}, \citenamefont {Padda}, \citenamefont {Torres},
  \citenamefont {Kar}, \citenamefont {Clarke}, \citenamefont {Musgrave},
  \citenamefont {Najmudin}, \citenamefont {Borghesi}, \citenamefont {Neely},\
  and\ \citenamefont {McKenna}}]{Proton_Accelaration_film}%
  \BibitemOpen
  \bibfield  {author} {\bibinfo {author} {\bibfnamefont {H.~W.}\ \bibnamefont
  {Powell}}, \bibinfo {author} {\bibfnamefont {M.}~\bibnamefont {King}},
  \bibinfo {author} {\bibfnamefont {R.~J.}\ \bibnamefont {Gray}}, \bibinfo
  {author} {\bibfnamefont {D.~A.}\ \bibnamefont {MacLellan}}, \bibinfo {author}
  {\bibfnamefont {B.}~\bibnamefont {Gonzalez-Izquierdo}}, \bibinfo {author}
  {\bibfnamefont {L.~C.}\ \bibnamefont {Stockhausen}}, \bibinfo {author}
  {\bibfnamefont {G.}~\bibnamefont {Hicks}}, \bibinfo {author} {\bibfnamefont
  {N.~P.}\ \bibnamefont {Dover}}, \bibinfo {author} {\bibfnamefont {D.~R.}\
  \bibnamefont {Rusby}}, \bibinfo {author} {\bibfnamefont {D.~C.}\ \bibnamefont
  {Carroll}}, \bibinfo {author} {\bibfnamefont {H.}~\bibnamefont {Padda}},
  \bibinfo {author} {\bibfnamefont {R.}~\bibnamefont {Torres}}, \bibinfo
  {author} {\bibfnamefont {S.}~\bibnamefont {Kar}}, \bibinfo {author}
  {\bibfnamefont {R.~J.}\ \bibnamefont {Clarke}}, \bibinfo {author}
  {\bibfnamefont {I.~O.}\ \bibnamefont {Musgrave}}, \bibinfo {author}
  {\bibfnamefont {Z.}~\bibnamefont {Najmudin}}, \bibinfo {author}
  {\bibfnamefont {M.}~\bibnamefont {Borghesi}}, \bibinfo {author}
  {\bibfnamefont {D.}~\bibnamefont {Neely}}, \ and\ \bibinfo {author}
  {\bibfnamefont {P.}~\bibnamefont {McKenna}},\ }\href {\doibase
  10.1088/1367-2630/17/10/103033} {\bibfield  {journal} {\bibinfo  {journal}
  {New Journal of Physics}\ }\textbf {\bibinfo {volume} {17}},\ \bibinfo
  {pages} {103033} (\bibinfo {year} {2015})}\BibitemShut {NoStop}%
\bibitem [{\citenamefont {Kar}\ \emph {et~al.}(2008)\citenamefont {Kar},
  \citenamefont {Borghesi}, \citenamefont {Bulanov}, \citenamefont {Key},
  \citenamefont {Liseykina}, \citenamefont {Macchi}, \citenamefont {Mackinnon},
  \citenamefont {Patel}, \citenamefont {Romagnani}, \citenamefont {Schiavi},\
  and\ \citenamefont {Willi}}]{plasmajet_kar}%
  \BibitemOpen
  \bibfield  {author} {\bibinfo {author} {\bibfnamefont {S.}~\bibnamefont
  {Kar}}, \bibinfo {author} {\bibfnamefont {M.}~\bibnamefont {Borghesi}},
  \bibinfo {author} {\bibfnamefont {S.~V.}\ \bibnamefont {Bulanov}}, \bibinfo
  {author} {\bibfnamefont {M.~H.}\ \bibnamefont {Key}}, \bibinfo {author}
  {\bibfnamefont {T.~V.}\ \bibnamefont {Liseykina}}, \bibinfo {author}
  {\bibfnamefont {A.}~\bibnamefont {Macchi}}, \bibinfo {author} {\bibfnamefont
  {A.~J.}\ \bibnamefont {Mackinnon}}, \bibinfo {author} {\bibfnamefont {P.~K.}\
  \bibnamefont {Patel}}, \bibinfo {author} {\bibfnamefont {L.}~\bibnamefont
  {Romagnani}}, \bibinfo {author} {\bibfnamefont {A.}~\bibnamefont {Schiavi}},
  \ and\ \bibinfo {author} {\bibfnamefont {O.}~\bibnamefont {Willi}},\ }\href
  {\doibase 10.1103/PhysRevLett.100.225004} {\bibfield  {journal} {\bibinfo
  {journal} {Phys. Rev. Lett.}\ }\textbf {\bibinfo {volume} {100}},\ \bibinfo
  {pages} {225004} (\bibinfo {year} {2008})}\BibitemShut {NoStop}%
\bibitem [{\citenamefont {Thomas}\ \emph {et~al.}(2018)\citenamefont {Thomas},
  \citenamefont {Joshi}, \citenamefont {Kumar},\ and\ \citenamefont
  {Philip}}]{Jinto_POP2018}%
  \BibitemOpen
  \bibfield  {author} {\bibinfo {author} {\bibfnamefont {J.}~\bibnamefont
  {Thomas}}, \bibinfo {author} {\bibfnamefont {H.~C.}\ \bibnamefont {Joshi}},
  \bibinfo {author} {\bibfnamefont {A.}~\bibnamefont {Kumar}}, \ and\ \bibinfo
  {author} {\bibfnamefont {R.}~\bibnamefont {Philip}},\ }\href {\doibase
  10.1063/1.5048834} {\bibfield  {journal} {\bibinfo  {journal} {Physics of
  Plasmas}\ }\textbf {\bibinfo {volume} {25}},\ \bibinfo {pages} {103108}
  (\bibinfo {year} {2018})},\ \Eprint
  {http://arxiv.org/abs/https://doi.org/10.1063/1.5048834}
  {https://doi.org/10.1063/1.5048834} \BibitemShut {NoStop}%
\bibitem [{\citenamefont {Kramida}\ \emph {et~al.}(2018)\citenamefont
  {Kramida}, \citenamefont {{Yu.~Ralchenko}}, \citenamefont {Reader},\ and\
  \citenamefont {{and NIST ASD Team}}}]{NIST_ASD}%
  \BibitemOpen
  \bibfield  {author} {\bibinfo {author} {\bibfnamefont {A.}~\bibnamefont
  {Kramida}}, \bibinfo {author} {\bibnamefont {{Yu.~Ralchenko}}}, \bibinfo
  {author} {\bibfnamefont {J.}~\bibnamefont {Reader}}, \ and\ \bibinfo {author}
  {\bibnamefont {{and NIST ASD Team}}},\ }\href@noop {} {}\bibinfo
  {howpublished} {{NIST Atomic Spectra Database (ver. 5.5.6), [Online].
  Available: {\tt{https://physics.nist.gov/asd}} [2018, July 4]. National
  Institute of Standards and Technology, Gaithersburg, MD.}} (\bibinfo {year}
  {2018})\BibitemShut {NoStop}%
\bibitem [{\citenamefont {Harilal}\ \emph {et~al.}(2002)\citenamefont
  {Harilal}, \citenamefont {Bindhu}, \citenamefont {Tillack}, \citenamefont
  {Najmabadi},\ and\ \citenamefont {Gaeris}}]{Harilal_plume_spliting}%
  \BibitemOpen
  \bibfield  {author} {\bibinfo {author} {\bibfnamefont {S.~S.}\ \bibnamefont
  {Harilal}}, \bibinfo {author} {\bibfnamefont {C.~V.}\ \bibnamefont {Bindhu}},
  \bibinfo {author} {\bibfnamefont {M.~S.}\ \bibnamefont {Tillack}}, \bibinfo
  {author} {\bibfnamefont {F.}~\bibnamefont {Najmabadi}}, \ and\ \bibinfo
  {author} {\bibfnamefont {A.~C.}\ \bibnamefont {Gaeris}},\ }\href {\doibase
  10.1088/0022-3727/35/22/307} {\bibfield  {journal} {\bibinfo  {journal}
  {Journal of Physics D: Applied Physics}\ }\textbf {\bibinfo {volume} {35}},\
  \bibinfo {pages} {2935} (\bibinfo {year} {2002})}\BibitemShut {NoStop}%
\bibitem [{\citenamefont {Harilal}(2001{\natexlab{a}})}]{HARILAL_spliting}%
  \BibitemOpen
  \bibfield  {author} {\bibinfo {author} {\bibfnamefont {S.}~\bibnamefont
  {Harilal}},\ }\href {\doibase https://doi.org/10.1016/S0169-4332(00)00837-0}
  {\bibfield  {journal} {\bibinfo  {journal} {Applied Surface Science}\
  }\textbf {\bibinfo {volume} {172}},\ \bibinfo {pages} {103 } (\bibinfo {year}
  {2001}{\natexlab{a}})}\BibitemShut {NoStop}%
\bibitem [{\citenamefont {Harilal}\ \emph {et~al.}(2006)\citenamefont
  {Harilal}, \citenamefont {O'Shay}, \citenamefont {Tao},\ and\ \citenamefont
  {Tillack}}]{Hari_spliting_sn}%
  \BibitemOpen
  \bibfield  {author} {\bibinfo {author} {\bibfnamefont {S.}~\bibnamefont
  {Harilal}}, \bibinfo {author} {\bibfnamefont {B.}~\bibnamefont {O'Shay}},
  \bibinfo {author} {\bibfnamefont {Y.}~\bibnamefont {Tao}}, \ and\ \bibinfo
  {author} {\bibfnamefont {M.}~\bibnamefont {Tillack}},\ }\href {\doibase
  10.1063/1.2188084} {\bibfield  {journal} {\bibinfo  {journal} {Journal of
  Applied Physics}\ }\textbf {\bibinfo {volume} {99}} (\bibinfo {year}
  {2006}),\ 10.1063/1.2188084},\ \bibinfo {note} {cited By 88},\ \Eprint
  {http://arxiv.org/abs/https://doi.org/10.1063/1.2188084}
  {https://doi.org/10.1063/1.2188084} \BibitemShut {NoStop}%
\bibitem [{\citenamefont {Mahmood}\ \emph {et~al.}(2010)\citenamefont
  {Mahmood}, \citenamefont {Rawat}, \citenamefont {Darby}, \citenamefont
  {Zakaullah}, \citenamefont {Springham}, \citenamefont {Tan},\ and\
  \citenamefont {Lee}}]{Plume_spliting_Fe_Al}%
  \BibitemOpen
  \bibfield  {author} {\bibinfo {author} {\bibfnamefont {S.}~\bibnamefont
  {Mahmood}}, \bibinfo {author} {\bibfnamefont {R.~S.}\ \bibnamefont {Rawat}},
  \bibinfo {author} {\bibfnamefont {M.~S.~B.}\ \bibnamefont {Darby}}, \bibinfo
  {author} {\bibfnamefont {M.}~\bibnamefont {Zakaullah}}, \bibinfo {author}
  {\bibfnamefont {S.~V.}\ \bibnamefont {Springham}}, \bibinfo {author}
  {\bibfnamefont {T.~L.}\ \bibnamefont {Tan}}, \ and\ \bibinfo {author}
  {\bibfnamefont {P.}~\bibnamefont {Lee}},\ }\href {\doibase 10.1063/1.3491410}
  {\bibfield  {journal} {\bibinfo  {journal} {Physics of Plasmas}\ }\textbf
  {\bibinfo {volume} {17}},\ \bibinfo {pages} {103105} (\bibinfo {year}
  {2010})},\ \Eprint {http://arxiv.org/abs/https://doi.org/10.1063/1.3491410}
  {https://doi.org/10.1063/1.3491410} \BibitemShut {NoStop}%
\bibitem [{\citenamefont {Harilal}(2001{\natexlab{b}})}]{HARILAL_C_He}%
  \BibitemOpen
  \bibfield  {author} {\bibinfo {author} {\bibfnamefont {S.}~\bibnamefont
  {Harilal}},\ }\href {\doibase https://doi.org/10.1016/S0169-4332(00)00837-0}
  {\bibfield  {journal} {\bibinfo  {journal} {Applied Surface Science}\
  }\textbf {\bibinfo {volume} {172}},\ \bibinfo {pages} {103 } (\bibinfo {year}
  {2001}{\natexlab{b}})}\BibitemShut {NoStop}%
\bibitem [{\citenamefont {{Gatsonis}}\ \emph {et~al.}(2004)\citenamefont
  {{Gatsonis}}, \citenamefont {{Byrne}}, \citenamefont {{Zwahlen}},
  \citenamefont {{Pencil}},\ and\ \citenamefont {{Kamhawi}}}]{TLP_3}%
  \BibitemOpen
  \bibfield  {author} {\bibinfo {author} {\bibfnamefont {N.~A.}\ \bibnamefont
  {{Gatsonis}}}, \bibinfo {author} {\bibfnamefont {L.~T.}\ \bibnamefont
  {{Byrne}}}, \bibinfo {author} {\bibfnamefont {J.~C.}\ \bibnamefont
  {{Zwahlen}}}, \bibinfo {author} {\bibfnamefont {E.~J.}\ \bibnamefont
  {{Pencil}}}, \ and\ \bibinfo {author} {\bibfnamefont {H.}~\bibnamefont
  {{Kamhawi}}},\ }\href {\doibase 10.1109/TPS.2004.835520} {\bibfield
  {journal} {\bibinfo  {journal} {IEEE Transactions on Plasma Science}\
  }\textbf {\bibinfo {volume} {32}},\ \bibinfo {pages} {2118} (\bibinfo {year}
  {2004})}\BibitemShut {NoStop}%
\bibitem [{\citenamefont {Kumar}\ \emph {et~al.}(2009)\citenamefont {Kumar},
  \citenamefont {Singh}, \citenamefont {Thomas},\ and\ \citenamefont
  {Sunil}}]{AjaiTLPJAP}%
  \BibitemOpen
  \bibfield  {author} {\bibinfo {author} {\bibfnamefont {A.}~\bibnamefont
  {Kumar}}, \bibinfo {author} {\bibfnamefont {R.~K.}\ \bibnamefont {Singh}},
  \bibinfo {author} {\bibfnamefont {J.}~\bibnamefont {Thomas}}, \ and\ \bibinfo
  {author} {\bibfnamefont {S.}~\bibnamefont {Sunil}},\ }\href {\doibase
  10.1063/1.3204946} {\bibfield  {journal} {\bibinfo  {journal} {Journal of
  Applied Physics}\ }\textbf {\bibinfo {volume} {106}},\ \bibinfo {pages}
  {043306} (\bibinfo {year} {2009})},\ \Eprint
  {http://arxiv.org/abs/https://doi.org/10.1063/1.3204946}
  {https://doi.org/10.1063/1.3204946} \BibitemShut {NoStop}%
\bibitem [{\citenamefont {Gurlui}\ \emph {et~al.}(2008)\citenamefont {Gurlui},
  \citenamefont {Agop}, \citenamefont {Nica}, \citenamefont {Ziskind},\ and\
  \citenamefont {Focsa}}]{gurlui_osc_PRE}%
  \BibitemOpen
  \bibfield  {author} {\bibinfo {author} {\bibfnamefont {S.}~\bibnamefont
  {Gurlui}}, \bibinfo {author} {\bibfnamefont {M.}~\bibnamefont {Agop}},
  \bibinfo {author} {\bibfnamefont {P.}~\bibnamefont {Nica}}, \bibinfo {author}
  {\bibfnamefont {M.}~\bibnamefont {Ziskind}}, \ and\ \bibinfo {author}
  {\bibfnamefont {C.}~\bibnamefont {Focsa}},\ }\href {\doibase
  10.1103/PhysRevE.78.026405} {\bibfield  {journal} {\bibinfo  {journal} {Phys.
  Rev. E}\ }\textbf {\bibinfo {volume} {78}},\ \bibinfo {pages} {026405}
  (\bibinfo {year} {2008})}\BibitemShut {NoStop}%
\bibitem [{\citenamefont {Eliezer}\ and\ \citenamefont
  {Hora}(1989{\natexlab{a}})}]{DL_S_Eliezier}%
  \BibitemOpen
  \bibfield  {author} {\bibinfo {author} {\bibfnamefont {S.}~\bibnamefont
  {Eliezer}}\ and\ \bibinfo {author} {\bibfnamefont {H.}~\bibnamefont {Hora}},\
  }\href {\doibase 10.13182/FST89-A29107} {\bibfield  {journal} {\bibinfo
  {journal} {Fusion Technology}\ }\textbf {\bibinfo {volume} {16}},\ \bibinfo
  {pages} {419} (\bibinfo {year} {1989}{\natexlab{a}})},\ \Eprint
  {http://arxiv.org/abs/https://doi.org/10.13182/FST89-A29107}
  {https://doi.org/10.13182/FST89-A29107} \BibitemShut {NoStop}%
\bibitem [{\citenamefont {Griffith}(1968)}]{Hotel_Threebody}%
  \BibitemOpen
  \bibfield  {author} {\bibinfo {author} {\bibfnamefont {W.}~\bibnamefont
  {Griffith}},\ }\href {\doibase 10.1017/S0022112068210510} {\bibfield
  {journal} {\bibinfo  {journal} {Journal of Fluid Mechanics}\ }\textbf
  {\bibinfo {volume} {31}},\ \bibinfo {pages} {826} (\bibinfo {year}
  {1968})}\BibitemShut {NoStop}%
\bibitem [{\citenamefont {Morse}\ and\ \citenamefont
  {Nielson}(1973)}]{Hotel_InvB}%
  \BibitemOpen
  \bibfield  {author} {\bibinfo {author} {\bibfnamefont {R.~L.}\ \bibnamefont
  {Morse}}\ and\ \bibinfo {author} {\bibfnamefont {C.~W.}\ \bibnamefont
  {Nielson}},\ }\href {\doibase 10.1063/1.1694445} {\bibfield  {journal}
  {\bibinfo  {journal} {The Physics of Fluids}\ }\textbf {\bibinfo {volume}
  {16}},\ \bibinfo {pages} {909} (\bibinfo {year} {1973})},\ \Eprint
  {http://arxiv.org/abs/https://aip.scitation.org/doi/pdf/10.1063/1.1694445}
  {https://aip.scitation.org/doi/pdf/10.1063/1.1694445} \BibitemShut {NoStop}%
\bibitem [{\citenamefont {Harilal}\ \emph {et~al.}(2011)\citenamefont
  {Harilal}, \citenamefont {Sizyuk}, \citenamefont {Hassanein}, \citenamefont
  {Campos}, \citenamefont {Hough},\ and\ \citenamefont
  {Sizyuk}}]{WavelengthHarilal}%
  \BibitemOpen
  \bibfield  {author} {\bibinfo {author} {\bibfnamefont {S.~S.}\ \bibnamefont
  {Harilal}}, \bibinfo {author} {\bibfnamefont {T.}~\bibnamefont {Sizyuk}},
  \bibinfo {author} {\bibfnamefont {A.}~\bibnamefont {Hassanein}}, \bibinfo
  {author} {\bibfnamefont {D.}~\bibnamefont {Campos}}, \bibinfo {author}
  {\bibfnamefont {P.}~\bibnamefont {Hough}}, \ and\ \bibinfo {author}
  {\bibfnamefont {V.}~\bibnamefont {Sizyuk}},\ }\href {\doibase
  10.1063/1.3562143} {\bibfield  {journal} {\bibinfo  {journal} {Journal of
  Applied Physics}\ }\textbf {\bibinfo {volume} {109}},\ \bibinfo {pages}
  {063306} (\bibinfo {year} {2011})},\ \Eprint
  {http://arxiv.org/abs/https://doi.org/10.1063/1.3562143}
  {https://doi.org/10.1063/1.3562143} \BibitemShut {NoStop}%
\bibitem [{\citenamefont {Hinnov}\ and\ \citenamefont
  {Hirschberg}(1962)}]{PhysRev.Recombinations}%
  \BibitemOpen
  \bibfield  {author} {\bibinfo {author} {\bibfnamefont {E.}~\bibnamefont
  {Hinnov}}\ and\ \bibinfo {author} {\bibfnamefont {J.~G.}\ \bibnamefont
  {Hirschberg}},\ }\href {\doibase 10.1103/PhysRev.125.795} {\bibfield
  {journal} {\bibinfo  {journal} {Phys. Rev.}\ }\textbf {\bibinfo {volume}
  {125}},\ \bibinfo {pages} {795} (\bibinfo {year} {1962})}\BibitemShut
  {NoStop}%
\bibitem [{\citenamefont {Maksimchuk}\ \emph {et~al.}(2000)\citenamefont
  {Maksimchuk}, \citenamefont {Gu}, \citenamefont {Flippo}, \citenamefont
  {Umstadter},\ and\ \citenamefont {Bychenkov}}]{ion_accelaration2}%
  \BibitemOpen
  \bibfield  {author} {\bibinfo {author} {\bibfnamefont {A.}~\bibnamefont
  {Maksimchuk}}, \bibinfo {author} {\bibfnamefont {S.}~\bibnamefont {Gu}},
  \bibinfo {author} {\bibfnamefont {K.}~\bibnamefont {Flippo}}, \bibinfo
  {author} {\bibfnamefont {D.}~\bibnamefont {Umstadter}}, \ and\ \bibinfo
  {author} {\bibfnamefont {V.~Y.}\ \bibnamefont {Bychenkov}},\ }\href {\doibase
  10.1103/PhysRevLett.84.4108} {\bibfield  {journal} {\bibinfo  {journal}
  {Phys. Rev. Lett.}\ }\textbf {\bibinfo {volume} {84}},\ \bibinfo {pages}
  {4108} (\bibinfo {year} {2000})}\BibitemShut {NoStop}%
\bibitem [{\citenamefont {Eliezer}\ \emph {et~al.}(1995)\citenamefont
  {Eliezer}, \citenamefont {Kolka},\ and\ \citenamefont
  {Szichman}}]{DL_accelaration_calc}%
  \BibitemOpen
  \bibfield  {author} {\bibinfo {author} {\bibfnamefont {S.}~\bibnamefont
  {Eliezer}}, \bibinfo {author} {\bibfnamefont {E.}~\bibnamefont {Kolka}}, \
  and\ \bibinfo {author} {\bibfnamefont {H.}~\bibnamefont {Szichman}},\ }\href
  {\doibase 10.1017/S0263034600009551} {\bibfield  {journal} {\bibinfo
  {journal} {Laser and Particle Beams}\ }\textbf {\bibinfo {volume} {13}},\
  \bibinfo {pages} {441} (\bibinfo {year} {1995})},\ \bibinfo {note} {cited By
  10}\BibitemShut {NoStop}%
\bibitem [{\citenamefont {Eliezer}\ and\ \citenamefont
  {Hora}(1989{\natexlab{b}})}]{ELIEZER_DL}%
  \BibitemOpen
  \bibfield  {author} {\bibinfo {author} {\bibfnamefont {S.}~\bibnamefont
  {Eliezer}}\ and\ \bibinfo {author} {\bibfnamefont {H.}~\bibnamefont {Hora}},\
  }\href {\doibase https://doi.org/10.1016/0370-1573(89)90118-X} {\bibfield
  {journal} {\bibinfo  {journal} {Physics Reports}\ }\textbf {\bibinfo {volume}
  {172}},\ \bibinfo {pages} {339 } (\bibinfo {year}
  {1989}{\natexlab{b}})}\BibitemShut {NoStop}%
\bibitem [{\citenamefont {Vuji{\v{c}}i{\'{c}}}\ \emph
  {et~al.}(1988)\citenamefont {Vuji{\v{c}}i{\'{c}}}, \citenamefont
  {Djurovi{\'{c}}},\ and\ \citenamefont {Halenka}}]{Assymmetry_broadening}%
  \BibitemOpen
  \bibfield  {author} {\bibinfo {author} {\bibfnamefont {B.~T.}\ \bibnamefont
  {Vuji{\v{c}}i{\'{c}}}}, \bibinfo {author} {\bibfnamefont {S.}~\bibnamefont
  {Djurovi{\'{c}}}}, \ and\ \bibinfo {author} {\bibfnamefont {J.}~\bibnamefont
  {Halenka}},\ }\href {\doibase 10.1007/BF01444427} {\bibfield  {journal}
  {\bibinfo  {journal} {Zeitschrift f{\"u}r Physik D Atoms, Molecules and
  Clusters}\ }\textbf {\bibinfo {volume} {11}},\ \bibinfo {pages} {119}
  (\bibinfo {year} {1988})}\BibitemShut {NoStop}%
\bibitem [{\citenamefont {Griem}(1964)}]{Griem}%
  \BibitemOpen
  \bibfield  {author} {\bibinfo {author} {\bibfnamefont {H.~R.}\ \bibnamefont
  {Griem}},\ }\href {www.scopus.com} {\emph {\bibinfo {title} {Plasma
  Spectroscopy}}}\ (\bibinfo {year} {1964})\ \bibinfo {note} {cited By
  :3232}\BibitemShut {NoStop}%
\bibitem [{\citenamefont {EL~Sherbini}\ \emph {et~al.}(2018)\citenamefont
  {EL~Sherbini}, \citenamefont {EL~Sherbini},\ and\ \citenamefont
  {Parigger}}]{Asymetry_sherbini}%
  \BibitemOpen
  \bibfield  {author} {\bibinfo {author} {\bibfnamefont {A.~M.}\ \bibnamefont
  {EL~Sherbini}}, \bibinfo {author} {\bibfnamefont {A.~E.}\ \bibnamefont
  {EL~Sherbini}}, \ and\ \bibinfo {author} {\bibfnamefont {C.~G.}\ \bibnamefont
  {Parigger}},\ }\href {\doibase 10.3390/atoms6030044} {\bibfield  {journal}
  {\bibinfo  {journal} {Atoms}\ }\textbf {\bibinfo {volume} {6}} (\bibinfo
  {year} {2018}),\ 10.3390/atoms6030044}\BibitemShut {NoStop}%
\bibitem [{\citenamefont {Oks}(2018)}]{Asymmetry_review}%
  \BibitemOpen
  \bibfield  {author} {\bibinfo {author} {\bibfnamefont {E.}~\bibnamefont
  {Oks}},\ }\href {\doibase 10.3390/atoms6030050} {\bibfield  {journal}
  {\bibinfo  {journal} {Atoms}\ }\textbf {\bibinfo {volume} {6}},\ \bibinfo
  {pages} {50} (\bibinfo {year} {2018})}\BibitemShut {NoStop}%
\end{thebibliography}

%

\end{document}